\documentclass[useAMS,usenatbib]{mn2e}
\usepackage{graphicx}
\usepackage{amsmath}
\usepackage{epstopdf}
\usepackage{grffile}
\usepackage{amssymb}
\usepackage{lscape}
\usepackage{url}
\newcommand{\Om}{\Omega_{\rm m}}
\newcommand{\Ob}{\Omega_{\rm b}}
\newcommand{\OL}{\Omega_{\Lambda}}

\newcommand{\powlt}{\alpha_{\rm LT}}
\newcommand{\normlt}{C_{\rm LT}}
\newcommand{\zevollt}{\gamma_{\rm z,LT}}
\newcommand{\scattlt}{\sigma_{lnL|T}}

\newcommand{\powmt}{\alpha_{\rm MT}}
\newcommand{\normmt}{C_{\rm MT}}
\newcommand{\zevolmt}{\gamma_{\rm z,MT}}
\newcommand{\scattmt}{\sigma_{lnT|M}}

\newcommand{\crate}{{\rm CR}}
\newcommand{\hratio}{{\rm HR}}

\newcommand{\mes}{{\rm mes}}

\newcommand{\dndcrdhr}{dn/d{\rm CR}/d{\rm HR}}

\newcommand{\dndcrdhrfrac}{\frac{dn}{d{\rm CR}\,d{\rm HR}}}

\newcommand{\obs}{\mathcal{O}}

\newcommand{\xc}{x_{c,0}}

\newcommand{\mdcb}{M_{\rm 200b}}
\newcommand{\mdcc}{M_{\rm 200c}}

\newcommand{\rccc}{R_{\rm 500c}}

\newcommand{\rdcb}{R_{\rm 200b}}

\newcommand{\nh}{{\rm N}_{\rm H}}

\title[XCLASS: catalogue construction and analysis]{The cosmological analysis of X-ray cluster surveys: II-~Application of the CR-HR method to the XMM archive}
\author[N. Clerc et al.]{N. Clerc$^1$\thanks{Present e-mail: nclerc@mpe.mpg.de (MPE/Garching)}, T. Sadibekova$^1$, 
M. Pierre$^1$, F. Pacaud$^2$, J.-P. Le F\`evre$^3$, C.~Adami$^4$, 
\newauthor B.~Altieri$^5$, I.~Valtchanov$^5$\\
$^{1}$Laboratoire AIM, CEA/DSM/IRFU/SAp, CEA Saclay, 91191 Gif-sur-Yvette, France.\\
$^{2}$Argelander-Institut f\"ur Astronomie, University of Bonn, Auf dem H\"ugel 71, 53121 Bonn, Germany.\\
$^{3}$CEA/DSM/IRFU/SEDI, CEA Saclay, 91191 Gif-sur-Yvette, France.\\
$^{4}$LAM, OAMP, Universit\'e Aix-Marseille \& CNRS, 38 rue Fr\'ed\'eric Joliot-Curie, 13388 Marseille 13 Cedex, France.\\
$^{5}$ESAC, Villafranca del Castillo, Spain.}

\begin{document}

\date{Accepted 2012 April 20. Received 2012 March 28; in original form 2011 September 20}

\pagerange{\pageref{firstpage}--\pageref{lastpage}} \pubyear{2002}

\maketitle

\label{firstpage}

\begin{abstract}
We have  processed 2774 high-galactic observations from the XMM archive (as
of May 2010) and extracted a serendipitous catalogue of some 850 clusters of galaxies
based on purely X-ray criteria, following the methodology developed for the
XMM-LSS survey. Restricting the sample to the highest signal-to-noise
objects (347 clusters), we perform a cosmological analysis using the X-ray
information only. The analysis consists in the modelling of the observed
colour-magnitude (CR-HR) diagram constructed from  cluster instrumental
count-rates  measured in the [0.5-2], [1-2] and [0.5-1] keV bands. A MCMC
procedure simultaneously fits the cosmological parameters, the evolution of
the cluster scaling laws and the selection effects.\\
Our results are consistent with the $\sigma_{8}$ and $\Omega_{m}$  values
obtained by WMAP-5 and point toward a negative evolution of the cluster
scaling relations with respect to the self-similar expectation. We are
further able to constrain the cluster fractional radius $\xc= r_c/\rccc$, to $\xc = 0.24
\pm 0.04$. This study stresses again the critical role of selection effects in
deriving cluster scaling relations, even in the local universe. Finally, we
show that CR-HR method applied to the eRosita all-sky survey - provided that
cluster photometric redshifts are available - will enable the determination
of the equation of state of the dark energy at the level of the DETF stage
IV predictions; simultaneously, the evolution of the cluster scaling-relations will be
unambiguously determined. \\
The XMM CLuster Archive Super Survey (XCLASS) serendipitous cluster catalogue is available online at:  
http://xmm-lss.in2p3.fr:8080/l4sdb/.
\end{abstract}


\section{Introduction}
Clusters of galaxies, the most massive bound objects in the universe, are  the direct products of the growth of cosmic structures.
Using cluster samples in cosmological analyses requires not only to  
span a large range of redshifts and masses. It is also mandatory to precisely  
understand how those objects have been selected and how the selection  
is related to the cosmological distribution of dark matter haloes, the only quantity handled by the theory.  
Indeed, understanding selection processes turned out to be one of the main  
challenges of today's cluster cosmology and is intimately related to our ability to adequately determine cluster scaling-relations along with the associated dispersion.  In this respect, X-ray surveys are potentially extremely powerful, given that the X-ray properties of the cluster population can be derived from {\em ab initio} models. Substantial efforts have been devoted  to assemble statistically significant cluster samples with the past generation of X-ray observatories:  \citep[e.g.][]{Scharf:1997p6157,  
Vikhlinin:1998p6138, Jones:1999p5930, Bohringer:2000p1456, Borgani:2001p4880, Bohringer:2004p1254, Burenin:2007p1251}. Ten years ago, XMM opened a new era in  
cluster surveys,  allowing us to access and to characterise  clusters,  much  fainter  than enabled by e.g., ROSAT. Nowadays, cluster serendipitous searches   in the  
XMM archive arouse a growing interest thanks to the wealth of pointed observations,  publicly available : The XCS survey \citep{Romer:2001p6300},  
launched a decade ago is now delivering its first X-ray selected  
catalogues of clusters   \citep{LloydDavies:2010p4565, Mehrtens:2011p4564}. Several  
other projects are being conducted such as SExclass  
\citep{Kolokotronis:2006p6304} and combined searches with Chandra  
archival data \citep{Peterson:2009p1459} or SDSS optical data  
\citep{Takey:2011p6467}.

This paper is the second of a series describing a novel approach to the cosmological  
interpretation of cluster number counts in X-ray cluster surveys.  
The first paper \citep[][hereafter paper I]{paperI} investigated the constraining power of a method  based on the analysis of instrumental X-ray observables, namely 
a count-rate (CR) and a hardness-ratio (HR) in well defined X-ray bands. The combination of the two quantities was shown   to  
reliably describe the surveyed cluster population:  the corresponding CR-HR  statistical distribution, which is analogous to a colour-magnitude diagram, can be  
fully predicted by an {\it ab initio} modeling involving the cosmology, cluster scaling-relations,  the survey  selection  
effects along with the XMM instrumental response.
In this paper, we present an independent analysis of 2774 high-galactic latitude observations  
from the XMM archive  having effective exposure times of 10 and 20~ks. Following a  
selection procedure adapted from the XMM-LSS survey \citep{Pierre:2007p3253},  
we detect 845 C1 galaxy clusters  
\citep{Pacaud:2007p250}, hence constituting the X-CLASS catalogue (XMM Cluster  
Archive Super Survey). We apply the CR-HR method to a subsample of  
347 clusters selected for their high signal-to-noise ratio over an effective geometrical area of 90~sq.deg. We devote special 
 care to the count-rate measurements and to  the   
derivation of the selection function associated to this heterogeneous archival data  
 
The structure of this paper is as follows. We first present the steps  
leading to the creation of the X-CLASS cluster sample (Sect.~\ref{data}). In a second part (Sect.~\ref{simulations}) we describe   the derivation of 
 the survey selection function and how we account for the  
presence of pointed clusters in the sample. In Sect.~\ref{cosmological_analysis} and~\ref{results} we present the results of the analysis of the CR-HR distribution.  We discuss our results in Sect.~\ref{discussion} and present some cosmological predictions for the eRosita all-sky survey. Summary and conclusion are gathered a  in Sect.~\ref{conclusion}.
%
%

\section{The X-CLASS catalogue}
\label{data}

The XMM CLuster Archive Super Survey (X-CLASS) is based on the analysis of archival observations from the XMM-Newton observatory. It is intended to provide a sample of several hundreds of clusters suitable for cosmological studies.
In this section, we describe the selection of the original data and the adopted  methodology for detecting and characterizing galaxy clusters.

	\subsection{Pre-selection of the XMM archival data}

As of May, 26th 2010, 7716 individual observations were listed in the XMM Science Archive system \citep{Arviset:2003p5561}. Out of these, we retrieved 2774 observations  selected as follows:
\begin{enumerate}
\item pointing center at high galactic latitude ($\vert b \vert \ge 20 \deg$) to minimize the effect of galactic absorption,
\item total exposure time greater than 5 ks,
\item all three detectors (MOS1, MOS2 and PN) in imaging mode and at least one of them in Full Frame mode,
\item pointing center not closer than 5 $\deg$ to the Small and Large Magellanic Clouds, and 2 $\deg$ away from M31,
\item public data (as of May, 26th 2010)
\end{enumerate}

\begin{figure}
	\includegraphics[width=84mm]{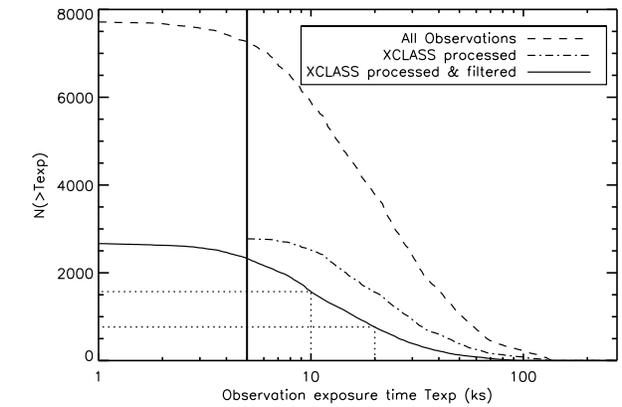}
 \caption{Cumulative exposure distribution  of the XMM archival observations. {\it Dashed line:} nominal exposure time for all 7716 observations.
 	{\it Dot-dashed line:} nominal exposure time for the 2774 retrieved  observations .
 	{\it Plain line:} effective ``clean" exposure time after processing and background flare removal.}
 \label{fig_texp_distrib} 
\end{figure}

\begin{figure}
	\includegraphics[width=84mm]{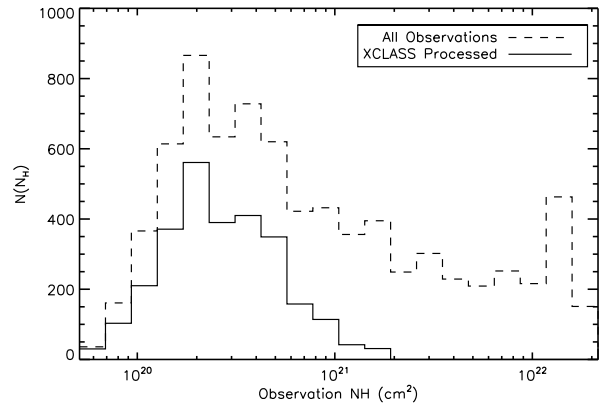}
 \caption{Distribution of hydrogen column densities,
 	{\it (dashed line:)} for the 7716  XMM archival observations,
	{\it (plain line:)} for the 2774 pre-processed observations.
		}
 \label{fig_nh_distrib} 
\end{figure}

\begin{figure*}
	\includegraphics[width=\linewidth]{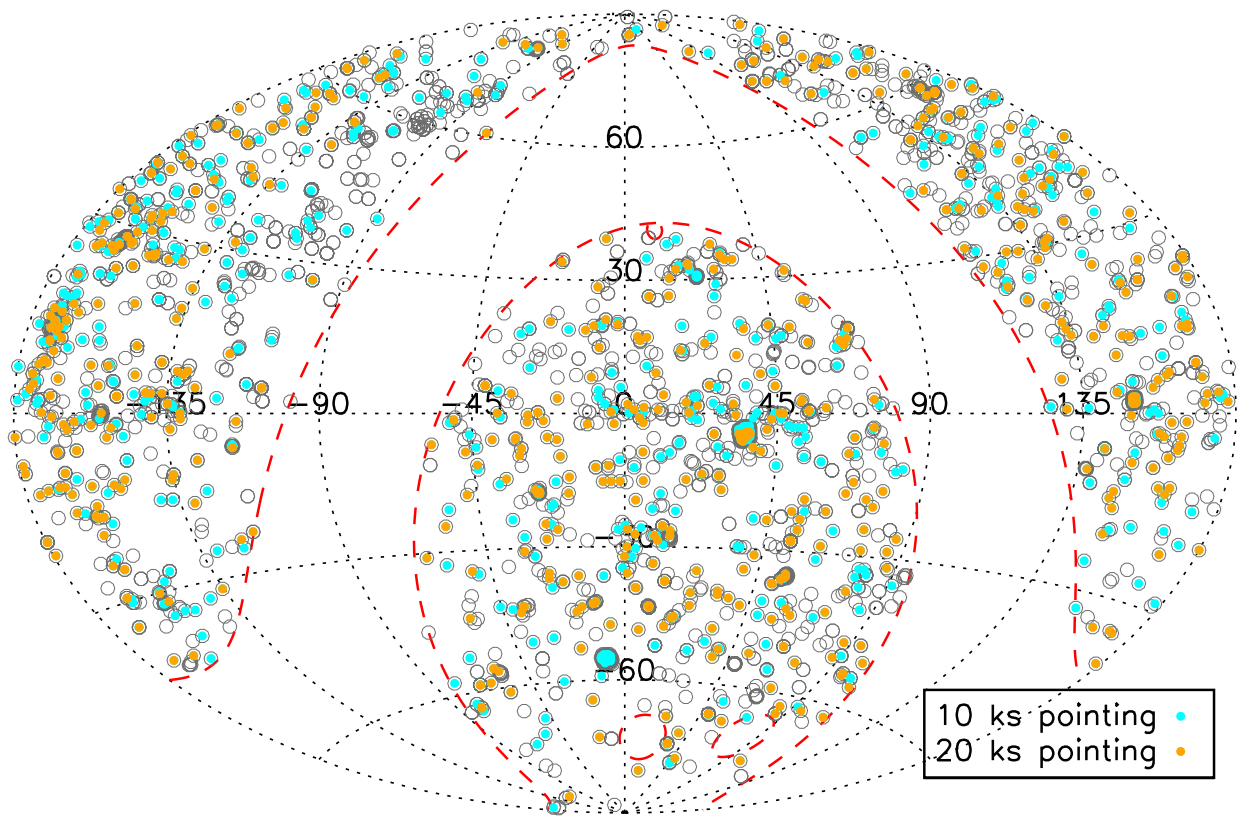}
 \caption{Sky location of the 2774 pre-processed XMM observations (equatorial coordinates).
 	Observations shorter than 10\,ks  (open circles) were not considered for the X-CLASS catalogue. The other pointings have been truncated to 10\,ks (blue points) and, if possible, to 20\,ks (gold points). Note that, by construction, a gold point (20\,ks pointing) always has a 10\,ks counterpart.
	The circle size is not representative of the XMM field of view ($\sim 30$\,arcmin diameter).}
 \label{fig_pointing_map} 
\end{figure*}

This selection contains in particular 92 observations from the 10~sq.deg. XMM--LSS survey \citep{Pierre:2007p3253}.
Figure~\ref{fig_texp_distrib} presents the statistics of the  exposure time distributions.
Figure~\ref{fig_nh_distrib} shows the distribution of hydrogen column density ($\nh$)  on the corresponding lines of sight. Column densities values were obtained through the Leiden/Argentine/Bonn $\nh$ maps \citep{Kalberla:2005p5566}. Processed pointings show a median $\nh$ of $\sim3.10^{20}$~cm$^{-2}$ and very few of them lie in regions above $10^{21}$~cm$^{-2}$.
We display on Fig.~\ref{fig_pointing_map} the sky distribution of the processed observations.  

	\subsection{Data processing and cluster detection}
	\label{data_processing}

		\subsubsection{Processing steps}
The individual Observation Data Files (ODF) pertaining to each observation were retrieved via the ESA Archive InterOperability system (XSA\footnote{http://xmm.esac.esa.int/xsa/}).
Our processing is   entirely based on the XMM-LSS pipeline \citep{Pacaud:2006p3257}  which  main steps are summarized below.

\begin{enumerate}

	\item Event lists are generated using XMM-SAS tasks {\verb emproc } and {\verb epproc } and filtered from proton and solar flares. This is achieved by creating the high energy events light curves ($12-14$\,keV for MOS and $10-12$\,keV for PN) and flagging out periods of high event rates (rates greater than $3~\sigma$ above the mean observation count-rate). Although adequate for removing short periods of high flares, this procedure may provide unsatisfactory results for  observations  having a   high mean particle background. The overall quality of each observation was then subsequently inspected by eye, and  some observations  discarded (Sect.~\ref{catalogue}).

	\item Clusters detected with XMM exposure times of $\sim$10-20\,ks constitute the most relevant population for cosmological studies \citep{Pierre:2007p3253, Pierre:2011p5484}; such exposure times  are well  above the XMM confusion limit. Further, the selection function of a survey consisting of homogeneous exposures  is  easier to handle.
	Using the {\it good time intervals} (GTI) resulting from the pre-processing, we thus build $10$\,ks and $20$\,ks ``chunks" (now denominated ``pointings") from the original exposures, starting at the beginning of the observation.
	From each observation 0, 1 or 2 {\it pointings} are extracted, each pointing having exactly a 10 or 20\,ks exposure time on the three detectors. The case of  ``0 pointing" occurs if one or more detector is insufficiently exposed, which corresponds to highly flared or problematic observations. Such observations are discarded from the analysis, and the total number of pointings entering the source detection process is 2409 (Table~\ref{table_pointings_numbers}). We display in Fig.~\ref{fig_pointing_map} the sky location of all pointings having a duration of exactly 10 or 20\,ks  and that subsequently undergo the source detection process. The archival GTI time used for cluster detection amounts thus to 24\,Ms, over the 40\,Ms GTI time available in total. In addition to these chunks, we  construct a ``full exposure" pointing containing the maximal GTI time from each observation, which will be used to obtain high signal-to-noise flux measurements for the detected clusters (Sect.~\ref{fluxmes}).

	\item For each observation, three images are created in the $[0.5-2]$\,keV band,  for the three EPIC detectors, which are in turn   co-added. The resulting image is subsequently filtered in the wavelet space, assuming a Poisson noise model  ({\verb mr_filter }, \citealt{Starck:1998p5573, Valtchanov:2001p6391}), and sources are extracted running {\scshape SExtractor} \citep{Bertin:1996p5580} over the filtered image. Only sources detected within the inner 13 arcmin radius FoV  are considered in the  subsequent analysis. 	
	
	\item Each detected source is further characterized by a maximum likelihood profile fitting procedure ({\scshape XAmin}, \cite{Pacaud:2006p3257}). Two source models are tested on each detection: (1) a point-like model using the analytical PSF  from the XMM calibration database, with the position held at the {\scshape SExtractor} output value, thus allowing two parameters to vary (the source count-rates on the MOS and PN detectors) ; and  (2) a $\beta$-model, convolved by the PSF, with five free parameters ((X,Y) position, core radius extent and   MOS and PN count-rates). A uniform background is added, whose level is such that the total number of photons in the data equals that of the modelled source plus background. We use the C-statisic \citep{Cash:1979p927} for quantifying the likelihood of the fits and, finally,  discriminating between the two types of sources.
\end{enumerate}

This procedure, when coupled to a representative set of simulations, readily enables the selection of sub-samples of X-ray extended sources with well characterised levels of completeness and contamination. It is thus perfectly suited to the analysis of large X-ray data sets such as the XMM archive. We note, however,  that the {\scshape XAmin} pipeline was originally designed to detect and discriminate between point-like and extended sources in the XMM-LSS ``empty'' cosmological fields \citep{Pierre:2007p3253}; consequently, because of the variety of astronomical objects present in the XMM archive (nearby galaxies, substructures in clusters, planets...) a subsequent human screening is necessary (see Sect.~\ref{catalogue}) in order to remove mis-interpreted detections.

\begin{table}
	\centering
\caption{\label{table_pointings_numbers}
Number of XMM archival observations handled for the present study. First line, all available observations. Second line, retrieved observations. Third line, number of usable observations after event filtering. Last line, final number of retained observations after discarding pathological cases (see. Fig.  \ref{fig_excluded_pointings})	
}
		\begin{tabular}{@{}lc@{}}
\hline
XMM observations (May, 26th 2010)	&	7716	\\
Observations retrieved from the XSA	&	2774	\\
Pointings "chunks" 10 + 20\,ks	&	2409 (1588 + 821)	\\
Pointings entering the cosmological analysis	&	1992 (1294 + 698)	\\
\hline
		\end{tabular}
\end{table}

	\subsubsection{Output parameters and source characteristics}

The procedures described above allow us to assign to each detected source a set of parameters characterising its properties:   position on the detector,  off-axis distance, sky  coordinates, count-rates in various energy bands,  plus three numbers related to the chosen fitting algorithm; (a) the detection likelihood (DET\_ML) gives the significance of the detection as compared to a pure background fluctuation; (b) the angular extent (EXT) is the apparent core radius of the best-fit $\beta$-model; (c) the extent statistic (EXT\_LIKE) compares the significances of the `extended model' and the `point-like model' and is thus called the source Extent Likelihood.  These parameters can be easily related to the simulations intended to assess the survey selection function (see Sect.~\ref{simulations}). Because of their relevance, these values are listed in the final cluster catalogue (App.~\ref{catalogue_details}).

\begin{figure}
	\includegraphics[width=84mm]{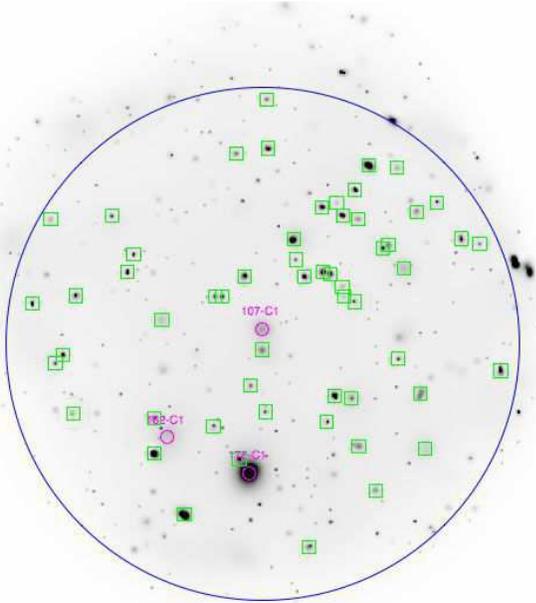}
 \caption{Example of wavelet filtered image with detected sources (ObsID: 0403072201).}
 \label{fig_xamin_exemple} 
\end{figure}

Following \citep{Pacaud:2006p3257}, we denote by ``C1" ,  sources characterised by EXT\_LIKE\,$>$\,33 and EXT\,$>$\,5\,arcsec : this corresponds to a sample of extended sources having a very low level of contamination by misclassified point-sources. We further show in Sect.~\ref{simulations} that this parameter combination can be applied regardless of the pointing intrinsic properties.
Fig.~\ref{fig_xamin_exemple} displays the pipeline detections over one XMM archival pointing (ObsID: 0403072201), containing three C1 candidates and $\sim 60$ point-like sources, most of them being AGNs.

%
%

	\subsection{Catalogue construction}
	\label{catalogue}
	
		\subsubsection{Removal of duplicates}
Only the high-quality C1 clusters are inserted into the final catalogue (App.~\ref{catalogue_details}). Because of the multiple overlaps between the archival observations, several sources are detected in more than one pointing. In particular, if an observation has been split in two pointings (10 and 20\,ks), almost all sources detected on the 10\,ks pointing are also found on the deeper one (26 over the 845 extended detections  were not in this case, most of them being close to the detection limit and four being nearby, bright clusters with a very peaked profile, mis-classified as point-sources). Furthermore, because of the presence of the CCD gaps and/or   of multiple maxima in the emission of widely extended sources, multiple detections of the same object occured.

We first associated sources closer than $20$ arcsec to each other. To decide which of the two sources has to be included in the final catalogue, we applied the following rules: If the two sources are on different pointings, the one lying on the deeper pointing prevails. If both detection lie on the same pointing or on different pointings having the same exposure time, the source with the lowest Extent Likelihood is discarded.
Each association was controlled by eye to avoid the matching of close, unrelated sources (e.g. a cluster and a background AGN). Note that positional differences of a few tens of arcsec are possible for extended sources located at the edge of the FoV, mainly because part of the emission is missing and because of the distorted shape of the PSF.
This procedure was thus repeated with larger correlation radii, until each  catalogue entry was  related to a  unique source.

		\subsubsection{Data screening and final selection}
\begin{figure*}
	\begin{tabular}{cc}
		\includegraphics[width=70mm]{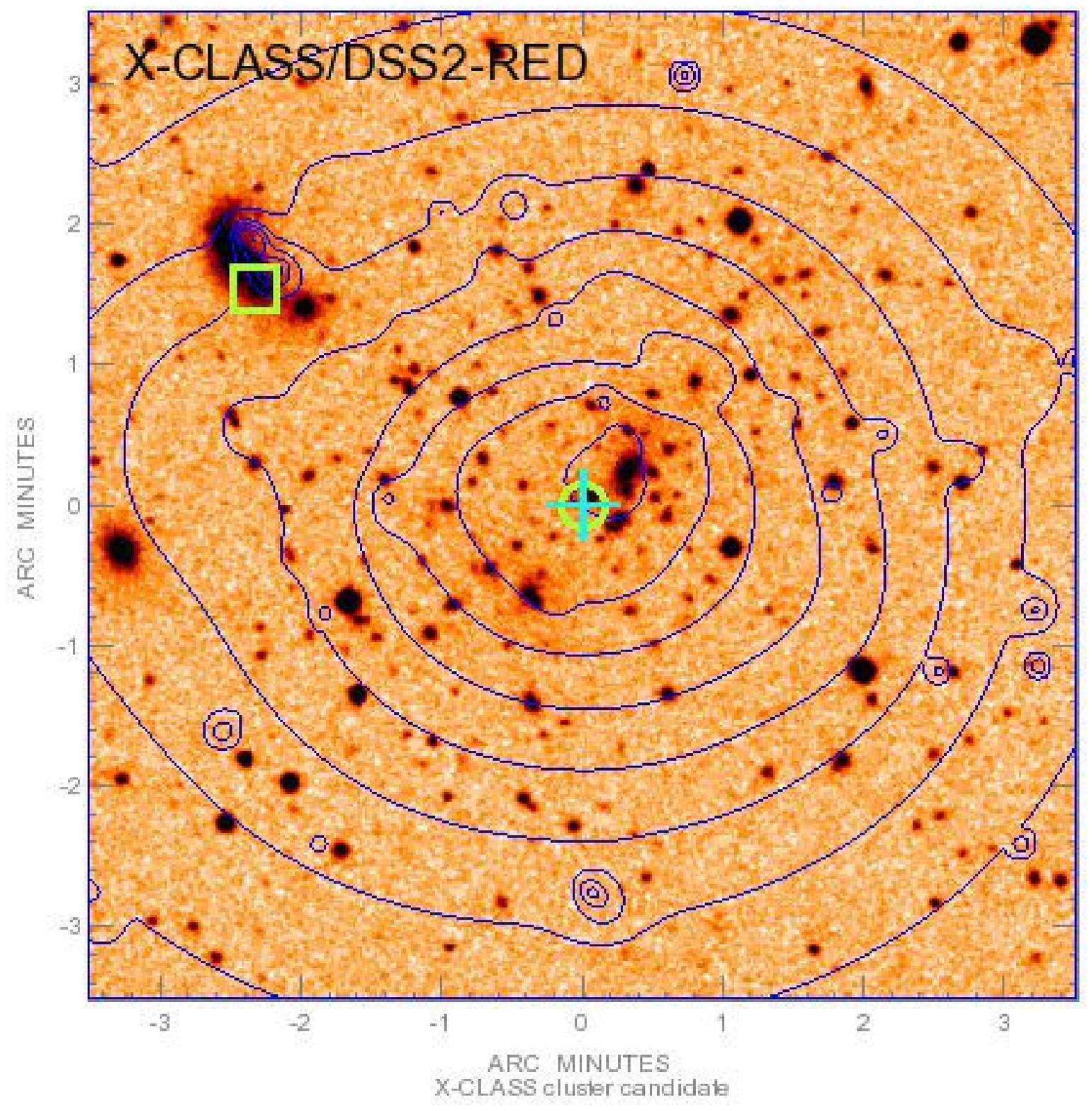} &
		\includegraphics[width=70mm]{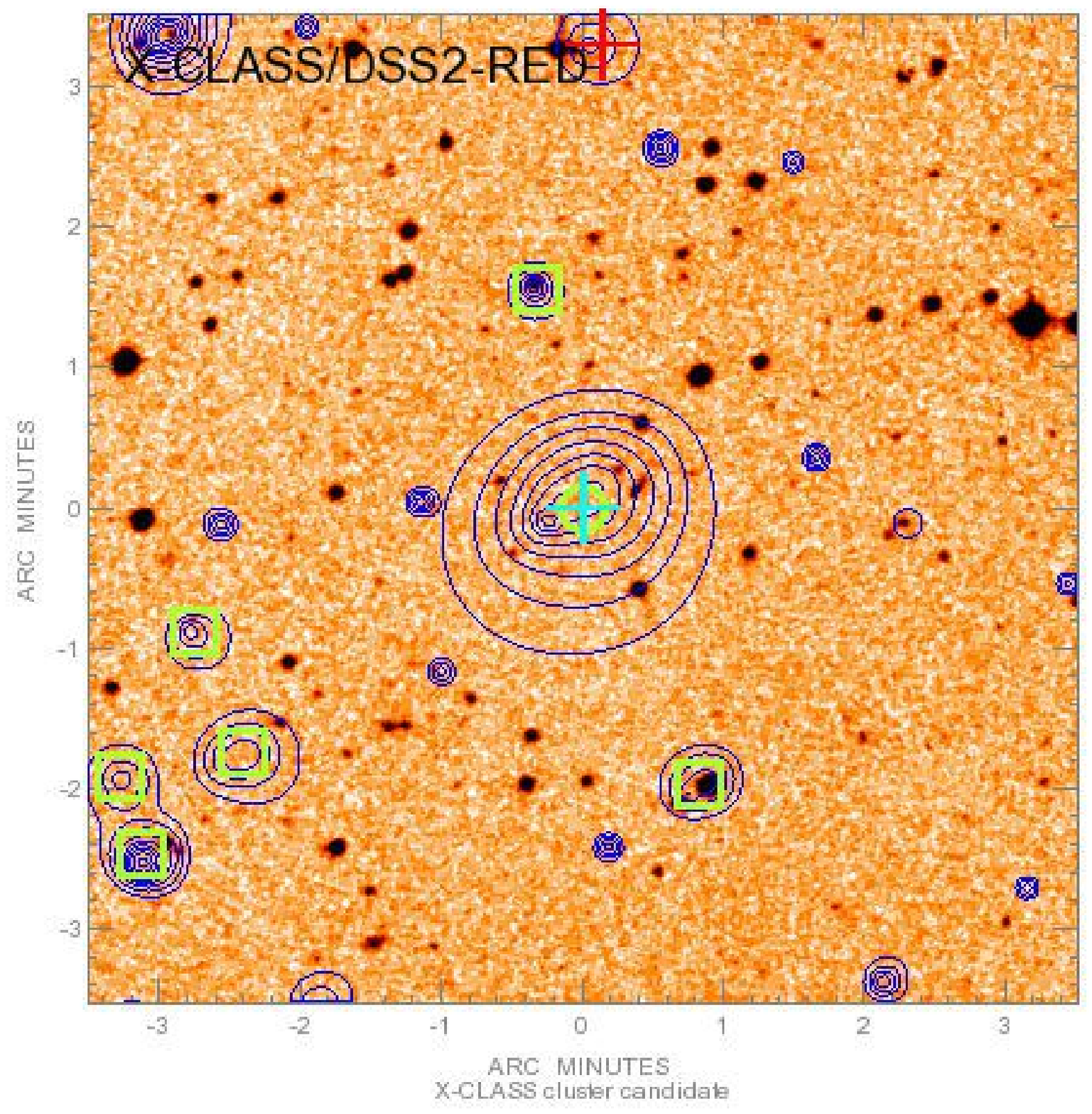} \\
		\includegraphics[width=70mm]{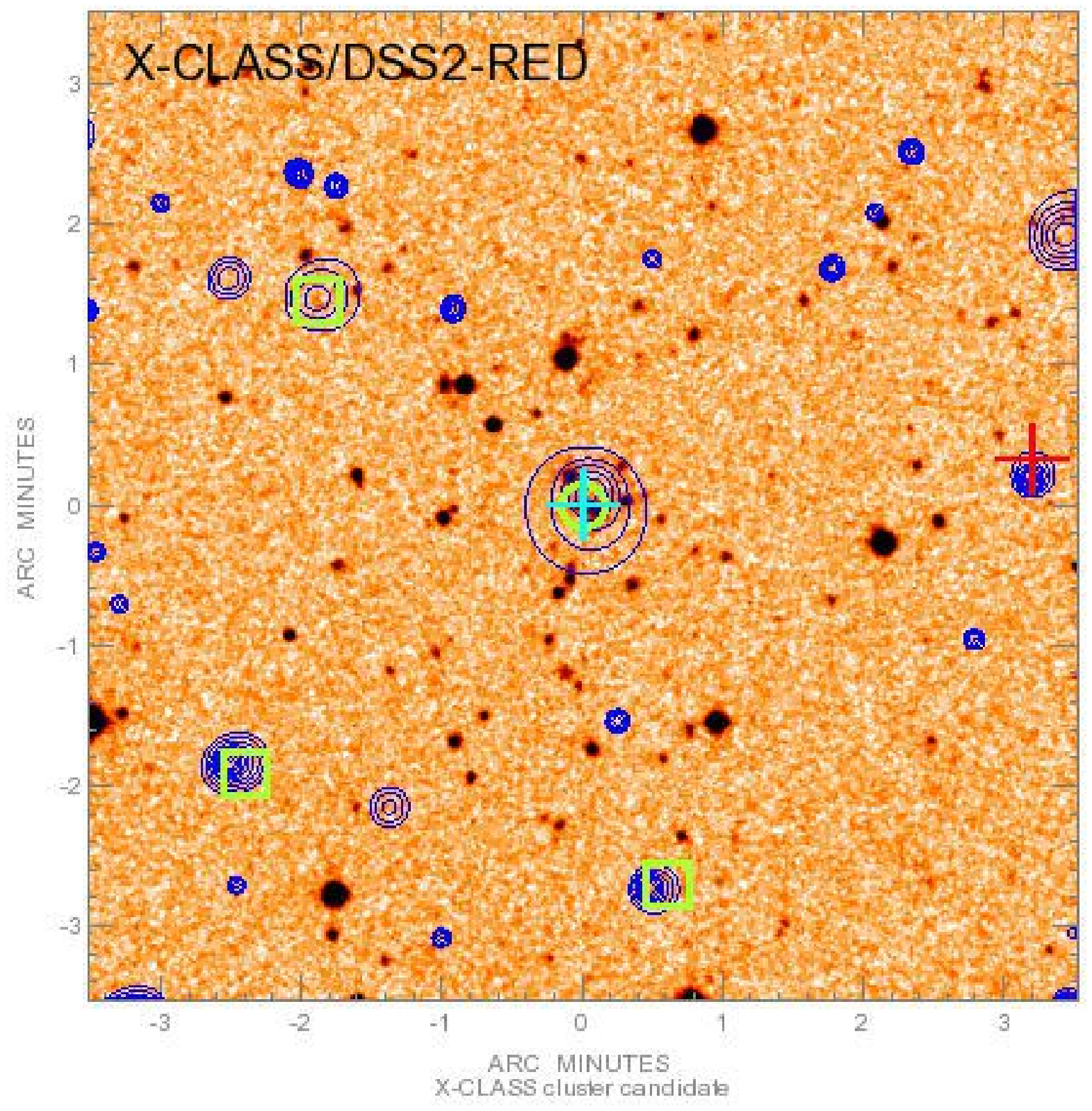} &
		\includegraphics[width=70mm]{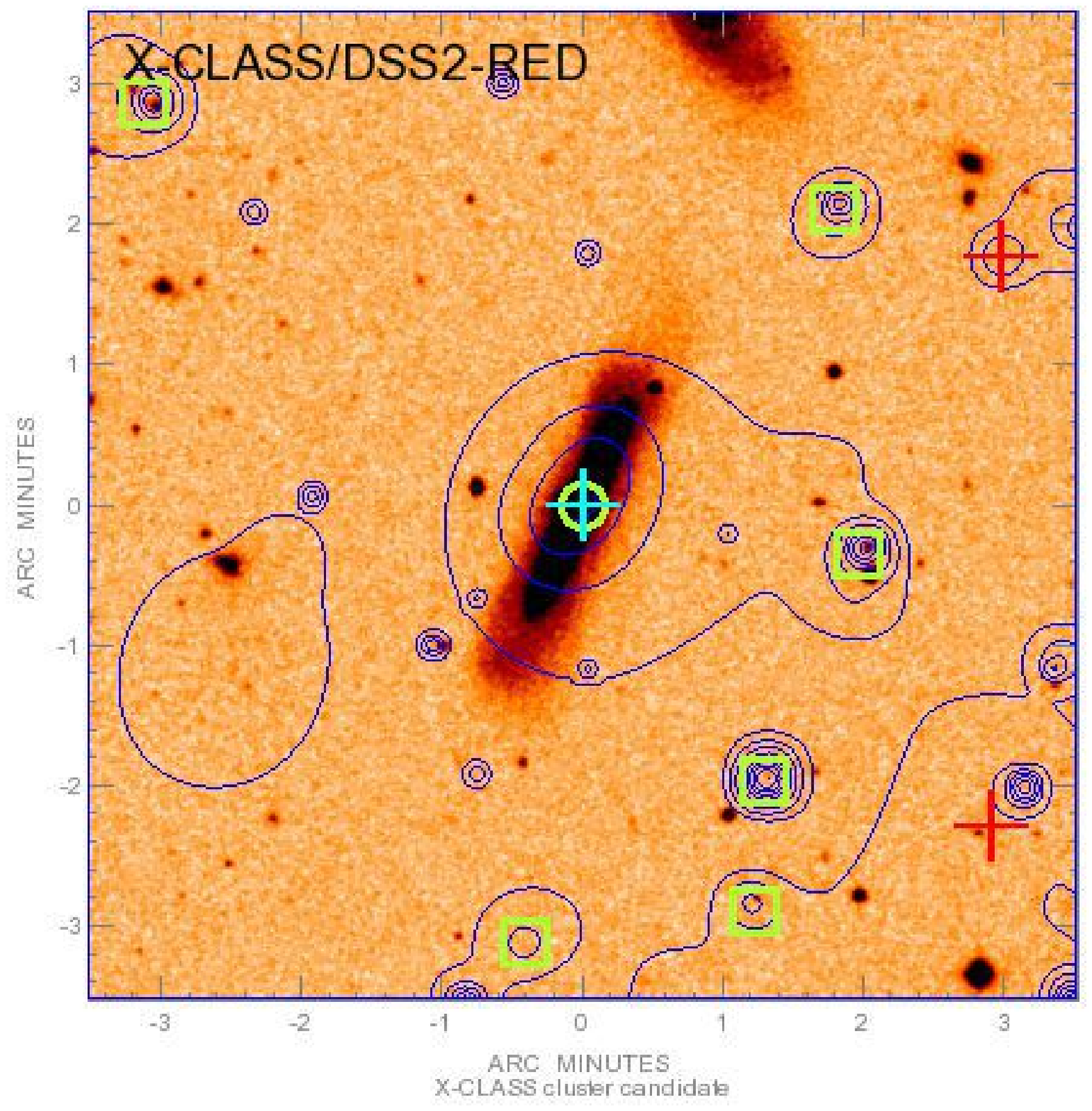}	\\
		\includegraphics[width=70mm]{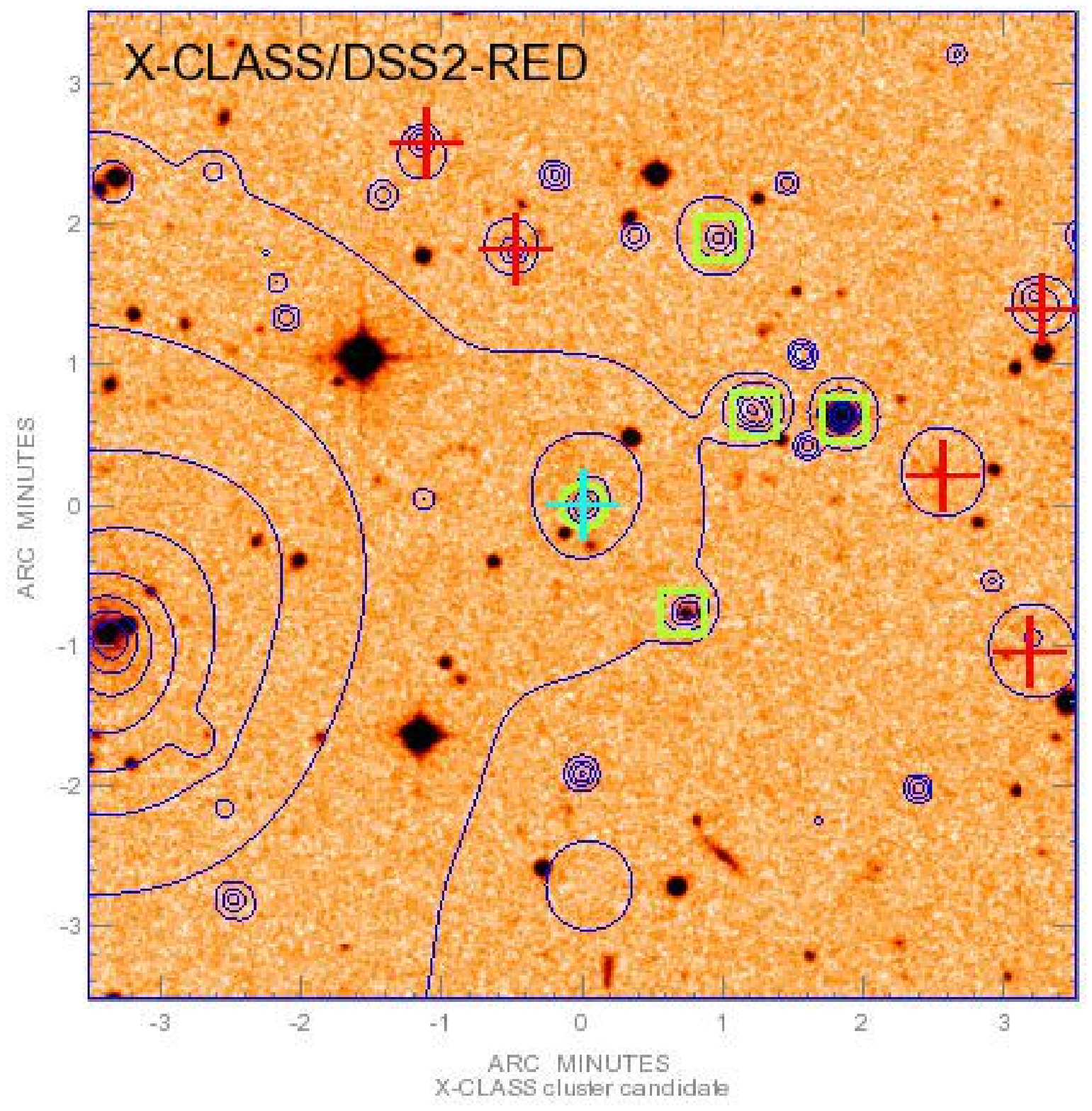}	&
		\includegraphics[width=70mm]{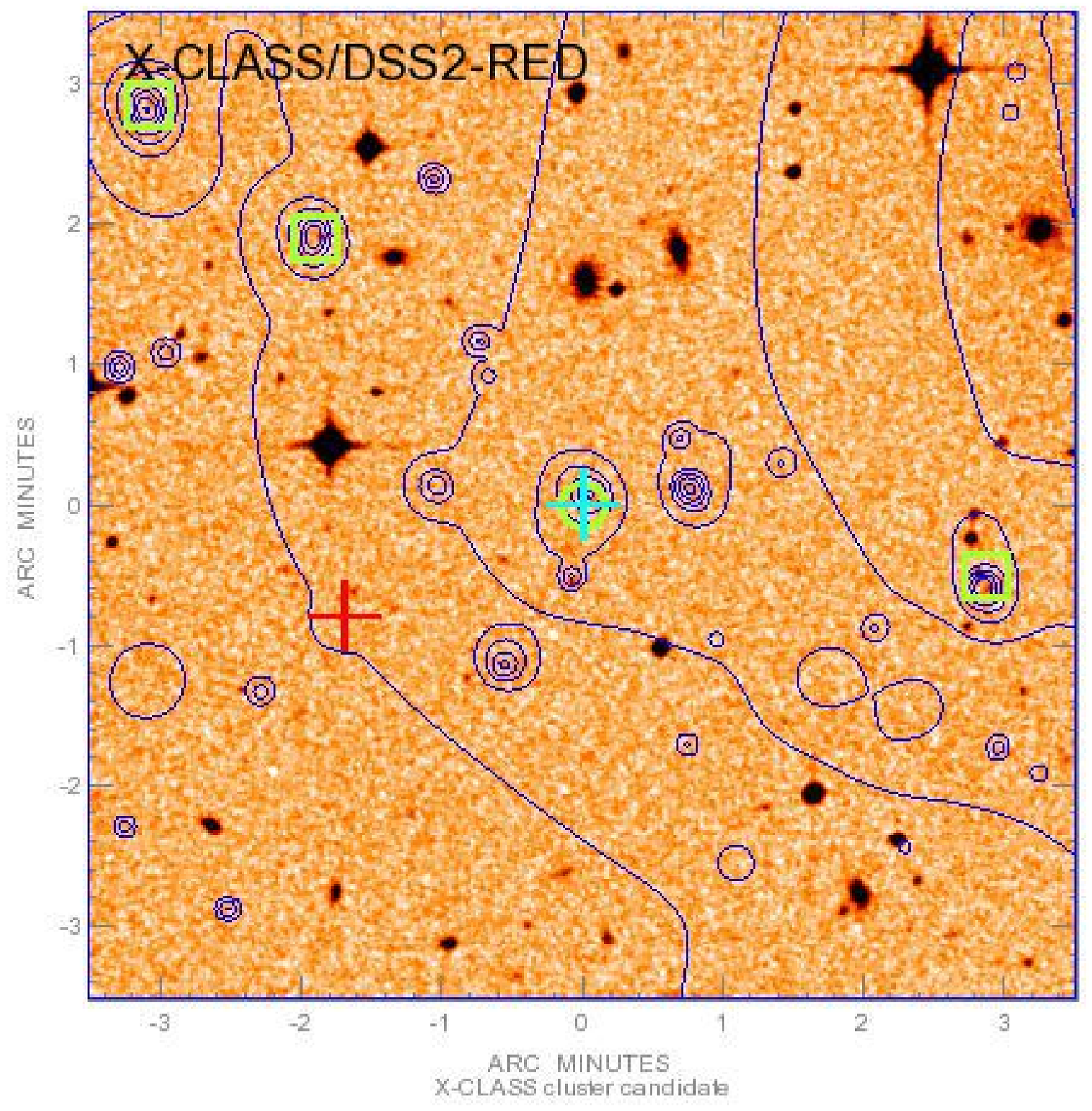}	\\
	\end{tabular}
 \caption{Illustration of the data screening classification (Sect.~\ref{catalogue}).
 From left to right, top to bottom : classification "$0<z<0.3$" (known cluster A2218 z=0.176), "$z>0.3$" (indicative redshift, not used in the analysis), "Dubious", "Nearby galaxy" (NGC 4634).
 Both bottom panels show `dubious' sources, likely being substructures in a close cluster or background clusters.
 Green circles indicate the position of C1 candidates and green boxes are others X-ray detections; the cyan cross indicates the centroid of the X-ray emission; red crosses stand for the first-pass sources that were found to have a likelihood detection smaller than 15 by {\scshape XAmin}.}
 \label{fig_class_sources} 
\end{figure*}

All remaining entries underwent a detailed screening based on optical data. For each putative cluster, we retrieved images from the Digitized Sky Survey (DSS) POSS-II on which we overlaid the X-ray contours. This step was mainly intended to remove extended sources, not relevant for our cluster catalogue: very large nearby clusters, halos of nearby galaxies, planets, unresolved double point-sources, and, possibly, saturated point-sources. For this purpose, the DSS images are sufficient.  During this process each source was assigned a quality flag by  two astronomers independently; the final decision  was made by a moderator upon the evaluators' comments.
False detections are classified as `point-like', `double source', `artefact' or `nearby galaxy'. Among the 1514 screened candidates, 234 X-ray detections were found to originate from nearby galaxies; 245 were classified as artefacts, the majority of them being found in the X-ray emission of large, pointed galaxy clusters (Fig~\ref{fig_class_sources}, bottom panels).
An additional `dubious' flag was assigned to sources for which the galaxy cluster nature is unclear: these mostly correspond to faint extended sources - at the C1 limit - with an overall compact emission.
As of Aug.~2010, the catalogue contains 845 C1 cluster candidates, 104 being classified as dubious (App.~\ref{catalogue_details})

%
%
	\subsection{Count-rate measurements in multiple bands}
	\label{fluxmes}

The {\scshape XAmin} pipeline is well suited to the count-rate measurement of moderately bright extended sources (up to a few thousands of photons). But in the case of sources occupying a large fraction of the detector or heavily contaminated by point-sources, hand-measurements are necessary in order to reach the accuracy required for an optimal application of the CR-HR method. We have thus developed a semi-interactive procedure to perform multi-band count-rate measurements and describe it below.

		\subsubsection{Methodology}
We have developed software allowing for the masking of selected contaminating sources (mostly background or host AGNs), a careful account of the background levels and a possible redefinition of the source centre. Following paper~I, we perform the measurements in the three different energy bands, well suited to the CR--HR analysis of the sample: [0.5-2], [1-2] and [0.5-1]\,keV.
Input for the procedure are   images and exposure maps for the three detectors in the given bands.
To correct for the masks, CCD gaps and detector borders, the source to be measured is assumed to be spherically symmetric, and count-rates are integrated in concentric annuli.
The initial X-ray center is the centroid determined by {\scshape XAmin}, but can be redefined by hand (e.g. if a CCD gap is masking part of the extended emission, and shifts the X-ray centroid). We define source count-rates as the mean number of source photons collected by the telescope during one second, corrected from vignetting (i.e. equivalent to {\it on-axis} measurement) and detector cosmetics (CCD gaps, etc.).
Using the detector exposure maps, we derive the mean count-rate of the source in each annulus and compute uncertainties, assuming Poisson noise. A control annulus, chosen sufficiently far away from the source, but close enough to account for local variations, provides the background estimate. Following \citep{Read:2003p3205}, we model the background by a sum of a vignetted component ({\it photon} background) and an flat {\it particle} background, each of them being described by one parameter. The uncertainties on these parameters are derived assuming Poisson noise, and propagated to the uncertainties on the individual source count-rates.

\begin{figure}
	\includegraphics[width=84mm]{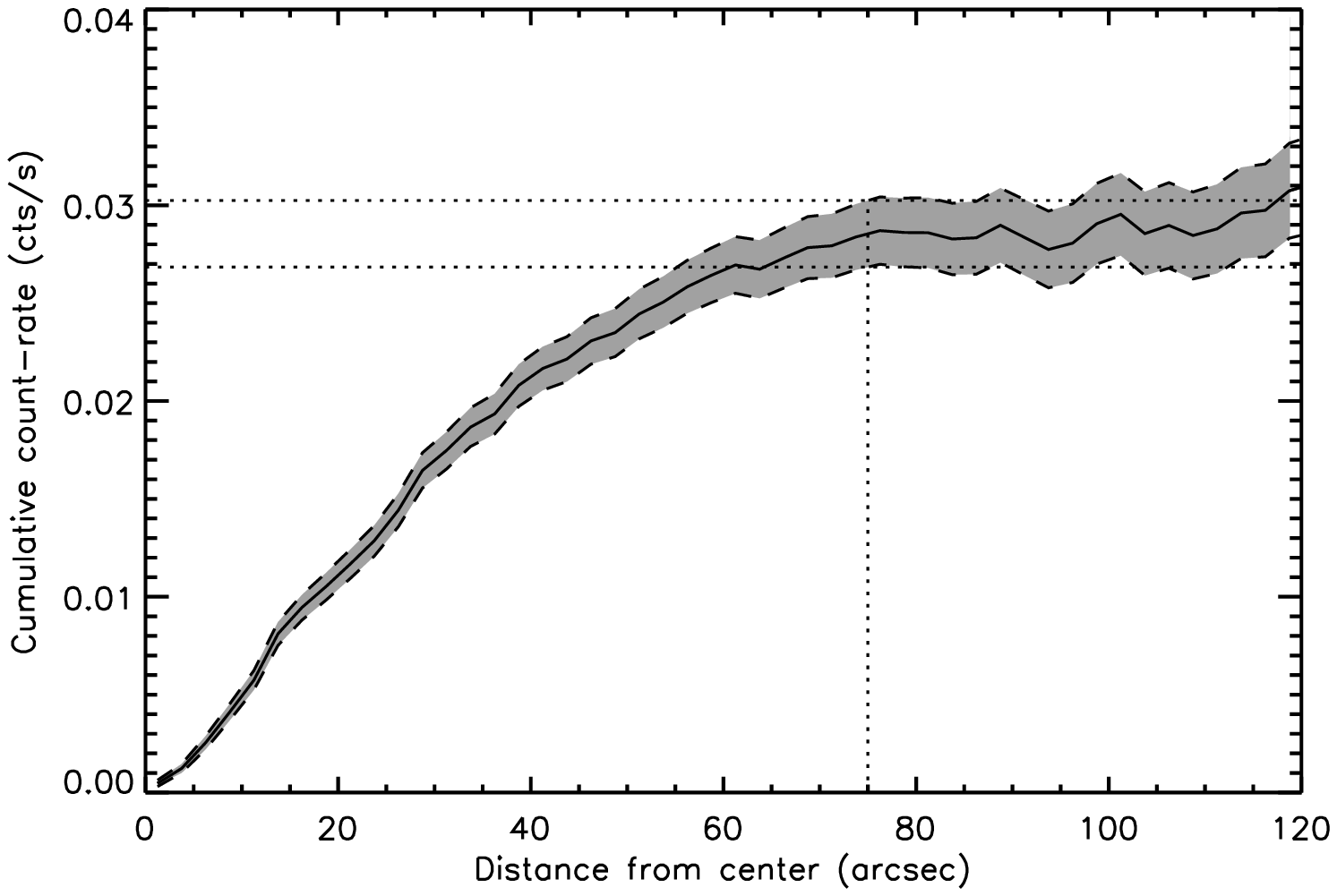}
 \caption{Example of a count-rate integration curve with associated $1\sigma$ error bars (Cluster tag 2094 in L4SDB). The vertical line indicates a S/N of one. Apparent fluctuations at large cluster-centric distances are due to uncertainties in the background subtraction.}
 \label{fig_gcurve_example} 
\end{figure}

Measurements are performed  on each detector, then summed up to provide a total count-rate. A count-rate growth curve  is then computed (Fig.~\ref{fig_gcurve_example}), as well as a surface brightness profile.
For each source, the masking areas, the source position, as well as the number and width of the annuli and the background area, are set by hand in the [0.5-2] keV energy band. These settings are then stored and used for automatically measuring count-rates growth curves in the two other bands [0.5-1] and [1-2]\,keV. Final measurements are always performed using the complete pointing exposure (i.e. not only the 10\,ks or 20\,ks data),  in order to maximise the signal-to-noise ratio.

Such a procedure presents the advantage to be model-independent and does not require any spatial fitting. In turn, only aperture count-rates are available, up to a limiting radius at which the source emission vanishes in the background. For each measurement, the software provides the radius at which the integrated count-rate value shows a signal-to-noise ratio equals to 1 (i.e.~compatible with background emission). The vertical dotted line displayed on Fig.~\ref{fig_gcurve_example} shows the position of this radius for a particular cluster measurement. In most cases, the   integration radius set manually is very close to the automatic guess by the software, except in the few cases where the measurements in consecutive annuli are noisy  (e.g. in presence of a mask or a CCD gap).

		\subsubsection{Assessment of the method and aperture corrections}
We quantitatively evaluated the validity of our measurement procedure by means of simulated XMM observations of galaxy clusters. We used the simulation set described in Section~\ref{simulations} which provides a representative range of observing conditions (in terms of background and exposure time) and of galaxy cluster sizes and fluxes. Table~\ref{table_simulation_summary} summarizes the set of simulated observations.
All simulated cluster profiles are azimuthaly symmetric $\beta$-models with fixed $\beta=2/3$ \citep{Cavaliere:1976p375}. The total count-rate and core radius are taken among discrete values close to those expected in the survey. Each of the 18,000 simulated observations is processed following the steps described previously.
We then applied selection criteria identical to those applied for the cosmological analysis, as outlined in Sect.~\ref{cosmological_analysis} and the count-rate of the $\sim 10,000$ selected clusters were automatically measured following the above procedure.  The size of the background annulus and the boundary radius for the integration were set according to the cluster input extent.

\begin{figure*}
	\begin{tabular}{c}
		\includegraphics[width=\linewidth]{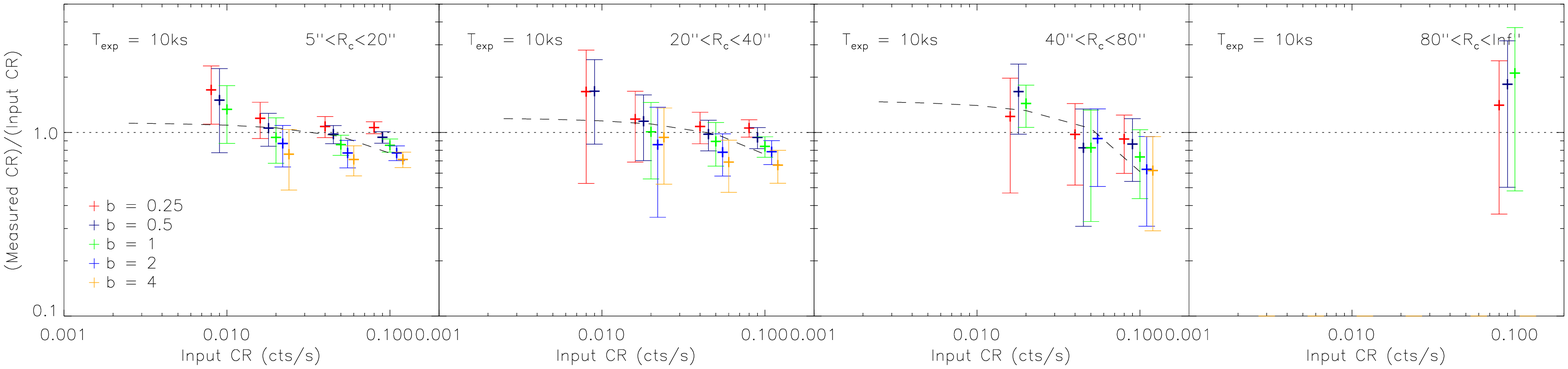} \\
		\includegraphics[width=\linewidth]{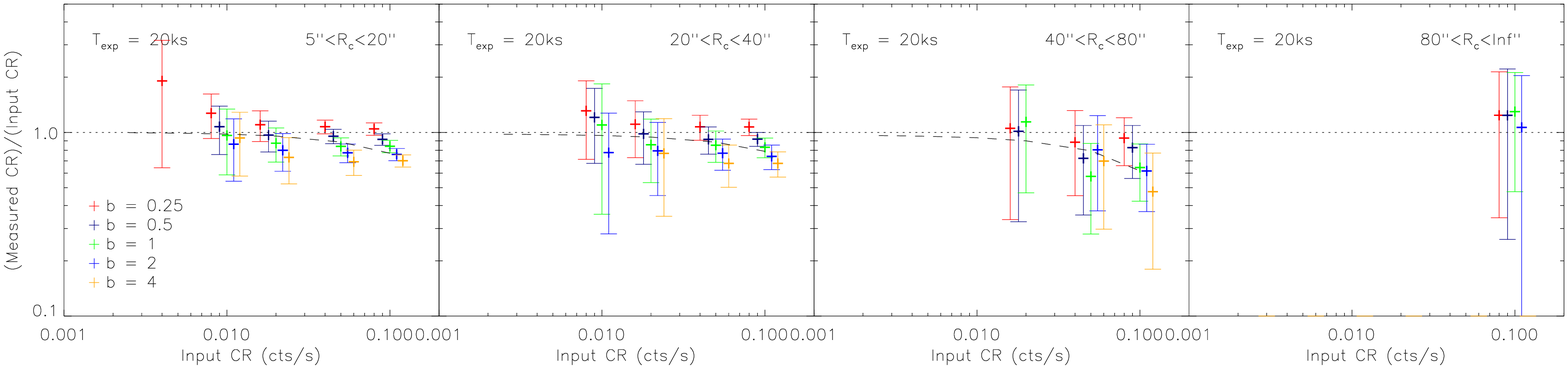} \\
	\end{tabular}
 \caption{Aperture correction for measurements of cluster count-rates.
 {\it Top: }10\,ks simulated pointings, {\it Bottom: }20\,ks simulated pointings.
 	In each panel only clusters having output core radii in the indicated range are included.
 Simulations were performed for various background levels as indicated by the colours. The data points have been horizontally shifted for clarity.
 	Measured count-rates are obtained by the aperture photometry described in Sect.~\ref{fluxmes}. Only simulated C1$^{+}$ clusters (matching the same criteria as for the cosmological subsample) are included here.
	The black, dashed line shows the best linear fit used for correcting the individual count-rate measurements.}
 \label{fig_fluxcalib} 
\end{figure*}

Fig.~\ref{fig_fluxcalib} presents the results obtained for both 10\,ks and 20\,ks simulated pointings.
Each panel shows the ratio of the measured count-rate over the true input value, as a function of measured core radii. The lack of statistics at large radii comes from the small number of extended sources detected in the simulations. The decreasing trend in this ratio as a function of input count-rate is explained by Eddington bias: at low fluxes, only clusters that pass the selection function criteria are measured. It artificially increases the mean value of the measured count-rate. Our method thus accounts for the statistical nature of the cluster sample.
These results show the overall accuracy of the count-rate measurement procedure. They are in agreement with the fact that sources that are more extended are less well measured and this effect is more pronounced for higher background levels.

To correct for the flux loss due to fixed aperture measurements, we fitted in each panel a linear relation of the form $\crate_{\mes} = a.\crate_{\rm input} + b$ (see Fig.~\ref{fig_fluxcalib}). We corrected individual cluster measurements by inverting this relation, thus providing an estimate of the true count-rate in the band of interest.

		\subsubsection{Filter combinations and N$_{H}$ values}
According to the choice of the XMM guest observers, the MOS and PN observations are obstructed by one of the three EPIC filters, namely {\scshape Thin1, Medium} and {\scshape Thick}. 
For the purpose of  applying the CR-HR method, cluster count-rates must be evaluated in a unique reference filter (THIN1 in our case). We thus need to apply some a posteriori correction for using observations performed with other filters.  Because filter transmission curves have different spectral dependences, these corrections are energy-dependent and we estimate them by means of empirical conversion relations: we form count-rate ratios obtained from a wide range of cluster spectra in the energy range of interest and fit a linear relation of the form:
\begin{equation}
	\label{equ_filter_formula}
	\frac{\rmn{CR}(X,\rmn{Thin/Thin/Thin})}{\rmn{CR}(X,\rmn{Filterset})} 
	= f \Big( \frac{\rmn{CR}(Y,\rmn{Filterset})}{\rmn{CR}(X,\rmn{Filterset})}  \Big)
\end{equation}
where $X$ and $Y$ are two different energy bands, the left-hand side representing the inverse of the filter attenuation in the considered energy band.  We made use of {\scshape Pimms 4.3}\footnote{http://heasarc.nasa.gov/Tools/w3pimms.html} and simulated XMM cluster count-rates on a grid of temperatures between 0.5 and 9\,keV and at different redshifts up to $z=1.5$. Count-rates were computed in our three reference bands  ([0.5-2], [0.5-1] and [1-2]\,keV).
We did not consider  clusters too cold and too distant since they are not retained by the C1 selection function of  \citet{Pacaud:2006p3257}.
Fifteen filter combinations out of the 27 possible ones are found in the 1992 pointings used in our analysis (Table~\ref{table_filter_statistique}), more than half of them being in the {\scshape Thin1-Thin1-Thin1} configuration.

We repeated the operation for typical galactic absorption values, ranging from $\nh = 10^{20}$ to $2.10^{21}\,$cm$^{-2}$ (see Fig.~\ref{fig_nh_distrib}). Table~\ref{table_filter_conv} shows an example of best-fitting values for the empirical correction.  The corrections are significant only if the {\scshape Thick} filter is used, which is consistent with the fact that the {\scshape Thin1} and {\scshape Medium} attenuations are comparable for  the chosen energy bands. The relative uncertainty of these conversions was found not to exceed a few percent.

\begin{table}
	\centering
		\begin{tabular}{@{}lllc@{}}
MOS1 Filter      &     MOS2 Filter     &     PN Filter    &     Nbr. Point.\\
\hline
Thin1      &     Thin1      &     Thin1      &     1063	\\
Medium      &     Thin1      &     Thin1      &     12	\\
Thin1      &     Medium      &     Thin1      &     12	\\
Medium      &     Medium      &     Thin1      &     168	\\
Thin1      &     Thick      &     Thin1      &     20	\\
Thin1      &     Thin1      &     Medium      &     23	\\
Medium      &     Thin1      &     Medium      &     5	\\
Medium      &     Medium      &     Medium      &     619	\\
Thick      &     Medium      &     Medium      &     1	\\
Medium      &     Thick      &     Medium      &     2	\\
Thick      &     Thick      &     Medium      &     12	\\
Thin1      &     Thin1      &     Thick      &     1	\\
Medium      &     Medium      &     Thick      &     6	\\
Medium      &     Thick      &     Thick      &     1	\\
Thick      &     Thick      &     Thick      &     47	\\
\hline
Total		&				&					&	1992	\\
\hline
		\end{tabular}
		\caption{\label{table_filter_statistique}
			 Distribution of filter configurations for the set of XMM archival pointings entering the scientific analysis.
			Only 5\% of those pointings have one or more detector observing with the Thick filter, which causes a $\sim$22\% diming at 1\,keV.
			}
\end{table}

\begin{table}
	\centering
		\begin{tabular}{@{}lllcc@{}}
MOS1 Filter	&	MOS2 Filter	&	PN filter	&	C$_0$	&	C$_1$	\\
		\hline
Medium  &  Thin1  &  Thin1  &   1.01  &  0.00	\\
Thin1  &  Medium  &  Thin1  &   1.01  &  0.00	\\
Medium  &  Medium  &  Thin1  &   1.03  &  0.00	\\
Thin1  &  Thick  &  Thin1  &   1.03  &   0.01	\\
Thin1  &  Thin1  &  Medium  &   1.00  &   0.00	\\
Medium  &  Thin1  &  Medium  &   1.02  &   0.00	\\
Medium  &  Medium  &  Medium  &   1.03  &  0.00	\\
Thick  &  Medium  &  Medium  &   1.04  &   0.01	\\
Medium  &  Thick  &  Medium  &   1.04  &   0.01	\\
Thick  &  Thick  &  Medium  &   1.06  &   0.02	\\
Thin1  &  Thin1  &  Thick  &   1.10  &   0.09	\\
Medium  &  Medium  &  Thick  &   1.14  &   0.09	\\
Medium  &  Thick  &  Thick  &   1.15  &   0.10	\\
Thick  &  Thick  &  Thick  &   1.17  &   0.12	\\
\hline
		\end{tabular}
		\caption{\label{table_filter_conv}
		Example of coefficients used for the empirical filter corrections :   bands $X$=[0.5-2]\,keV and $Y$=[0.5-1]\,keV   and $\nh=4.10^{20}$\,cm$^{-2}$ are considered here. 
		The correction is modeled by a linear relation of the form $x_{\rm corr}= {\rm C}_0.x_{\mes} + {\rm C}_1.y_{\mes}$, where 
		$x_{\mes}$ and $y_{\mes}$ are count-rates measured with the same particular set of filters, respectively in bands $X$ and $Y$ and $x_{\rm corr}$,  the corrected count-rate as it would be measured in  band $X$ with the highest transmission set of filters ({\scshape Thin1-Thin1-Thin1}).
		These conversions were found to be accurate at the few percent level for clusters with temperatures ranging from 0.5 to 9\,keV and redshifts out to 1.5, as long as their location in the temperature--redshift plane is covered by the C1 selection (see text).}
\end{table}

%
%

\section{The X-CLASS survey selection function}
	\label{simulations}

This section describes the steps leading to the construction of the final cluster sample used in the cosmological analysis.
We  present the list of   retained XMM observations and subsequently  describe the  image simulations leading to the cluster selection function. We finally expose the method we have developed for evaluating the statistical bias due to the presence of numerous pointed clusters in the XMM archive.

	\subsection{ The cosmological subsample}
The cluster sample described in Sect.~\ref{data} contains 845 C1 sources detected in a homogeneous way in the XMM archival data. In order to perform the cosmological analysis by means of the CR-HR method, we extracted a high signal-to-noise ratio subsample as follows.

\begin{figure*}
	\begin{tabular}{ccc}
		\includegraphics[width=0.3\linewidth]{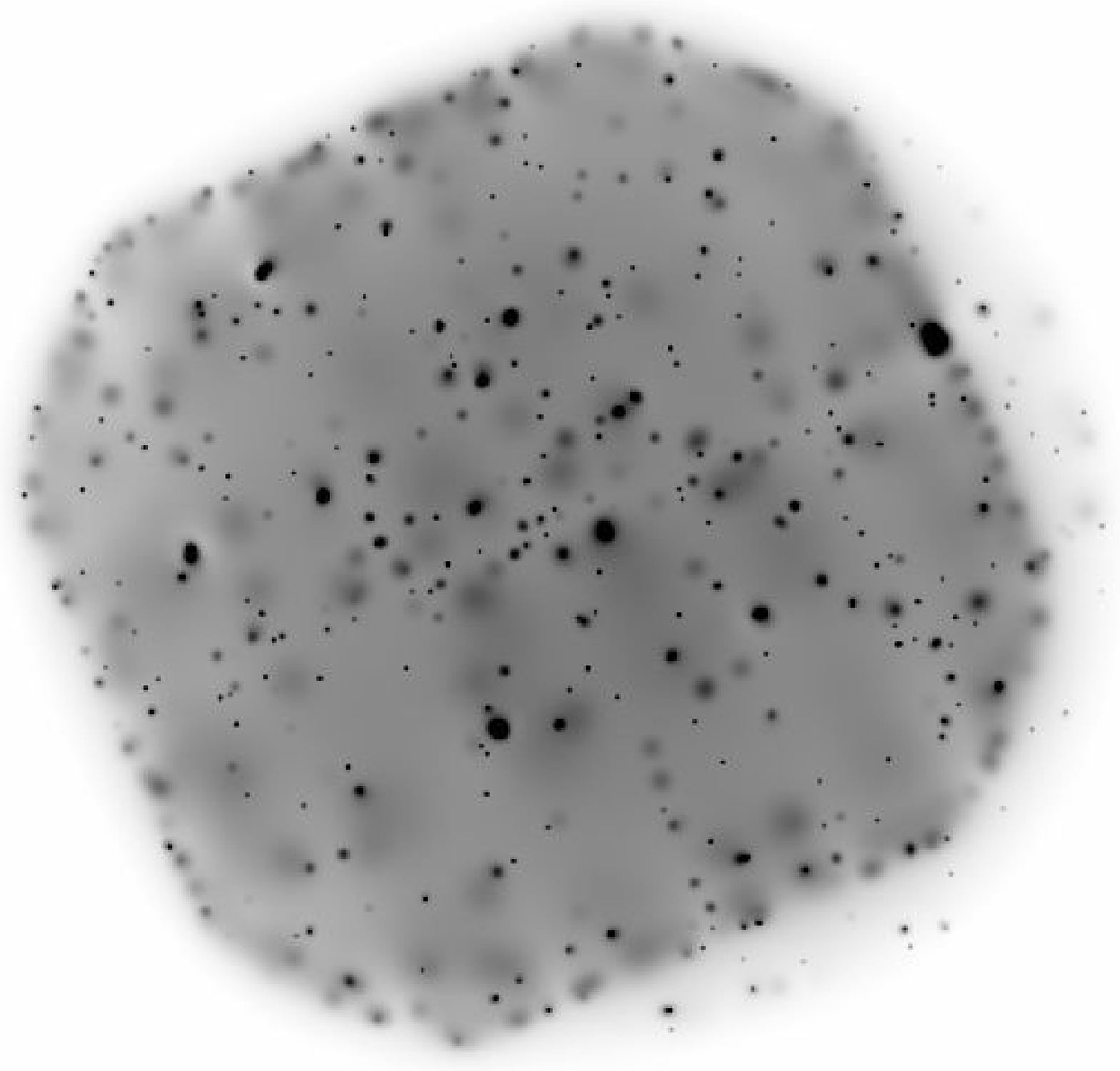} &
		\includegraphics[width=0.3\linewidth]{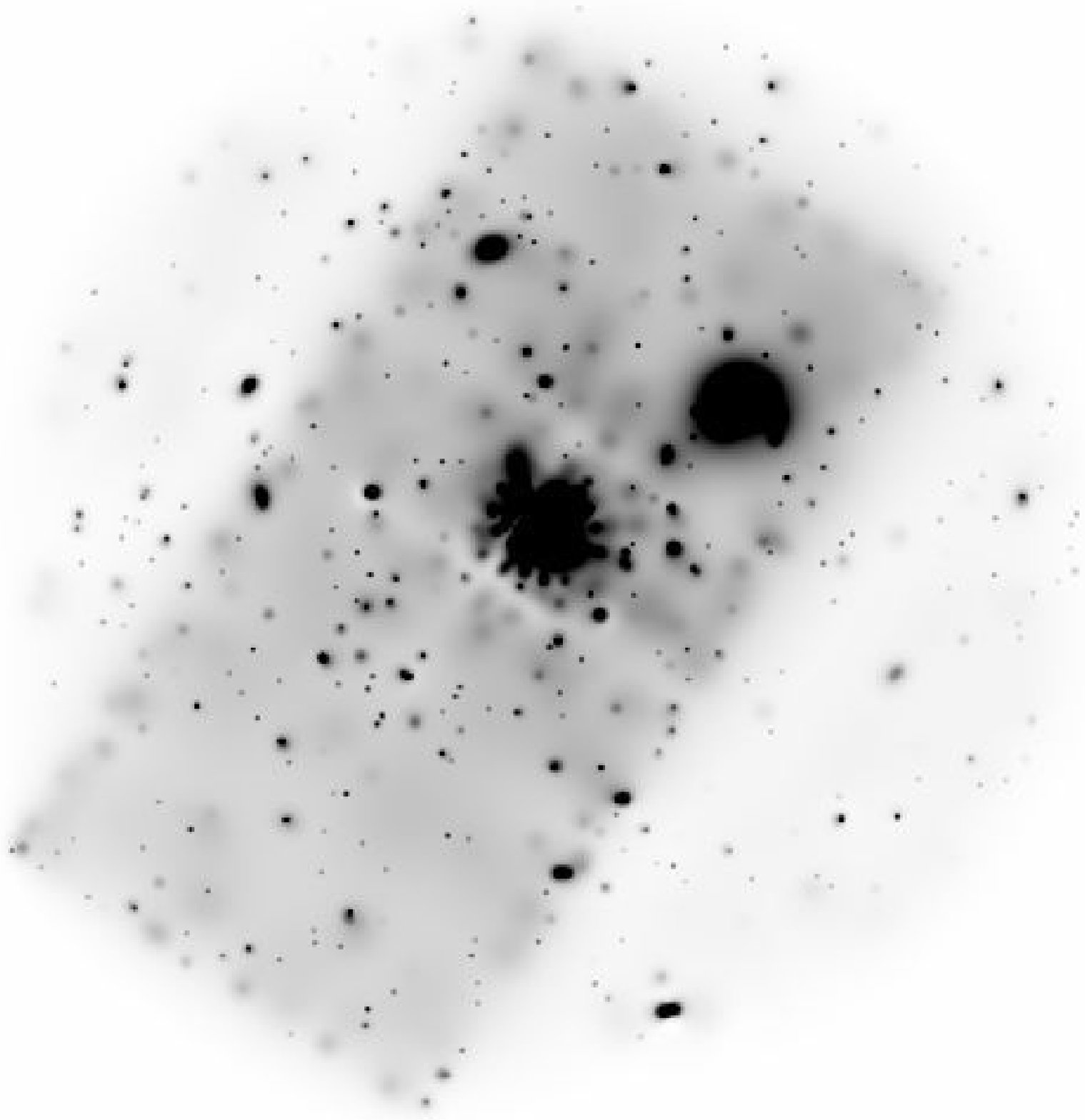} &
		\includegraphics[width=0.3\linewidth]{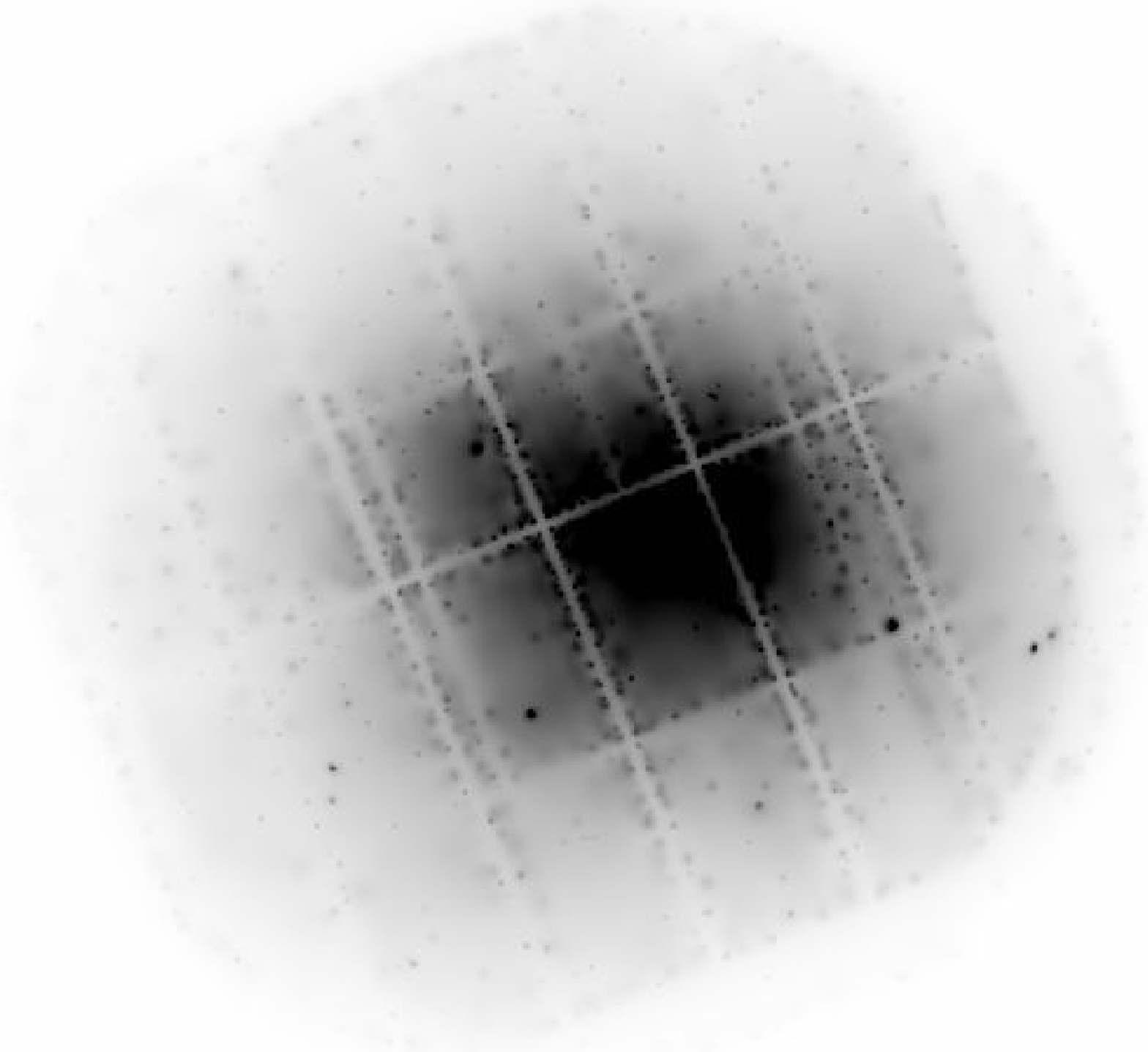}\\
	\end{tabular}
 \caption{Three examples of pointings excluded from the cosmological analysis (images have been filtered by wavelets).
 	{\it Left panel: }this pointing shows a high background on the PN detector which has not been optimally filtered by the pre filtering procedure due to long continuous periods of background flares (ObsId: 0039140101).
 	{\it Central panel: }the PN detector was working in Large Window mode thus not collecting photons from the entire field of view (ObsId: 0083150401).
	{\it Right panel :}observation of Coma extending over a large part of the field of view thus preventing the detection of background sources (ObsId: 0300530301).}
 \label{fig_excluded_pointings} 
\end{figure*}

We first selected a more homogeneous data set, i.e. by excluding  pointings (1) with a high background, (2) with one or more detectors not being in full-frame mode and (3) those centered on very nearby, luminous clusters  (see examples on Fig.~\ref{fig_excluded_pointings}).   For this purpose,  we inspected the 2409 pointings  by eye. 
In the end, the surveyed area used for the cosmological fits consists of 1992 pointings (Table~\ref{table_pointings_numbers})

We defined a sub-class of galaxy clusters called C1$^{+}$ by selecting all sources having an Extent Likelihood above 40 and an extent greater than 5 arcsec.
Only C1$^{+}$ sources within 10\,arcmin off-axis of their parent pointing and not flagged as dubious were considered.  Defining CR as the [0.5-2]\,keV measured count-rate and HR as the ratio between the [1-2]\,keV and [0.5-]\,keV count-rates, we imposed clusters to have $0.009<CR<0.5$\,cts/s and $0.05<HR<2$.
Finally, the cosmological subsample consists of 347 clusters.

\begin{table}
		\caption{\label{table_sources_numbers}
		Number of detected clusters in the X--CLASS database. Clusters entering the cosmological analysis are C1$^{+}$ sources not classified as `dubious', within 10 arcmin off-axis and having $0.009<CR<0.5$\,cts/s (count-rate in [0.5-2]\,keV) and $0.05<HR<2$ (ratio [1-2]\,keV / [0.5-1]\,keV). }
	\centering
		\begin{tabular}{@{}lllcc@{}}
		\hline
C1 sources detected in the 2409 X--CLASS pointings	&	845	\\
C1$^{+}$ sources :		&		745	\\
 - classified as `dubious',		&	74	\\
 - within 10\,arcmin off-axis, 	&	630\\
 - entering the cosmological analysis	&   347	\\
		\hline
		\end{tabular}

\end{table}

	\subsection{Description of the simulations}

		\subsubsection{Principle}

We use an updated version of {\scshape InstSimulation} \citep{Valtchanov:2001p6391, Pacaud:2006p3257} to generate fake XMM observations, taking into account the main characteristics of the XMM EPIC instruments. In particular, an analytic vignetting model and a detector mask are superimposed to the simulated sources, as well as instrumental and photon backgrounds.
The shape and off-axis dependency of the PSF is modeled by using the {\scshape Medium} model from the XMM calibration files.
Apart from the peculiar observations of bright, saturated sources, or extremely extended sources, the simulation set captures the most important features impacting the detection and the characterization of the sources. We performed two sets of simulations:  one with point-like sources only, and the other with clusters and point-like sources.

		\subsubsection{Point-like sources only}
The first set of simulations without extended sources serves as a test for contamination and for parametrizing the background level on the instruments.
Point-like sources are distributed across the field of view, following a sampled $\mathrm{log}N\mathrm{-log}S$ taken from \citep{Moretti:2003p1679} in the [0.5-2\,keV] band. Conversion from flux to count-rate is performed assuming a constant MOS to PN count-rate ratio, regardless of the source spectral distribution.
The flux lower bound is chosen accordingly to the exposure time so as to give $\sim 2$ photons on-axis (i.e.~below the XMM detection limit). Non-resolved AGN photon background is added following values from \citep{Read:2003p3205}, then corrected from the estimated fraction of  AGNs resolved by the pipeline. This background component is vignetted, thus showing a strong off-axis dependence. We finally add the non-vignetted particle background component using the standard values from \citet{Read:2003p3205} multiplied by an arbitrary factor $b=0.25$, $0.5$, $1$, $2$ or $4$ so as to investigate the impact of pointing-to-pointing background variations on the detection efficiency. For each background value, 540 pointings are simulated both at $10$ and $20$\,ks exposure times.

Each of these pointings is processed by the {\scshape XAmin} pipeline described in Sect.~\ref{data}, exactly in the same way as real observations. In particular, an extended model fit is performed over each detected point-source to evaluate the contamination of the cluster sample.

		\subsubsection{Extended source simulations}
Similar simulations are performed by adding clusters assumed to be spherically symmetric sources and defined by a  $\beta=2/3$ profile;  apparent core radii $r_c$ range from $10$ to $100$ arcsec and total count-rates   from  $2.5.10^{-3}$ to $0.1$ \,cts/s. From 4 to 8 simulated clusters are injected in each pointing, depending on their angular size, and we avoided source overlap by defining exclusion sectors in the XMM field of view.
In total, some 87000 extended sources were simulated over more than 18000 pointings (Table~\ref{table_simulation_summary}).
In order to reproduce the effects of point source contamination, a population of point-like sources was added as described in the previous section.

Each pointing is processed by the pipeline and positions of the input extended sources are correlated with the output  positions within a $37$ arcsec radius. In case of multiple matches, the detected source with the highest Extent Likelihood is chosen as the best matching counterpart and all others are discarded. Fig.~\ref{fig_simulations} (top panel) shows three examples of simulated pointings in different observing conditions, along with the corresponding pipeline results.

\begin{table}
	\centering
			\caption{\label{table_simulation_summary}
			Summary of the extended-source simulations in XMM images. Last column indicates the number of simulated clusters out to an off-axis of 10\,arcmin.
			Simulations were performed for exposure times of 10 and 20\,ks and five background levels ($b$=0.25, 0.5, 1, 2 and 4). The total number of simulated and processed pointings is 18140 and the total number of clusters amounts to $\sim 87000$.
			}
		\begin{tabular}{@{}cccccccc@{}}
Input count-rate	&	\multicolumn{4}{c}{Input core radius}			&&			\\
($10^{-2}$\,cts/s)&	10"		&	20"			&	50"		&		100"	&&	Total		\\
\hline
{\it 0.25}		&          450    &      450    &     750    &      240		&--&	1890		\\
{\it 0.5}		&          450    &      450    &     750    &      240		&--&	1890		\\
{\it 1.0}		&          450    &      450    &     750    &      240		&--&	1890		\\
{\it 2.0}		&          450    &      450    &     750    &      240		&--&	1890		\\
{\it 5.0}		&          100    &      100    &     150    &      240		&--&	590			\\
{\it 10.0}		&          100    &      100    &     150    &      240		&--&	590			\\
\hline		\end{tabular}

\end{table}

	\subsection{Analysis of the simulations -   selection criteria}

		\subsubsection{Contamination by spurious and point-like sources}
Following \citet{Pacaud:2006p3257} we report for each detected source its location in a two-parameter space  Extent/Extent-Likelihood, as shown in Fig.~\ref{fig_simulations}, middle panels. In this figure, green symbols represent point-like sources, magenta symbols are for extended sources and red points stand for spurious detections, i.e. detections in the point-only simulations which are not associated to input sources within the 6" correlation radius.
Fig.~\ref{fig_simulations} shows the good stability of the C1$^{+}$ criterion across the range of exposure times and background levels, in terms of contamination by point-like sources and spurious detections. From our simulations, we expect the number of contaminating sources (i.e. point-sources interpreted as extended sources) not to exceed one in every 300 pointings for normal observing conditions (10\,ks, low background).

		\subsubsection{Efficiency of the extended sources detection}
\begin{figure*}
	\begin{tabular}{ccc}
		$T_{exp}=10$\,ks, $b=1$	&
		$T_{exp}=20$\,ks, $b=1$	&
		$T_{exp}=10$\,ks, $b=4$	\\
		\includegraphics[width=0.3\linewidth]{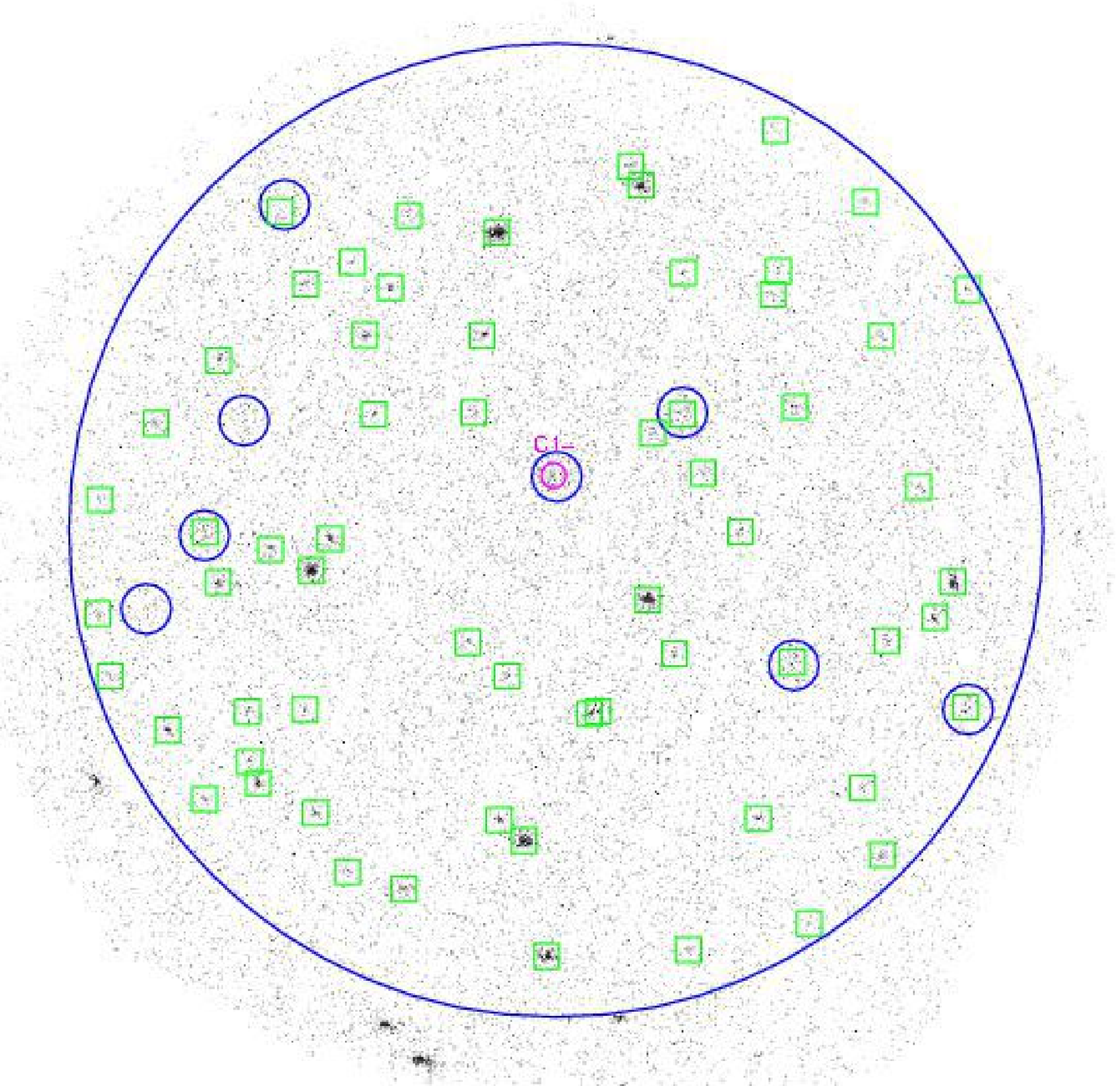} &
		\includegraphics[width=0.3\linewidth]{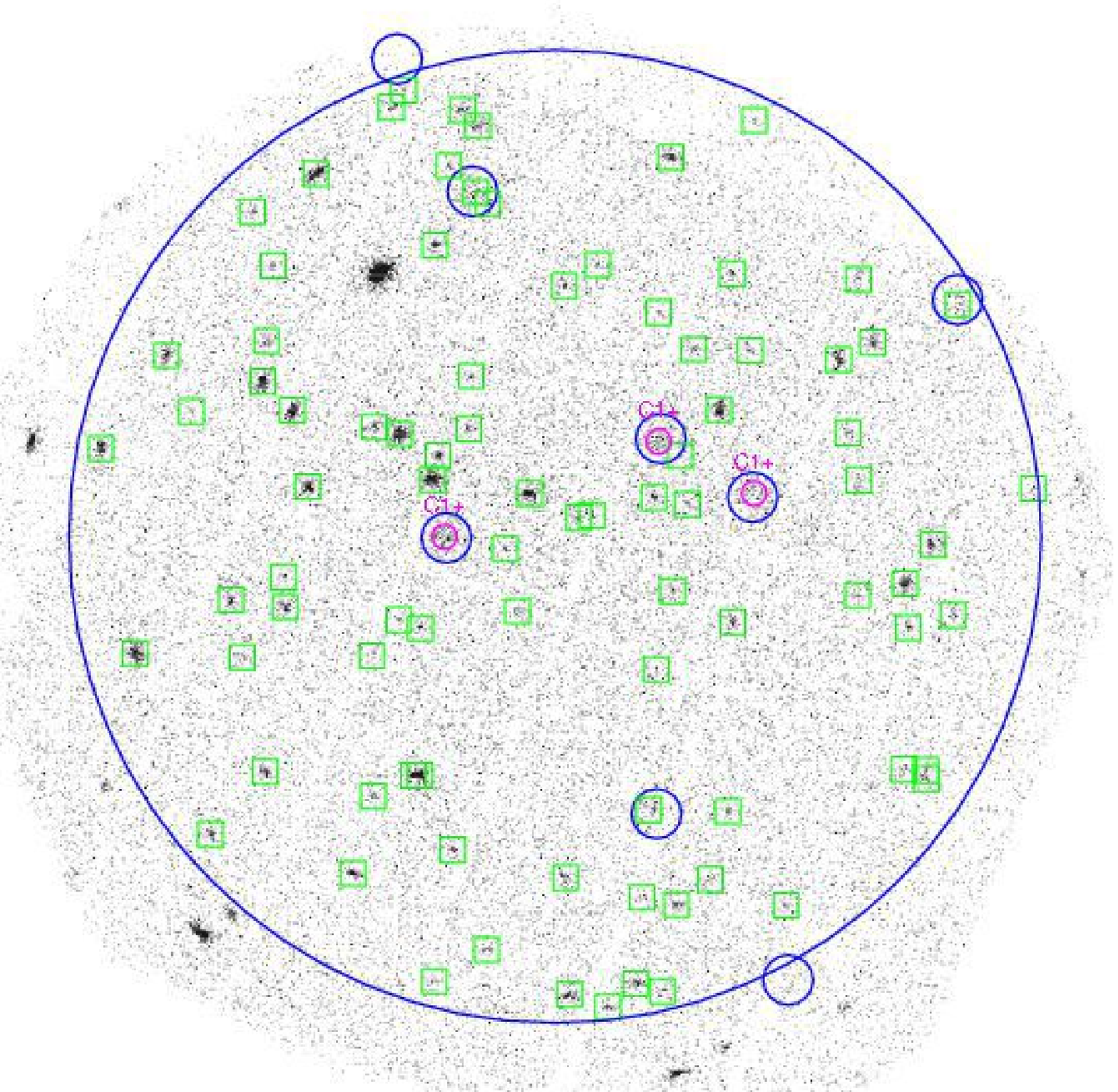} &
		\includegraphics[width=0.3\linewidth]{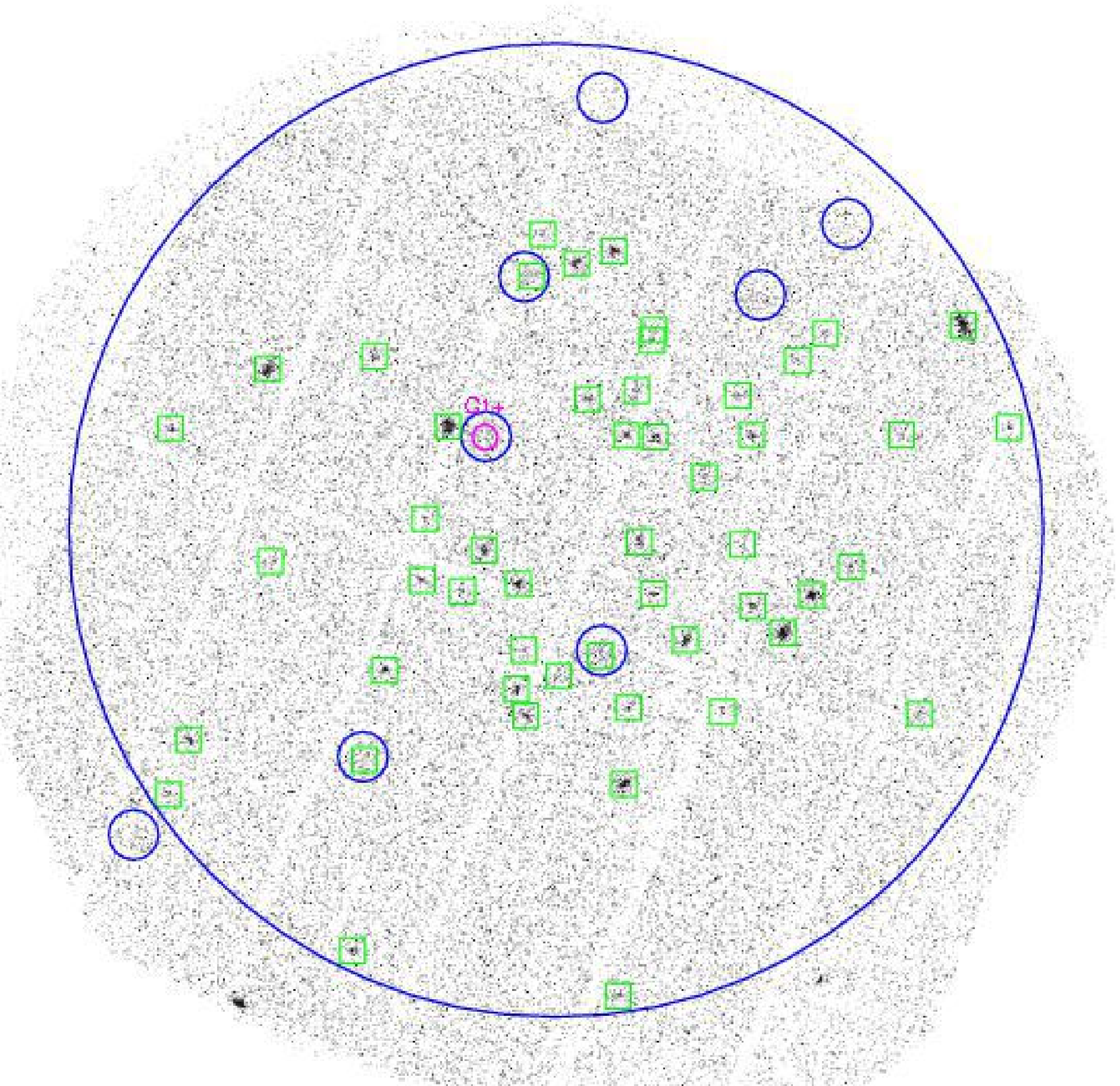} \\
		\includegraphics[width=0.3\linewidth]{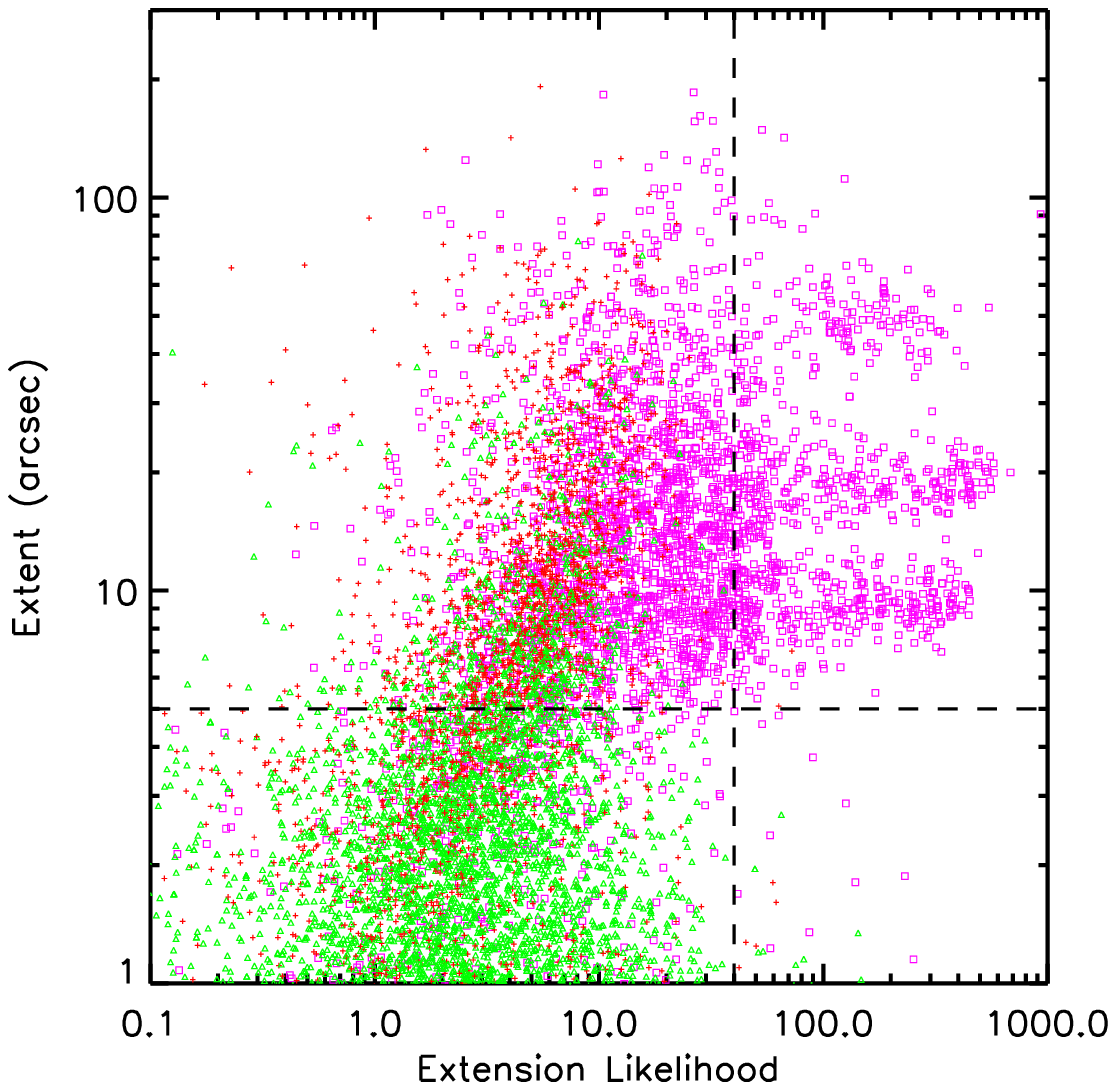} &
		\includegraphics[width=0.3\linewidth]{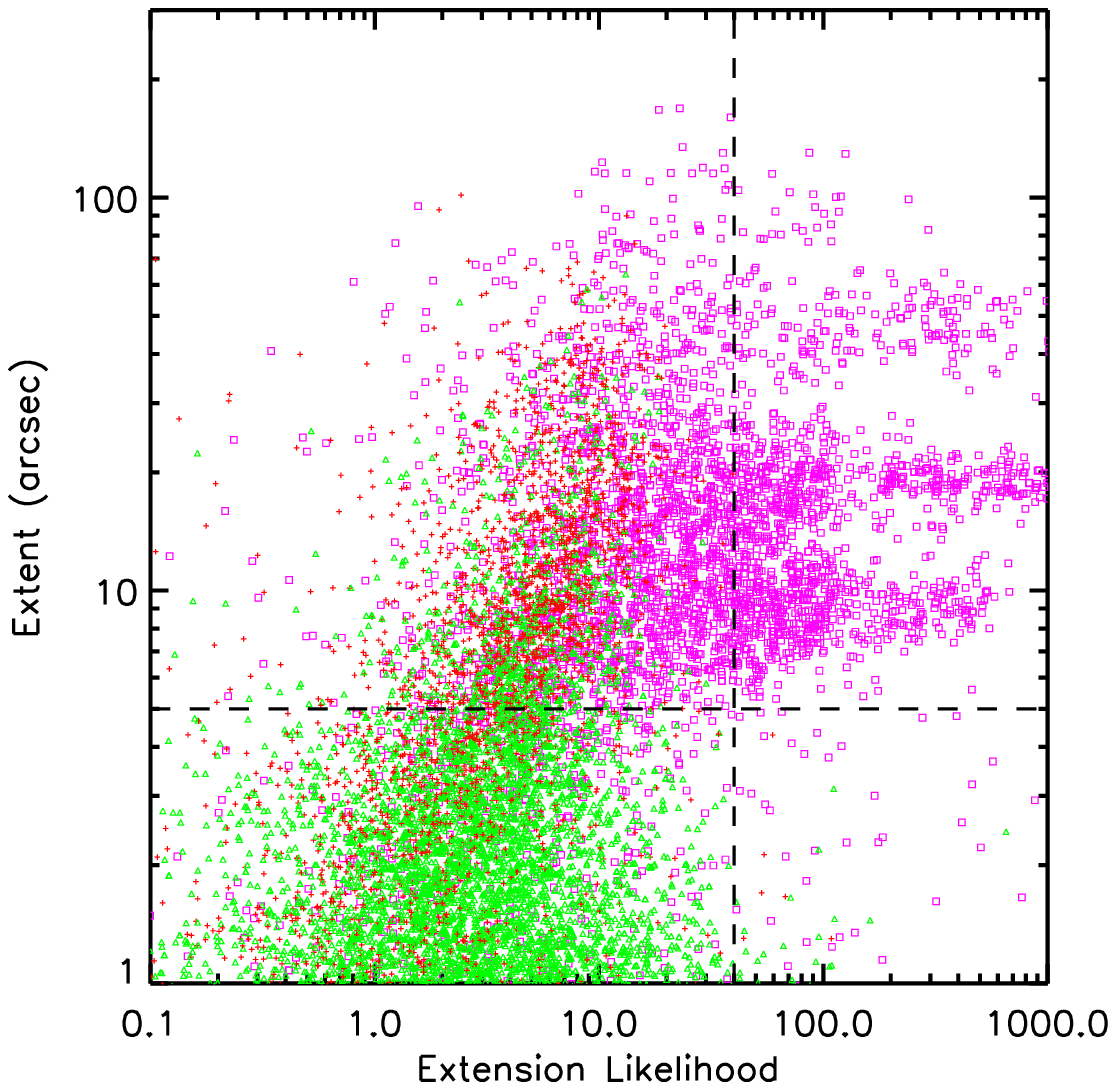} &
		\includegraphics[width=0.3\linewidth]{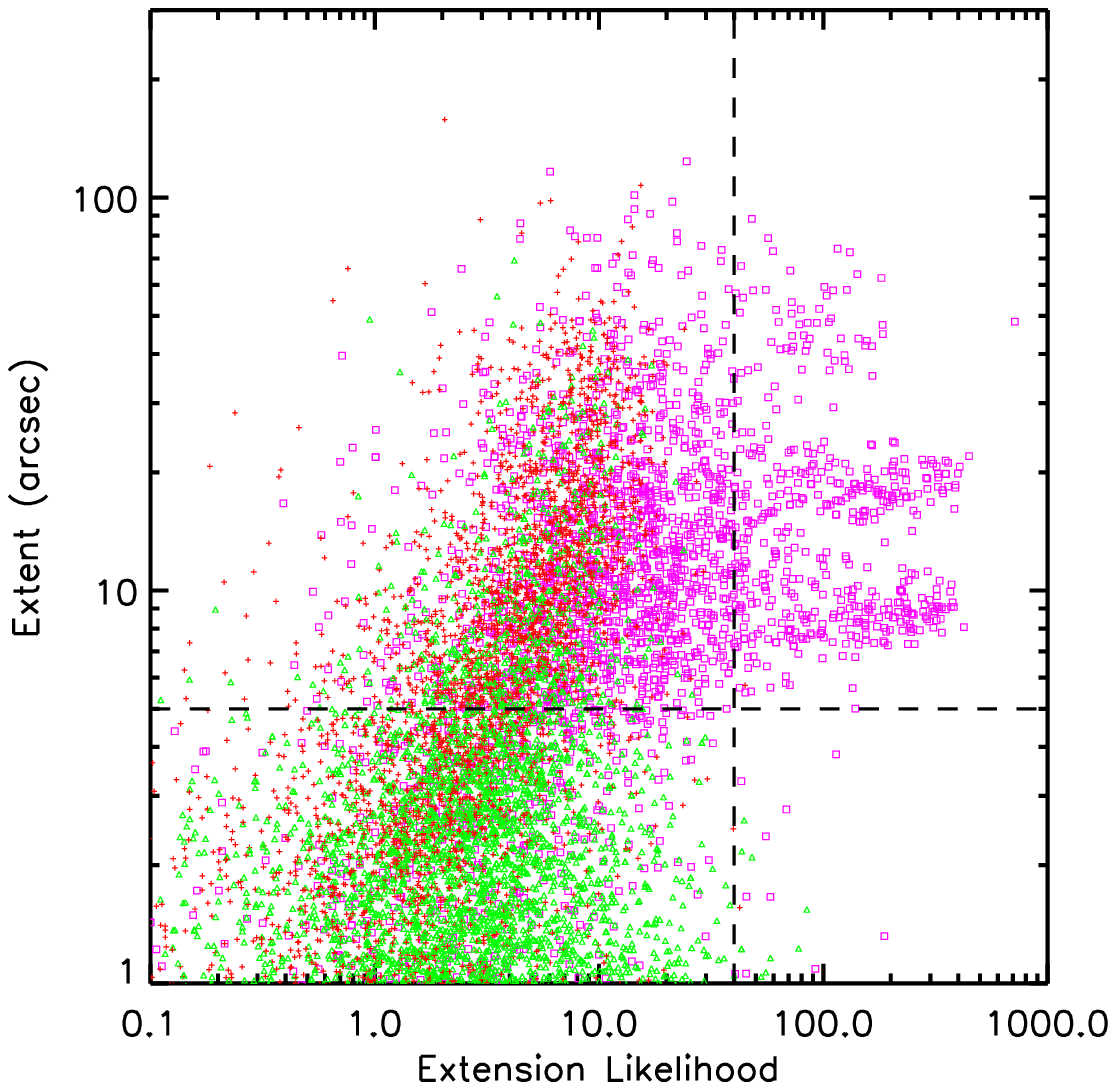} \\
		\includegraphics[width=0.3\linewidth]{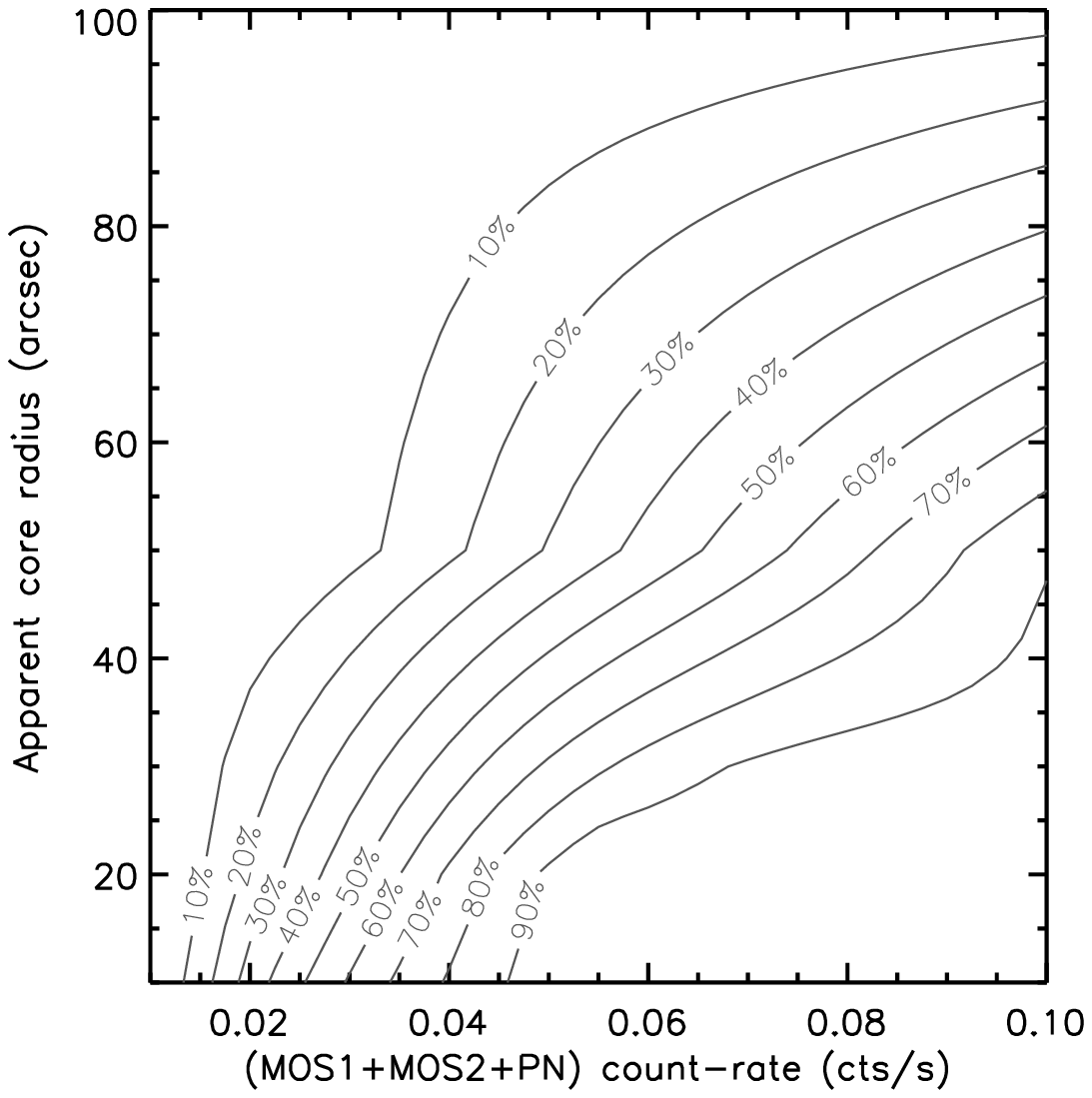} &
		\includegraphics[width=0.3\linewidth]{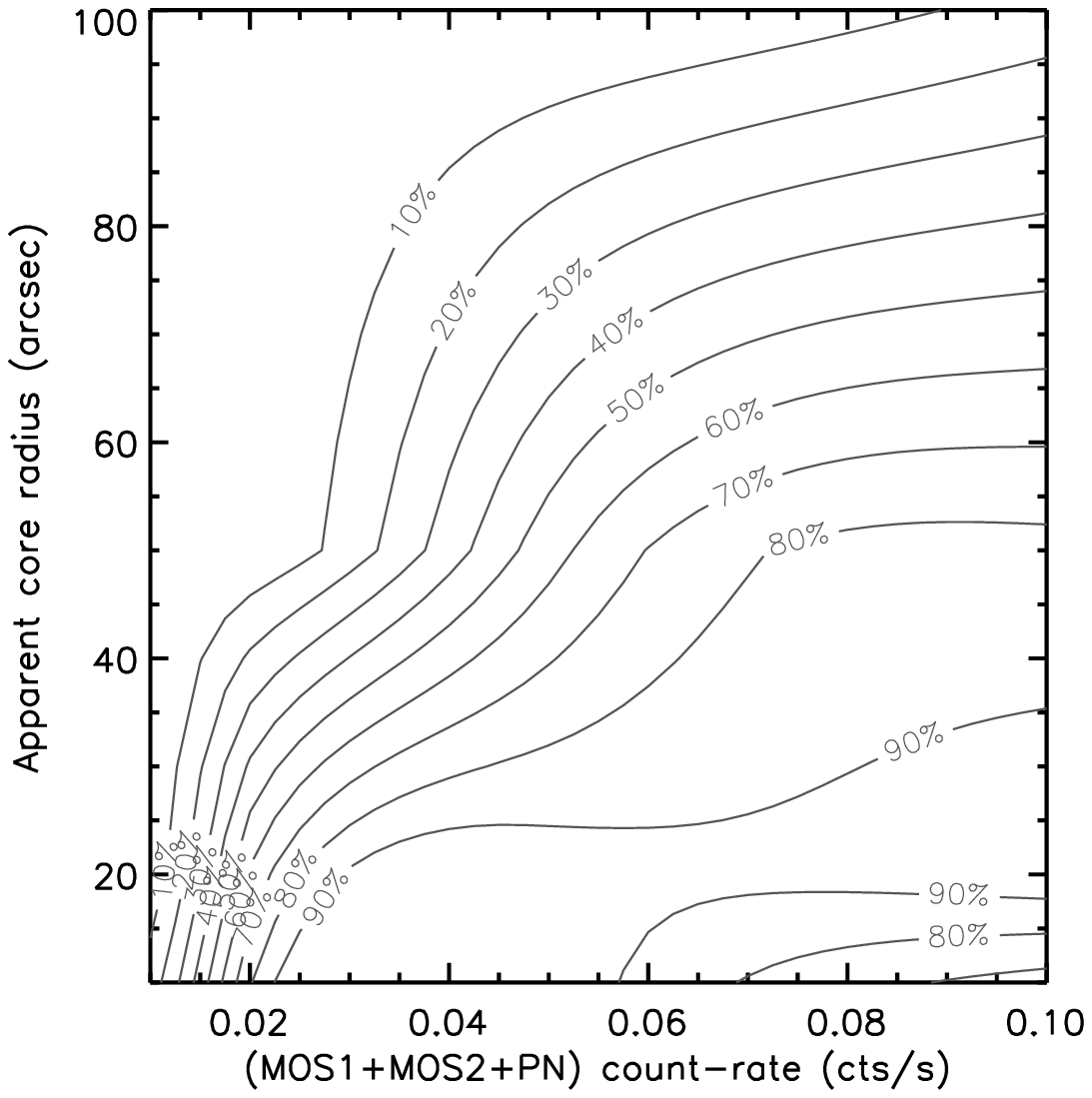} &
		\includegraphics[width=0.3\linewidth]{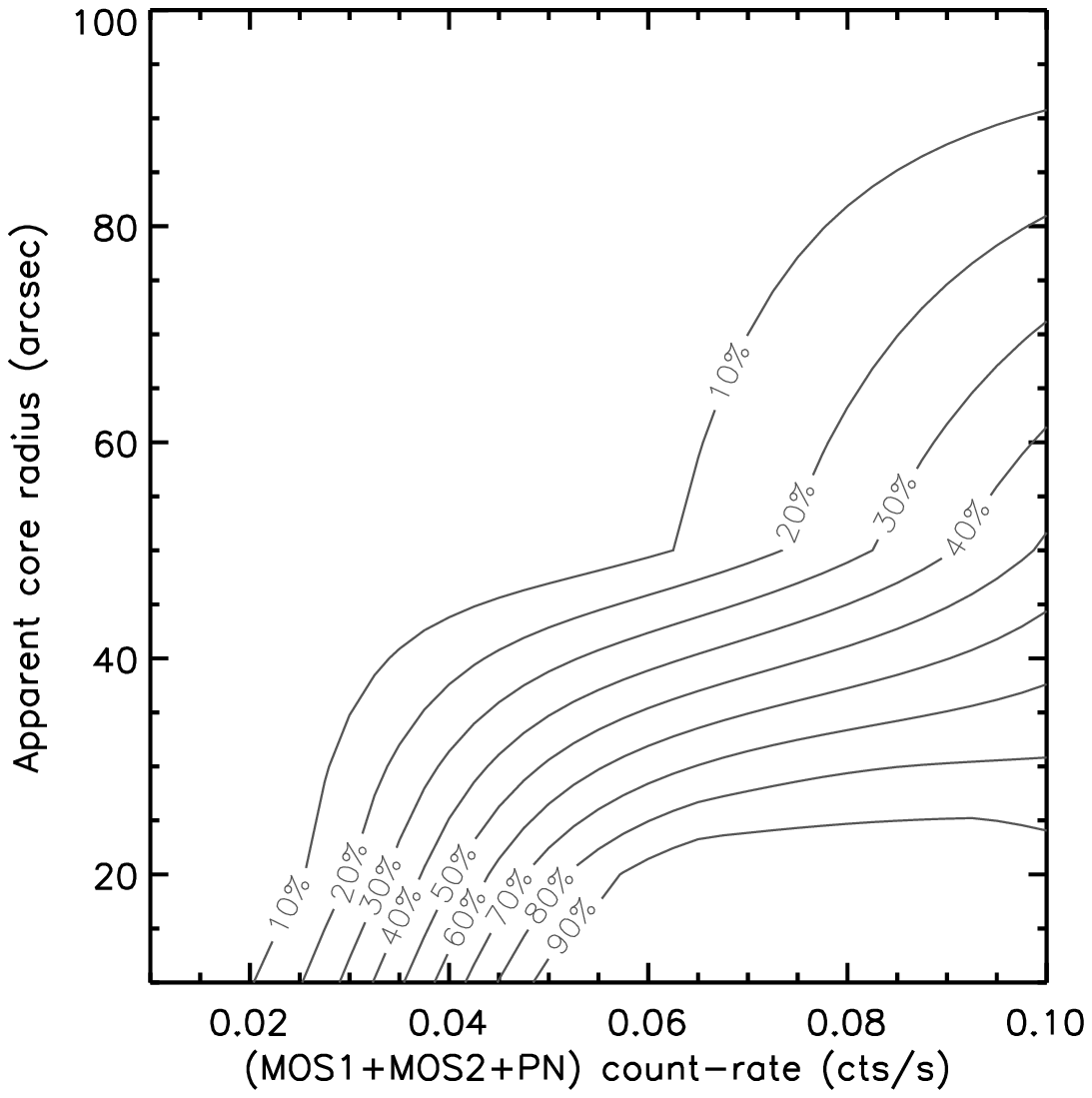} \\
	\end{tabular}
 \caption{The X--CLASS selection function. We present the results of simulations performed for three pointing configurations (different exposure times and background levels) among the 10 simulated configurations.
 	{\it Top: }Example of simulated XMM observations. Blue circles show the locations of simulated 20\,arcsec core-radius clusters and magenta circles denote detections classified as C1$^{+}$ by the pipeline. Green boxes are for the remaining, unclassified sources (including point sources). The radius of the large blue circle is 13\,arcmin.
	{\it Middle: }Distribution of detected sources in the Extent-Likelihood/Extent plane. Recovered clusters are in magenta, point-like sources in green and spurious detections in red. Vertical and horizontal lines delimit the C1$^{+}$ classification (EXT\_ML\,$>$\,40 and EXT\,$>$\,5); the discreetness of the Extent distribution reflects the input core radius values (Table \ref{table_simulation_summary})	{\it Bottom: }Corresponding detection probability for the C1$^{+}$ sources as a function of the input [0.5-2]\,keV count-rate and the input core-radius. 
 		$[2.5.10^{-3},5.10^{-3},0.01,0.02,0.05,0.1]$
		}
 \label{fig_simulations} 
\end{figure*}

We then derive the C1$^{+}$ detection efficiency by taking all sources in the 0--10\,arcmin off-axis range.
Bottom panels of Fig.~\ref{fig_simulations} display the probability of detecting a C1$^{+}$ cluster as a function of its total input count-rate and its input core radius, as derived from our simulations.
From these results, it clearly appears that the selection is not flux-limited, but rather surface-brightness limited. These curves also reveal the expected increase in efficiency from 10 to 20\,ks and for lower background levels.
The sharp decrease observed in the 20\,ks selection for high count-rates ($\sim$0.06--0.1\,cts/s) and small core radii ($<$\,20\,arcsec) indicates that these sources are identified as point-like sources by the pipeline; such objects, however, are unlikely to appear in real observations.
We derived similar  probability functions for the ten simulated configurations (10 and 20\,ks exposures and five background levels).

Then, for any given pointing in the survey, we estimate its background parameter ($b$). To this purpose, we measure local background estimates at several locations on the detectors and compare them to the values found in the simulated observations for which $b$ is known. We finally derive the selection function of each  XMM observation entering the cosmological analysis.

	\subsection{The survey geometrical area}

For the cosmological analysis, we consider only sources within a 10\,arcmin radius around each pointing center, thus removing regions of the detectors where the point spread function has an elongated shape and the vignetting factor is greater than 50\%. Because of the multiple overlaps between pointings, we estimate the net area  by means of a Monte-Carlo integration. If two or more pointings of the same exposure time overlap, the intersecting area is equally distributed between those pointings. If one 20 ks pointing and one 10 ks pointing overlap, the intersecting area is fully attributed to the deeper pointing. This set of rules is thus compatible with the procedure applied for removing duplicate detections in the catalogue (Sect.~\ref{data}).
The net total area covered by the 1992 pointings is 90.3\,deg$^2$.

	\subsection{Correcting for the target bias in the XMM archive}
	\label{pointing_bias}
	
	In all cluster serendipitous surveys to date, it has always been implicitly assumed that discarding the central target of the considered pointings (along with subtracting the corresponding survey area) would not statistically affect the studied cluster population. It is not the purpose of the present paper to verify these past assumptions, but in the case of  serendipitous surveys based on the XMM archive, this hypothesis is questionable.  Among the 7716 archival observations available by May 2010, 1008 of them pertain to pointed observations of galaxy clusters.  Out of the  347 clusters selected for the present cosmological analysis, 92 of them are central targets (within 3 arcmin offaxis). One cannot simply discard them or include them (or ignore the complete pointing) in the statistical analysis,  because the process of target selection from guest observers  is highly subjective as well as motivated by practical constraints. This is particularly true for distant clusters, as only the brightest ones could be observed.\\
	In this section, we describe  the method that we have developed to account for the target selection bias and correct for its impact in the cosmological analysis. This bias is hard to model from first principles as it depends on the history of the XMM observing programs. Basically, we  make use of the fact that pointed galaxy clusters are preferentially located at the center of the XMM field of view and split our sample in two subsamples to apply a joint correction.

		\subsubsection{Off-axis source distribution}
\begin{figure}
	\includegraphics[width=84mm]{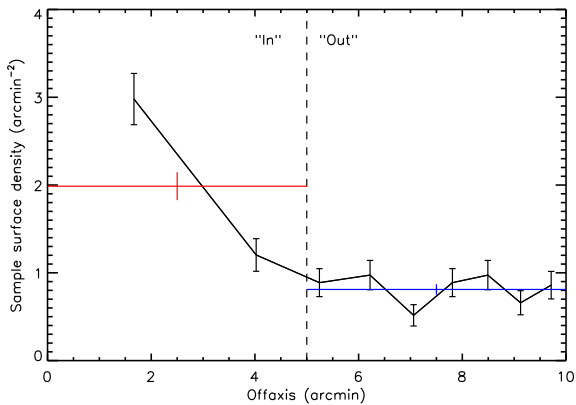}
 \caption{Surface density of clusters as a function of off-axis distance, for the selected subsample used in the cosmological analysis (black curve).
 	The red and blue points are the average surface densities in the inner [0-5]\,arcmin and the outer [5-10]\,arcmin regions respectively. The factor $\sim 2$ between the two values is mainly caused by the presence of pointed clusters in the archival data.}
 \label{fig_surface_density} 
\end{figure}

Fig.~\ref{fig_surface_density} shows the distribution of detected clusters as a function of off-axis distance. The excess of sources in the 0--5\,arcmin range is conspicuous, as well as the stabilization at higher radii. There is a factor of 2 between the density of sources in the `outer' regions of the detectors and the `inner' regions. Part of this difference stems from the better sensitivity in the central part of the XMM FoV   but a also from a number of pointed clusters.
We label by `in' and `out' the corresponding two subsamples and make the hypothesis that all pointed clusters are found in the `in' sample.

		\subsubsection{Bias model}
\begin{figure}
	\includegraphics[width=84mm]{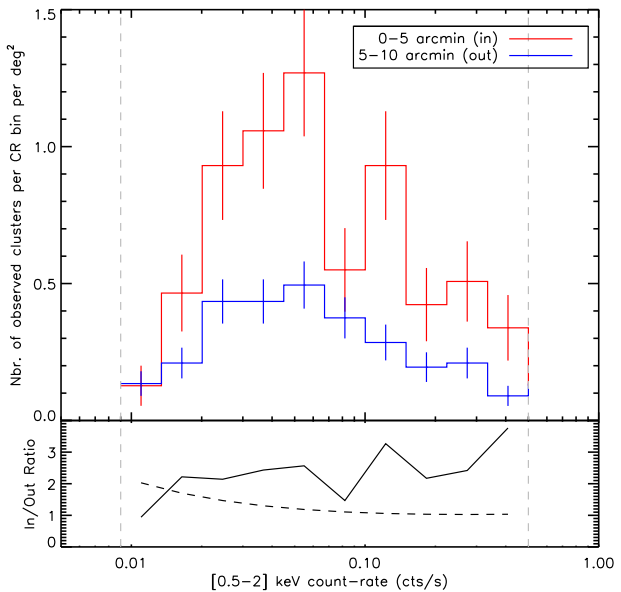}
 \caption{{\it Top panel: }count-rate distribution of the sources found within 5 armin off-axis (red) and between 5 and 10 arcmin off-axis (blue) for the cosmological subsample containing 347 clusters (all count-rates are rescaled to their on-axis values). 
 	{\it Bottom panel:} the plain curve shows the ratio between the two histograms shown in the top panel. The dashed line is the ratio that one would expect from the sensitivity gradient only.}
 \label{fig_obs_bias} 
\end{figure}

We display in upper panel of Fig.~\ref{fig_obs_bias} the count-rate distribution of sources in both subsamples per unit sky area. According to  Fig.~\ref{fig_surface_density}, there is a factor $\sim$2 between both distributions, but the the excess is not uniformly spread over the range of count-rates. For instance the factor 3 excess around CR\,$\sim$\,0.15\,cts/s corresponds to clusters with fluxes $\sim\,1-2.10^{-13}$\,ergs/cm$^2$/s typical of those found in {\it ROSAT} serendipitous surveys: 160d \citep{Vikhlinin:1998p1674}, 400d \citep{Burenin:2007p1251}, WARPS \citep{Jones:1998p1610} and SHARC \citep{Adami:2000p0000,Romer:2000p1587} -- see \citet{Piffaretti:2010p1548} for a thorough compilation of {\it ROSAT} cluster catalogues. Clusters from the {\it ROSAT} All-Sky Survey are more than ten times brighter on average and are thus excluded from our cosmological sample (limited to $CR<0.5$\,cts/s). 

\begin{figure}
	\includegraphics[width=84mm]{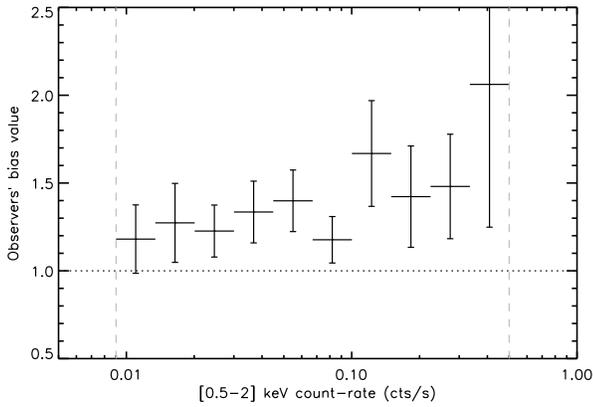}
 \caption{Bias factor $F_j$ for cluster counts (see App~\ref{bias_details}) due to the presence of pointed clusters in XMM archival data.
 $F_j$ is computed in 10 count-rate bins and evaluated by comparing the density of clusters in the inner [0-5]\,arcmin to the [5-10]\,arcmin density, taking into account the higher sensitivity around the EPIC optical center.
		The bias value is the ratio between the observed number of sources in a given bin and the number expected for a sample without pointed clusters. We use it as an empirical correction for the final sample of 347 clusters selected for the cosmological analysis.}
 \label{fig_obs_bias_result} 
\end{figure}

Because of the finite number of clusters of given flux across the entire sky, the sample dubbed `out' does not exactly reflect the cluster population as it lacks all clusters being pointed.
As both subsamples derive from the same parent distribution, we use a single parameter for the inner excess and the outer dearth of clusters. We detail in App.~\ref{bias_details} our procedure to infer its value, taking into account the effective areas of both the inner and outer part of the XMM field of views. All count-rate bins $j$ are treated separately and we compute a bias factor $F_j$ (Fig.~\ref{fig_obs_bias_result}) whose value represents the ratio between the observed number of clusters in the considered bin and the actual expected number of clusters if no object were pointed. By definition, $F_j$ is always greater or equal to one.

%
%
\section{The cosmological analysis}
\label{cosmological_analysis}
This section presents the analysis performed with the subsample of 347 C1$^{+}$ clusters selected over our effective area of 90\,deg$^2$ from the XMM archival data. We first show the resulting CR--HR diagram, which is the sole observable quantity used in the cosmological analysis. We then describe its modeling from first principles, taking into account a cosmological model, X-ray cluster scaling-laws and the various selection effects affecting the sample. We finally show the results obtained by a Monte-Carlo Markov Chain sampling.

	\subsection{Sample CR--HR distribution}
\begin{figure}
	\includegraphics[width=84mm]{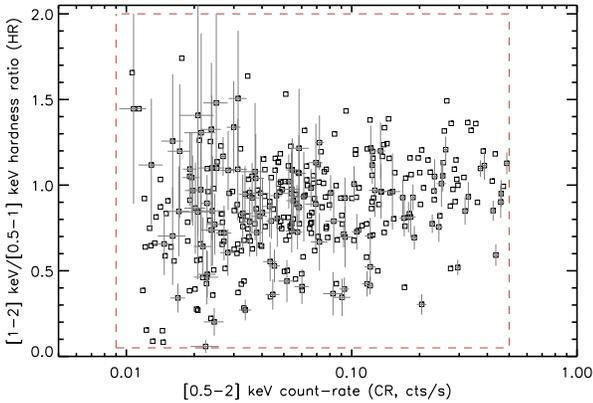}
 \caption{CR-HR diagramme for the 347 clusters  pertaining to the cosmological subsample. Not all error bars are displayed in order to ease visualization. The dashed box delimitates the region used for the cosmological fit.}
 \label{fig_obs_crhr} 
\end{figure}

Following Sect.~\ref{data}, count-rates for all 347 clusters entering the analysis have been measured in three energy bands: [0.5-2], [0.5-1] and [1-2]\,keV. These values have been corrected from  flux loss due to the finite aperture measurement and, if necessary, converted in the {\scshape Thin1} filter set.
We compute the hardness ratio of each cluster by dividing the [1-2]\,keV count-rate by the [0.5-1]\,keV measurement and report its value in a CR--HR plane, where CR stands for the wide band measurement.
Fig.~\ref{fig_obs_crhr} shows the distribution of clusters in this diagram, along with associated error bars. The distribution is spread over the range of count-rates peaking around 0.3-0.4\,cts/s.

	\subsection{Modelling the CR-HR distribution of sources}
We describe the main steps of the computation of the CR-HR distribution of clusters, starting from the halo mass function and using the survey selection function. These steps are more thoroughly detailed in paper~I.
	
		\subsubsection{Halo mass and redshift distribution}
We assume a $\Lambda$CDM cosmological model and a flat Universe ($\Om+\OL=1$) with no-evolving dark energy ($w=-1$). Starting from the primordial, scale-invariant power spectrum with slope $n_s=0.961$ \citep{Dunkley:2009p3181} we make use of the \citet{Eisenstein:1998p912} fitting formula for the transfer function to obtain the linear power-spectrum. We use the \citet{Tinker:2008p1554} fit to the mass function and obtain the comoving density of haloes per mass interval $d\mdcb$ about $\mdcb$ at redshift $z$, where $\mdcb$ is the mass within a radius $\rdcb$ inside which the mean density is 200 times the mean density of the Universe at that redshift. We convert this distribution into the sky-projected density of clusters per redshift slice. Only $\Om$ and $\sigma_8$ (normalization of the power spectrum at a scale $R=8 h^{-1}$\,Mpc) are let free in the analysis. All other parameters are held at their WMAP-5 value \citep{Dunkley:2009p3181}, in particular $H_0=72.4$\,km/s/Mpc.

		\subsubsection{Cluster emissivity and extent}
We assign to each cluster three quantities describing its X-ray emission: its redshift $z$, its plasma temperature $T$ (X-ray spectral temperature) and its bolometric luminosity $L_X$ integrated over its complete extent. We consider a mean metallicity of 0.3\,$Z_{\odot}$.
The conversion from cluster mass to temperature and luminosity is achieved thanks to scaling relations of the form:

\begin{equation}
	\label{equ_mt}
   \frac{\mdcc}{10^{14} h^{-1} {\rm M}_{\odot}}  = 10^{\normmt} \Big(\frac{T}{4 {\rm keV}}\Big)^{\powmt} E(z)^{-1} (1+z)^{\zevolmt}
\end{equation}
\begin{equation}
	\label{equ_lt}
   \frac{L_X}{10^{44} {\rm ergs/s/cm}^2}  = 10^{\normlt} \Big(\frac{T}{4 {\rm keV}}\Big)^{\powlt} E(z) (1+z)^{\zevollt}
\end{equation}	
To account for the intrinsic scatter in those relations, we assume two parameters $\scattmt$ and $\scattlt$ constant over the entire redshift, mass and temperature ranges considered in our analysis. Throughout this work we will use the $\mdcc$--T relation from \citet{Arnaud:2005p392} derived for their hot cluster sample (see Table~\ref{table_parameters}).

As discussed in paper~I, a reasonable choice for the emissivity profile of clusters is a $\beta$-model \citep{Cavaliere:1976p375} with $\beta=2/3$ and a core radius $r_c$ scaling with $\rccc$, parametrized by $\xc= r_c/\rccc$ at all redshifts and masses. This parameter is critical as it enters in the selection function describing the observed population of clusters.

		\subsubsection{Instrumental model and measurement errors}
The next steps consist in converting $z$, $T$ and $L_X$ into observable quantities and fold the cluster distribution into the selection function in order to obtain a CR--HR diagram for a given set of model parameters. Ideally, one should compute this distribution separately on each pointing and sum up their independent contributions to derive the complete CR--HR diagram. In order to avoid excessively large computational times we decided to group pointings by similar $N_H$ and background values.

We use APEC spectral models with a metallicity of 0.3\,$Z_{\odot}$ along with EPIC XMM response matrices to obtain count-rates in the three bands of interest. They represent the expected number of source events per second impacting the telescope cameras mounted with {\scshape Thin1} filters.
The [0.5-2]\,keV count-rate as well as the cluster apparent core-radii are then derived. Finally,  the sample selection function is used to compute the expected distribution of clusters.
Measurement errors are included by convolving the CR--HR distribution with an error model using the statistical uncertainties from the count-rate measurements.

To fully model the cluster population we add a supplementary step to the methodology presented in paper~I: the CR--HR model distribution is multiplied by the bias value $F(CR)$ as computed in Sect.~\ref{simulations} (see Fig.~\ref{fig_obs_bias_result}) to account for the excess of sources due to pointed observations.

	\subsection{Likelihood and MCMC sampling method}
		
				Given a set of parameters, the expression for the likelihood $L$ is expressed following e.g.~\citet{Cash:1979p927} by dividing the CR--HR two-dimensional space in narrow bins such that each bin contains at most one cluster. Assuming Poisson statistics in each bin, we can write:
\begin{eqnarray}\nonumber
\label{equ_likelihood}
	{\rm ln}\,\mathcal{L} &=& \sum_i {\rm ln} \Big(\dndcrdhrfrac(\crate_i,\hratio_i) \Big) \\
	&& - \int_{\crate_{\rm min}}^{\crate_{\rm max}} \int_{\hratio_{\rm min}}^{\hratio_{\rm max}} \dndcrdhrfrac \,d\crate \, d\hratio
\end{eqnarray}	
		where we have neglected the constant term including the size of the bins as we will consider likelihood ratios only.
In the equation above, the sum runs over the 347 selected clusters. The integral is computed for $0.09 \leq \crate \leq 0.5$ and $0.05 \leq \hratio \leq 2$ and simply represents the expected number of clusters within this CR--HR region.
The modeled number density contains the effect of measurement errors and pointed cluster bias and includes the selection function and its variations over the surveyed area, thus is as close as possible to the real CR--HR distribution of sources.
We summarize in Table~\ref{table_parameters} the choice of parameters and priors we made for the different cases studied.

Confidence intervals and mean values for the parameters being studied are obtained via the Bayesian formalism and are computed from the posterior distribution of parameters given the CR--HR diagram and the underlying model. In this study, we limit ourselves to a maximum of 5 free parameters, for which a Monte-Carlo Markov Chain (MCMC) likelihood exploration becomes competitive with a grid-based computation. We use a custom Metropolis-Hastings sampler that produces MCMC chains. After a so-called {\it burn-in} period, the chain reaches a stationary state representative of the actual posterior parameters distribution.
The `jump' function is taken as a multivariate gaussian distribution with covariance matrix $C_{\mu \nu}$ computed with (see~paper~I for a more thorough description and references therein):

\begin{equation}
 F_{\mu \nu} = C_{\mu \nu}^{-1} = \sum_i \frac{1}{\obs_i} \frac{\partial \obs_i}{\partial \theta_{\mu}} \frac{\partial \obs_i}{\partial \theta_{\nu}}
\end{equation}
where $\obs_i$ stands for the binned density $\dndcrdhr$ and ${\bmath \theta}$ is the set of varying parameters.
The PDF `jump' function from the current parameter set ${\bmath \theta}_n$ to the new one ${\bmath \theta}$ thus writes : 
\begin{equation}
 p( {\bmath \theta} | {\bmath \theta}_n ) \propto {\rm exp}\Big[-\frac{1}{2}\, {}^T({\bmath \theta}-{\bmath \theta}_n) F_{\mu \nu} ({\bmath \theta}-{\bmath \theta}_n) \Big]
\end{equation}

For each chain we extract the mean value of the sampled parameters and compute the associated highest density intervals. Such an interval contains $(1-\alpha)100\%$ of the posterior probability and ensures that the posterior density within the interval is always greater than outside. We choose $\alpha=0.32$ which, in the case of a normal distribution, corresponds to the 1-$\sigma$ boundaries of the distribution. The MCMC analysis is performed using the R-package BOA \citep{Smith:2007p5859}. Unless otherwise stated, best-fit results are quoted using the mean of the posterior distributions (and not the maximum likelihood estimate).

	\section{Results}
	\label{results}

Our methodology, as presented in paper~I, assumes that local scaling laws are known, we thus focus on their evolution and on the cosmological parameter determination.
Different expressions for the local scaling laws are found in the current literature. This is particularly true for the L--T relation \citep[e.g.][]{Arnaud:1999p398,Branchesi:2007p6150,Maughan:2011p5987}, likely because of the different populations being taken into account \citep{Pratt:2009p322,Mittal:2011p6134}, but also because of different selection effects.
However, our cluster sample is not quite large enough to allow for a simultaneous fit of the local scaling laws, of their evolution and of cosmology.
We thus proceed with a step by step approach to select a local L--T that matches well our sample, focusing on two relations from \citet{Pratt:2009p322}. We use these relations since they are well suited to the mass range of our sample, contrary to, e.g.~\citet{Mantz:2010p1258} scaling relations which have been derived for much more massive clusters.
We first set the cosmology to the WMAP-5 values and select a `best' relation by comparison to our data and to published log$N$-log$S$ from the literature.
We then release cosmological parameters to perform an enlarged fit of our data and finally show that the selected L--T still adequately describes our clusters.
In all cases, we always assume a flat $\Lambda$CDM Universe.

\begin{table*}
	\centering
			\caption{\label{table_parameters} List of parameters used in this work.
		The cosmological parameters are from WMAP-5 \citep{Dunkley:2009p3181}. Numbers in brackets indicate the uniform priors applied in the MCMC fitting procedure. When fixed, $\Om$ and $\sigma_8$ are held at their WMAP-5 values, namely  $\Om=0.249$ and $\sigma_8=0.787$.
		`ALL' and `NCC' refer to the corresponding L--T relations from \citet{Pratt:2009p322}.}
		\begin{tabular}{@{}ccl@{}}
Parameter 	&	Fixed value or [prior range]	&	Description		\\
\hline
$\Om$								&		$[0.09-1]$	&\\
$\OL$								&		$1-\Om$		&(Flat Universe)\\
$\Ob$								&		0.043		&	\\
$\sigma_8$							&		$[0.05-2]$		&	\\
$n_{s}$								&		0.961		&	\\
$h$									&		0.72		&	\\
\hline
$\powmt$							&		1.49			&$M-T$ power-law index						\\
$\normmt$ 							&		0.46			&$M-T$ logarithmic normalization			\\
$\zevolmt$							&		$[-4,4]$		&$M-T$ evolution index				\\
$\scattmt$							&		0.1				&$M-T$ constant logarithmic dispersion			\\
\hline
$\powlt$							&		`ALL': 2.7, `NCC':2.9			&$L-T$ power-law index						\\
$\normlt$							&		`ALL': 0.52, `NCC':	0.40		&$L-T$ logarithmic normalization		\\
$\zevollt$							&		$[-5,3]$			&$L-T$ evolution index				\\
$\scattlt$							&		0.3 or 0.7		&$L-T$ constant logarithmic dispersion			\\
\hline
$\xc$								&		$[0-0.9]$		&$\beta$-model core radius scaling wrt. $\rccc$	\\
\hline
		\end{tabular}
\end{table*}

		\subsection{Fixed cosmology, fixed local scaling laws}
		
			\subsubsection{Constraints from the CR--HR distribution}
We first fit $\zevolmt$ and $\zevollt$, the parameters governing the non-self-similar behaviour of the M--T and L--T relations, as well as $\xc$, the cluster size. Other parameters are set according to Table~\ref{table_parameters}, in particular $\Om$ and $\sigma_8$ which are held at their WMAP-5 value and we assume that scaling relations are perfectly known.
For the temperature to luminosity conversion we consider the numerical values of \citet{Pratt:2009p322} using two of their $L_1-T_1$ relation without core excision. One of them has been computed for all clusters present in their sample (`ALL') while the other (`NCC') excludes all clusters showing high central gas density and thus hosting a cool core.
As the intrinsic scatter in these relations depends on the population of clusters in the sample, we use two test values, $\scattlt=0.3$ and $0.7$; these values correspond to the lowest and the highest scatter found in \citet{Pratt:2009p322}. We emphasise here that we allow us to somewhat generalise their results since, formally, scatter values of 0.3 and 0.7 are associated to the `NCC' and 'ALL' samples respectively.

The fit results with associated uncertainties are quoted in Table~\ref{table_mcmc_3par}, along with an indication of the relative likelihood for each test case.
The Cash statistics $C=-2{\rm ln}\,\mathcal{L}$ is evaluated at the mean of the posterior distribution.
As changing from one L--T relation to the other involves 3 parameters in our model ($\powlt$, $\normlt$, $\zevollt$), the difference between $C$ and the (unknown) minimal value $C_{\min}$ obtained when all 6 parameters are let free behaves as a $\chi^2$ with 3 degrees of freedom \citep{Cash:1979p927}. This allows us to put a lower boundary to the $\chi^2$ of all four fits, e.g.:
\begin{eqnarray}\nonumber
\chi^2_{{\rm ALL},0.7} &=& C_{{\rm ALL},0.7} - C_{\min} \\
\nonumber
	&	>& C_{{\rm ALL},0.7} - C_{{\rm NCC},0.7} =  8.4
\end{eqnarray}

Because the probability for $\chi^2_3$ to be greater than 8 is $\sim 5$\%, it indicates that our data prefers the `NCC' scaling relation with a large intrinsic scatter to the three other scaling laws.

\begin{table*}
	\centering
			\caption{\label{table_mcmc_3par}
			 Best-fit values for the evolutionary parameters $\zevolmt$, $\zevollt$ and the geometrical scaling factor $\xc$ for a fixed cosmology. Quoted results are the mean and 68\% confidence intervals obtained by fitting the X--CLASS CR--HR distribution, while cosmological parameters are held fixed at their WMAP-5 value and only those 3 parameters are varied.
			 The $C = -2 \ln {\mathcal{L}_{max}}$ values are computed at the location of the best-fit parameters and differences are quoted relative to that obtained for `NCC' with large scatter (2nd column).
`ALL' and `NCC' refer to the corresponding L--T relations from \citet{Pratt:2009p322}			}
			
		\begin{tabular}{@{}lcccc@{}}
Local L--T:		&	\multicolumn{2}{c}{NCC}					&	\multicolumn{2}{c}{ALL}						\\
$\scattlt$:		&	0.3					&		0.7				&	0.3					&	0.7		\\
\hline
$\zevolmt$	&	 $0.60 \pm 0.15$		&	 $0.32 \pm 0.13$		&	$0.17 \pm 0.18$				&	$-0.13^{+0.16}_{-0.12}$			\\
\\
$\zevollt$	&	$-1.23 \pm 0.41$		&	$-1.30^{+0.54}_{-0.37}$	&	$-2.25^{+0.61}_{-0.48}$		&	$-2.06^{+0.56}_{-0.43}$			\\
\\
$\xc$		&	 $0.17 \pm 0.02$		&	 $0.26 \pm 0.03$		&	$0.27 \pm 0.02$				&	$0.39 \pm 0.04$					\\
\hline
$C - C_{{\rm NCC},0.7}$
			&		$7.9$				&			($0$)			&			$9.1$					&	$8.4$							\\
		\end{tabular}
\end{table*}

	\subsubsection{Comparison with published log$N$-log$S$ distributions}
\begin{figure*}
	\begin{tabular}{cc}
	\includegraphics[width=84mm]{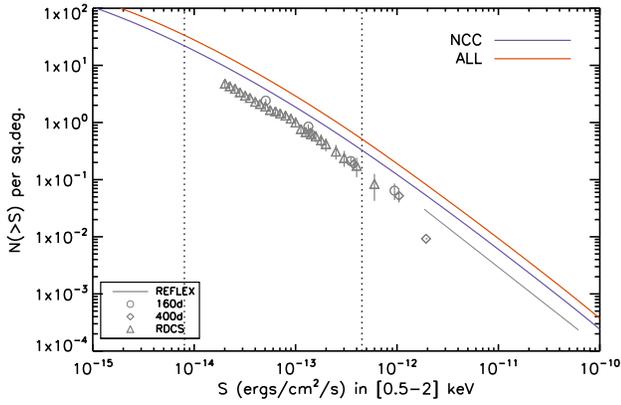} &
		\includegraphics[width=84mm]{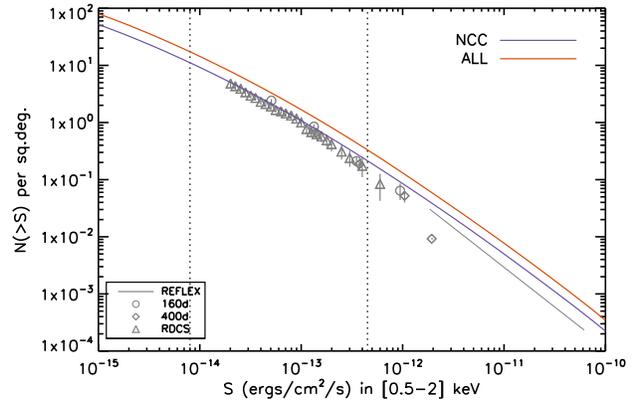}
		\end{tabular}
 \caption{The cluster log$N$-log$S$. The blue and red curves correspond to our predictions assuming the local `NCC' and `ALL' L--T relations respectively, both taken with $\scattlt=0.7$ and are computed for the WMAP-5 cosmology. {\it Left panel:} Self-similar evolution is assumed for the scaling laws ($\zevolmt=\zevollt=0$). {\it Right panel:}  Evolution as inferred from our 3-parameter best fit on the X--CLASS CR--HR distribution ($\zevolmt$ and $\zevollt$,  Table~\ref{table_mcmc_3par}). Data points correspond to observations from various surveys (see text). The vertical lines indicate the approximate flux range of our sample. 
}
 \label{fig_lognlogs_3par} 
\end{figure*}

We cross-check  the outcome of our 3-parameter fit by comparing the resulting cluster log$N$-log$S$ to that from other surveys. We show on Figure~\ref{fig_lognlogs_3par} the log$N$-log$S$ computed with two local L--T (`NCC' and `ALL', $\scattlt=0.7$). We display the result with and without the $(1+z)^{\gamma}$ evolution factor in the scaling laws. In the former case, we used our best-fit parameters $\zevolmt$ and $\zevollt$ from Table~\ref{table_mcmc_3par}. We note that the value of $\xc$ has no impact on the predicted log$N$-log$S$  since such a distribution is meant to be flux limited.
The REFLEX data correspond to \citet{Bohringer:2001p1461} best-fit power-law converted into a [0.5-2]\,keV log$N$-log$S$ using a constant factor calculated for a APEC plasma at $z=0$ with $T=5$\,keV. RDCS values are from \citet{Rosati:1998p3234} and 160d and 400d correspond respectively to \citet{Vikhlinin:1998p1674} and \citet{Burenin:2007p1251}.

From this figure it turns out that the `ALL' relation predicts too many clusters compared to observations from other authors, particularly for the brightest, most massive, nearby clusters. Conversely, the `NCC' scaling relation (also assuming $\scattlt=0.7$) is less discordant and our proposed evolution nicely fits the low-flux log$N$-log$S$ from the RDCS and the 400d surveys.  

	\subsection{Free $\Om$ and $\sigma_8$, fixed local scaling law}
	\label{result_5par}

\begin{figure*}
	\includegraphics[width=\linewidth]{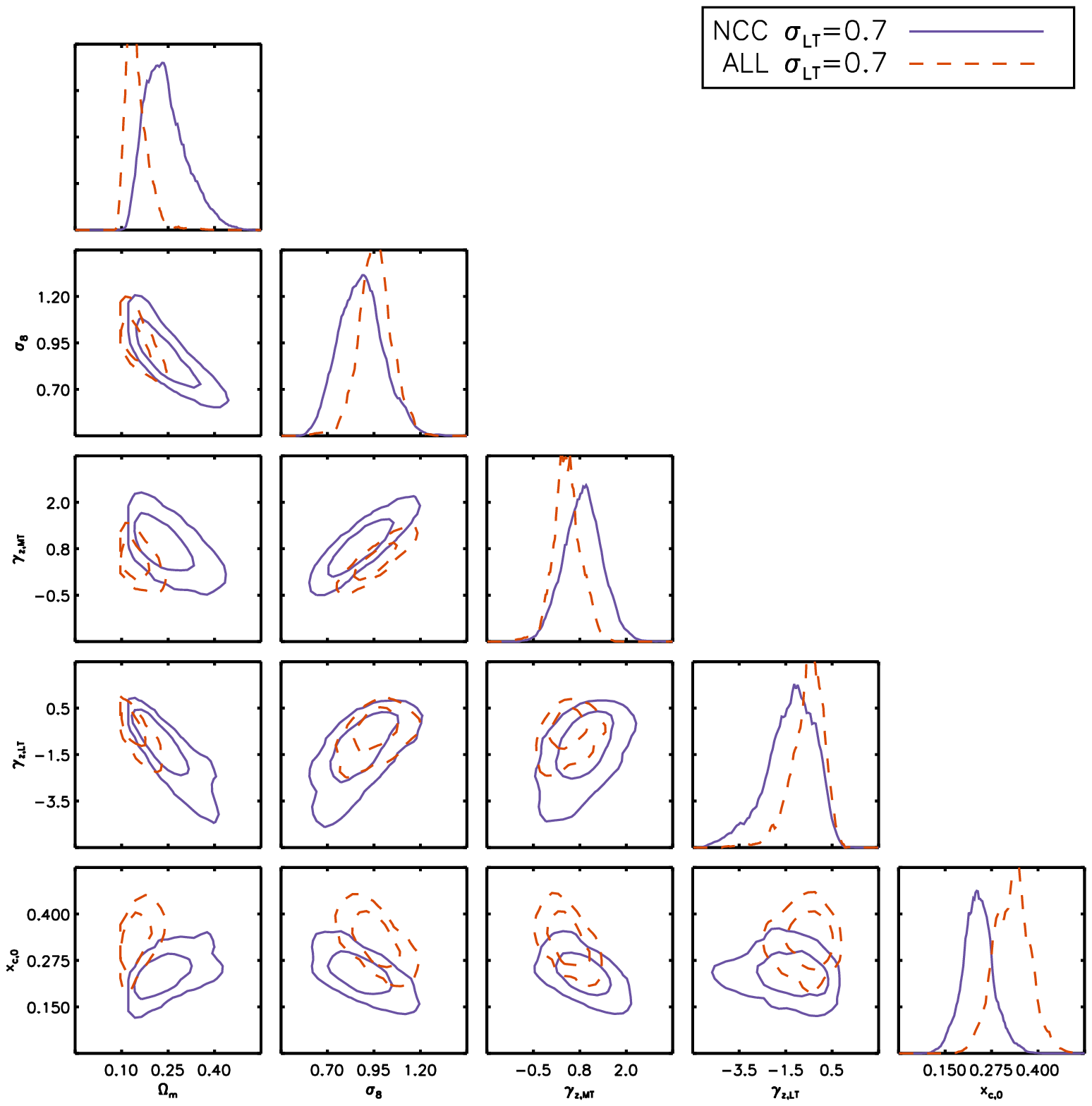}
 \caption{Posterior distribution for the five parameters fitted using the X--CLASS CR--HR distribution.
 	The local L--T relation is `NCC' or `ALL' with a logarithmic dispersion $\scattlt=0.7$.
	Diagonal panels represent the one-dimensional marginal distributions for each parameter (normalized to unit area), sub-diagonal panels show the two-dimensional contours enclosing 68\% and 95\% of the marginalized posterior distribution.
 	}
 \label{fig_mcmc_5par} 
\end{figure*}

\begin{figure}
	\includegraphics[width=\linewidth]{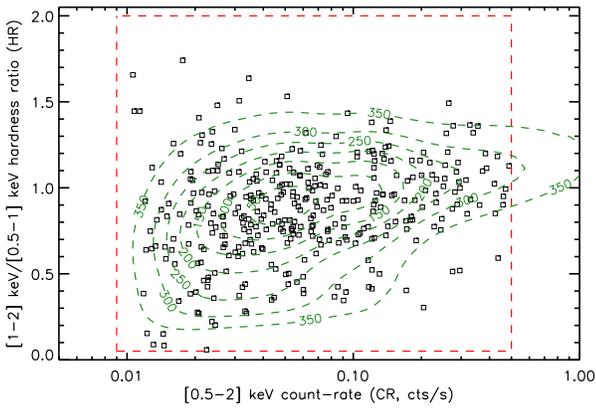}
 \caption{Best fit model (green dashed contours) overlaid on the CR--HR diagram data points.
 Five parameters are fitted assuming the local L-T relation from `NCC' and $\scattlt=0.7$.
 	Contour labels indicate the predicted number of clusters enclosed by the corresponding contours.
	The red box shows the region in which the fit is performed.
	Measurement errors and bias due to pointed clusters have been included in the computation of the green contours.}
 \label{fig_obs_compar} 
\end{figure}

We now relax $\Om$ and $\sigma_8$ while fitting the CR--HR diagram, in addition to $\zevolmt$, $\zevollt$ and $\xc$, thus allowing 5 parameters to vary in total.
We consider the `NCC' relation with $\scattlt=0.7$ as our reference L--T relation and keep it fixed in the analysis.
Figure~\ref{fig_mcmc_5par} shows the resulting posterior distribution obtained from the MCMC chains. The mean value and associated 1-$\sigma$ error bars for each parameter are :
\begin{eqnarray}\nonumber
	\Om=0.24^{+0.04}_{-0.09}, \\ \nonumber
	\sigma_8 = 0.88^{+0.10}_{-0.13}, \\ \nonumber
	\zevolmt = 0.83^{+0.45}_{-0.56}, \\ \nonumber
	\zevollt=-1.3^{+1.3}_{-0.7},\\ \nonumber
	 \xc=0.24 \pm 0.04.
\end{eqnarray}
We note that the values for the three last parameters are consistent with the previous results from the 3-parameter fit.
Figure~\ref{fig_obs_compar} illustrates the good agreement between the observed CR--HR distribution and the best-fit model.  Using the best-fit model we predict a total amount of 369 clusters in the region where the fit is performed (red dashed box on Fig~\ref{fig_obs_compar}) which is comparable to the 347 clusters actually present in the analysis.
We note the presence of a few outliers that will be discussed in the next section.

\begin{figure}
	\includegraphics[width=\linewidth]{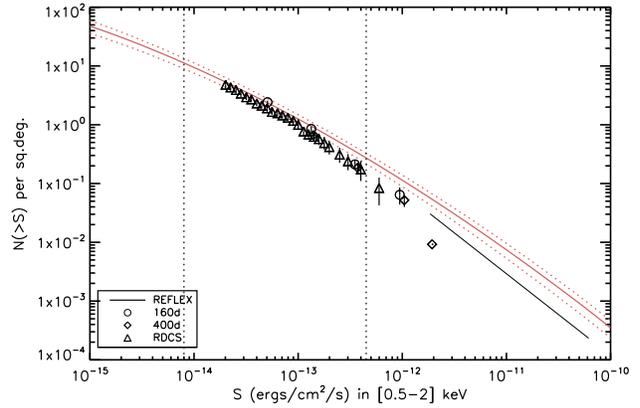}
 \caption{The cluster log$N$-log$S$.
 	The plain red curve shows the modeled distribution computed assuming the local L--T relation of 'NCC' with $\scattlt=0.66$. 
	Cosmological parameters as well as the non-similar evolution of scaling laws have been adjusted so as to match the CR--HR distribution of sources in the X--CLASS sample (best-fit parameters quoted in Sect.~\ref{result_5par}). The red dotted curves are computed by propagating the uncertainties on these parameters in the log$N$-log$S$ computation. The vertical lines indicate the approximate flux range of our sample.
	}
 \label{fig_lognlogs_5par} 
\end{figure}

We show on Figure~\ref{fig_lognlogs_5par} the cluster log$N$-log$S$ predicted by this set of best-fit parameters, along with the 1-$\sigma$ lower- and upper-boundaries obtained by propagating the posterior covariance matrix from the MCMC analysis.
The log$N$-log$S$ agrees well with the reference log$N$-log$S$ curves, especially in the range $10^{-14}-2.10^{-13}$\,ergs/s/cm$^2$. This interval roughly corresponds to the count-rate region probed by our CR--HR diagram (assuming an average flux conversion factor of $9.10^{-13}$\,ergs/s/cm$^{-2}$ per cts/s). The high- and low-flux ends of this curve are not probed by our data points but rather rely on the validity of the assumed model, and in particular on the fact that scaling laws behave as Equs.~\ref{equ_mt} and~\ref{equ_lt} at all redshifts. Such an extrapolation disagrees with REFLEX data points, as is the case in Fig.~\ref{fig_lognlogs_3par} where cosmological parameters are held at their WMAP value. Changes in the shape of scaling laws as a function of cluster properties and/or a disagreement between the different selection functions of various surveys may explain this discrepancy.

%
%
\section{Discussion}
\label{discussion}

	\subsection{Cosmological parameters}
Our analysis indicates values for $\Om$ and $\sigma_8$ of $0.24$ and $0.88$  respectively with a $\sim 25\%$ and $\sim 15\%$ accuracy.
These constraints are compatible with the most recent measurements from the CMB \citep{Dunkley:2009p3181, Larson:2010p179} and BAO observations \citep{Percival:2010p6104, Blake:2011p5018}.
They are also in agreement with the most recent   studies of  X-ray selected clusters
\citep{Vikhlinin:2009p1250, Henry:2009p4428,Mantz:2010p1259}, Sunyaev-Zeldovich selected clusters \citep{Vanderlinde:2010p6481, Sehgal:2011p5963} and optically selected clusters \citep{Rozo:2010p6386}.

Our results have been obtained for fixed local scaling relations, in particular the `NCC' L--T relation was taken from \citet{Pratt:2009p322}, with a constant logarithmic scatter of 0.7. We note that their relation has been derived for an assumed $\Lambda$CDM cosmology ($\Om=0.3$,$\OL=0.7$,$h=0.70$) and strictly speaking it should be converted for each tested cosmology. However, we checked that the correponding correction from $\Om=0.3$ to $\Om=0.24$ on the L--T normalisation amounts to less than 5\% and thus neglected this correction in the analysis.

We checked that assuming the L--T relation from \citet{Arnaud:1999p398} (converted in the WMAP-5 cosmology) leads to compatible constraints on $\Om$ and $\sigma_8$.
Interestingly, the `ALL'  ($\scattlt =0.7$) from \citet{Pratt:2009p322} predicts $\Om=0.15 \pm 0.04$ and $\sigma_8 = 0.96 \pm 0.08$, in disagrement with the other works cited above. Fig~\ref{fig_mcmc_5par} compares the confidence contours obtained for both scaling relations and illustrates the marginal agreement of the posterior distributions.
In all cases (`NCC', `ALL' and \citealt{Arnaud:1999p398}) we tested two different values for the intrinsic scatter $\scattlt$ (0.3 and 0.7) and noticed little change in the $(\Om,\sigma_8)$ constraints. This result is compatible with  \citet{Sahlen:2009p298} who find very little degeneracy between $\scattlt$ and cosmological parameters.

\begin{figure}
	\includegraphics[width=84mm]{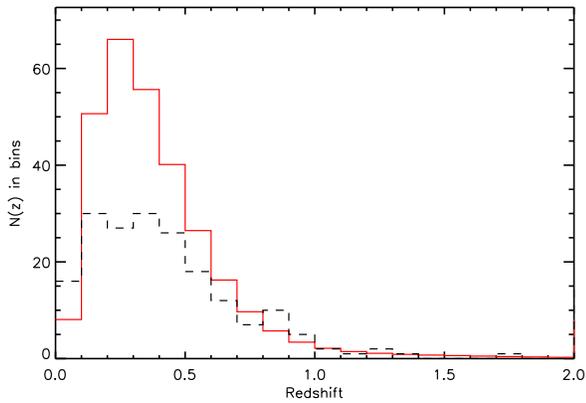}
 \caption{Predicted redshift distribution for the ``C1$^{+}$" clusters in the X--CLASS cosmological subsample (red solid histogram). The redshift distribution of  clusters in our sample with known redshifts from the literature is indicated by the black dashed line.}
 \label{fig_dndz_predicted} 
\end{figure}

From the best set of parameters (Sect~\ref{result_5par}), we infer the redshift distribution of  our cosmological cluster sample (Fig.~\ref{fig_dndz_predicted}) for the pure C1$^{+}$ selection (with no bias from pointed clusters). It shows ({\it a posteriori}) that  the distribution peaks around $z_{med} \sim 0.3-0.4$ with only 8.4 C1$^{+}$ clusters  beyond $z=1$. The redshift histogram of clusters with known redshifts from the literature clearly illustrates the complexity of this bias in our sample. All sources in the first $0<z<0.1$ bin  have a redshift, and there is a small excess of pointed clusters between $z=0.8$ and $z=1$.
In total, 188 clusters out of the 347 ones selected for the CR--HR analysis have a redshift (66 of them being flagged as `tentative' in our database, see Sect.~\ref{catalogue_details}).
However, including this partial redshift information in the MCMC analysis requires a precise knowledge of the associated selection process which is currently out of reach.

	\subsection{Cluster X--ray profiles}
\begin{figure}0
	\includegraphics[width=84mm]{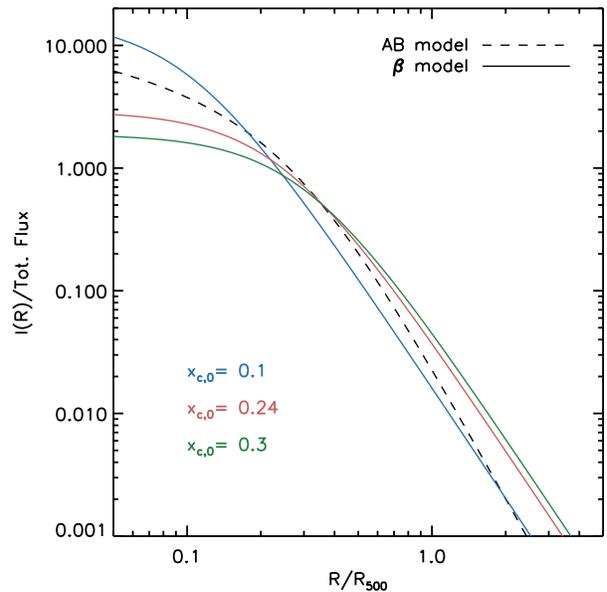}
 \caption{Comparison between the AB model from \citep{Croston:2008p776, Piffaretti:2010p1548} and the $\beta$-model used in this analysis ($\beta = 2/3$).
	These curves show surface brightness profiles normalized to the same total flux.
	The red curve corresponds to a $\beta$ profile computed with the best-fit parameter $\xc$ from the MCMC analysis.}
 \label{fig_compar_AB-beta} 
\end{figure}

	To account for the physical extent of clusters entering the survey selection function, we have introduced the $\xc$ parameter linking the X-ray cluster extent to the cluster mass ($\xc= r_c/\rccc$).
	Contrary to other studies, we did not assume a fixed physical size for the core radius  \citep{Pacaud:2007p250, Sahlen:2009p298} nor a size distribution  \citep{Burenin:2007p1251}. 
	In all configurations we investigated, this parameter is constrained at the $10-20$\% level and is found to have a value of $0.24 \pm 0.04$ in the best-fit model.
For a cluster of mass $\mdcc = 10^{14} h^{-1}$\,M$_{\odot}$ at redshift $z=0.4$,   $\rccc$ is $ \sim 0.6$\,Mpc,  and our result suggests a physical core radius of about 150\,kpc, indeed typical of those found in other cluster studies at a similar redshift \citep[e.g.][]{Pacaud:2007p250, Vikhlinin:1998p1674, Burenin:2007p1251}.
As shown on Fig.~\ref{fig_compar_AB-beta}, there is a reasonable agreement between a $\beta$-profile computed with our value of $\xc$ and the AB-model used in \citep{Piffaretti:2010p1548},  based on   the local gas density profiles  of the REXCESS clusters \citep{Croston:2008p776}.

	\subsection{Scaling-laws evolution}

We parametrised the redshift evolution of each scaling law by two factors of the form $(1+z)^{\gamma}$ such that $\gamma \neq 0$ indicates a departure from self-similar evolution (Equ.~\ref{equ_mt} and ~\ref{equ_lt}).
The modelling of the CR--HR distribution includes self-consistently selection effects and the evolution of cluster scaling relations which is a key point of such analyses \citep[e.g.][]{Pacaud:2007p250, Short:2010p388, Mantz:2010p1259}.

\begin{figure*}
	\begin{tabular}{cc}
	\includegraphics[width=84mm]{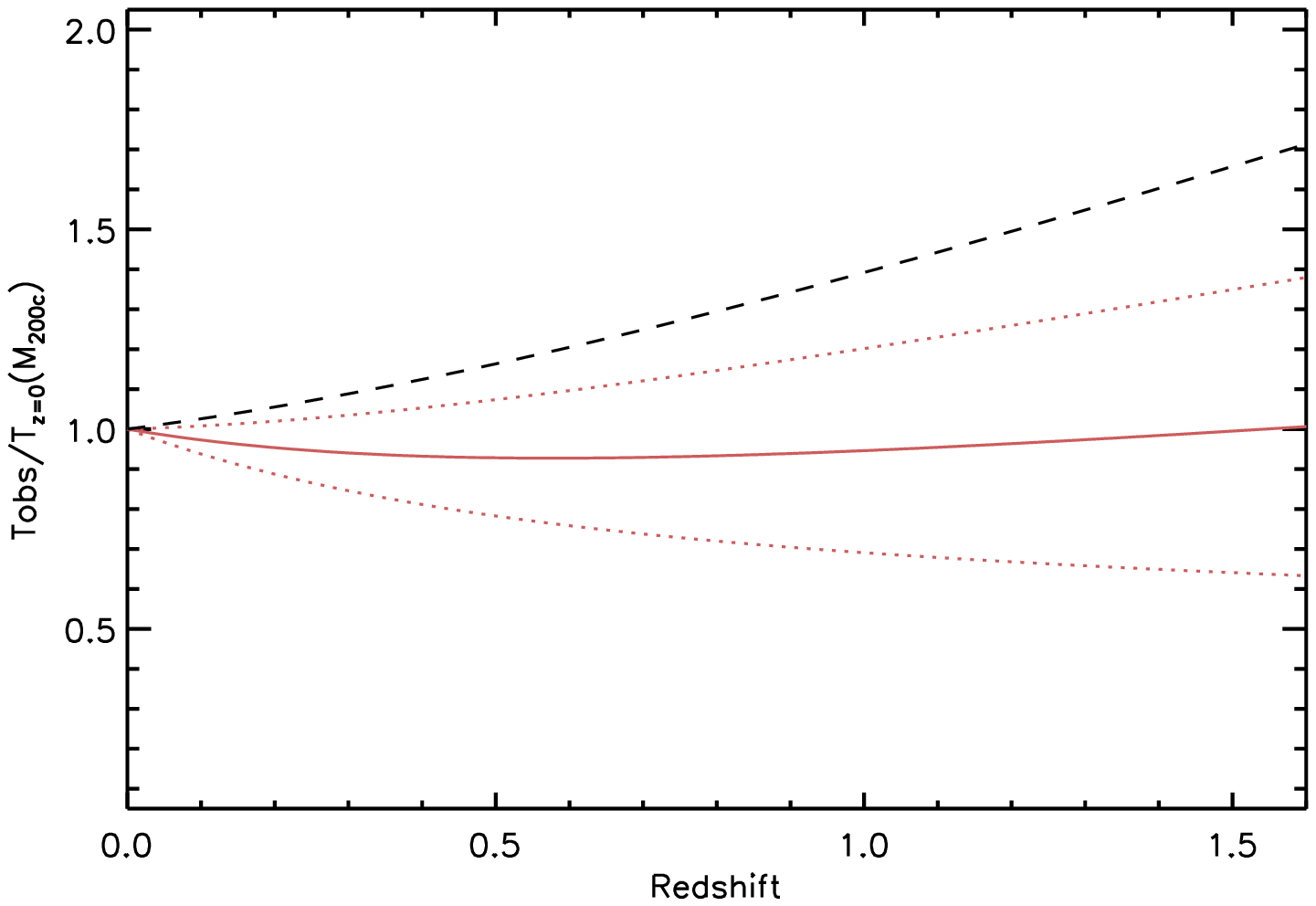}	&
	\includegraphics[width=84mm]{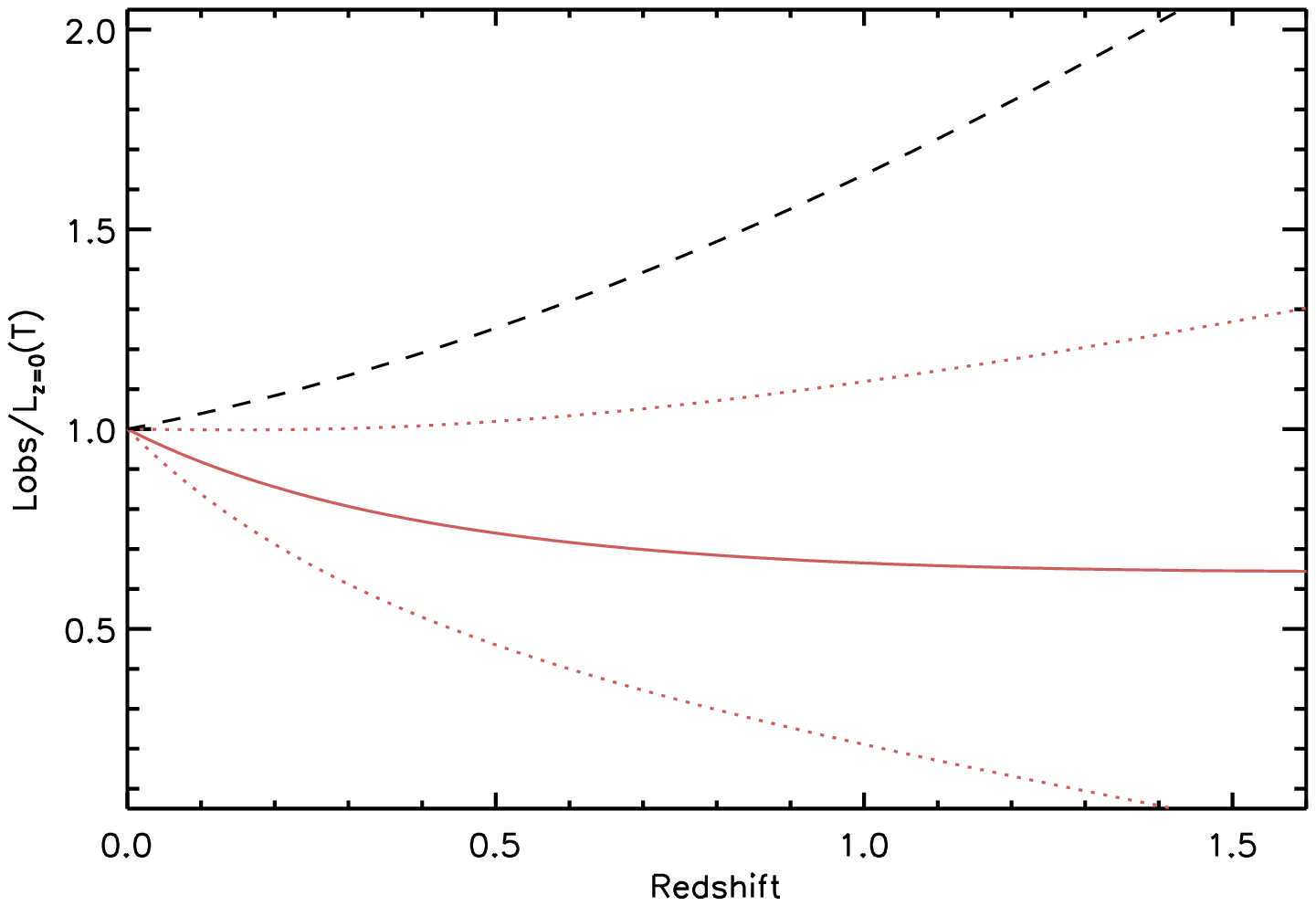}	\\
	\end{tabular}	
 \caption{Scaling-law evolution as predicted by our model and  parametrised by  equations \ref{equ_mt} and \ref{equ_lt}. The solid red line shows  the best-fit model ($\Om=0.24$, $\zevolmt=0.8$, $\zevollt=-1.3$)  with the dotted line indicating the 68\% confidence boundaries.
 The black dashed line shows the self-similar expectation ($\zevollt = \zevolmt =0$). }
 \label{fig_evolution_sl} 
\end{figure*}

We illustrate in Fig.~\ref{fig_evolution_sl} the net effect of the combined factors $E(z)$ and $(1+z)^{\gamma}$ in the evolution of scaling laws, using the best-fit values from our analysis. In this figure, the T--M relation has been computed by inverting Eq.~\ref{equ_mt} and is expressed in terms of the crticial mass $\mdcc$. The upper and lower boundaries have been computed using the covariance matrix of $\Om$ (which enters $E(z)$), $\zevolmt$ and $\zevollt$ output of our MCMC chain.

Our fit to the X--CLASS CR--HR diagram indicates a quasi non-evolving T--$\mdcc$ relation (equivalently, a negative evolution relative to the self-similar expectation), meaning that a cluster with a given mass $\mdcc$ shows approximately the same temperature at all redshifts.
Our data also indicates a negative evolution of the L--T relation, below the self-similar expectation,  a result that has been found in simulations of \citet{Kay:2007p6124} where feedback by AGN and stars is included in cluster simulations, and in   \citet{Short:2010p388} for their preheating model of cluster evolution.
These trends are also observed by \citet{Ettori:2004p6146} for the M--T and L--T relations but are in conflict with studies from \citet[e.g.][]{Kotov:2005p6390, Branchesi:2007p6150}.
However, the comparison between these studies is hampered by the different selection processes entering different cluster samples, which can have a tremendous effect on the derived evolution \citep{Pacaud:2007p250}.

Finally, we note that our results do not firmly exclude self-similar evolution  both in the M--T and the in L--T relations. The relatively large uncertainties on $\zevolmt$ and $\zevollt$ are due to the absence of redshifts  and to the degeneracies with cosmological parameters inherent to the CR--HR analysis (Fig.~\ref{fig_mcmc_5par}).
We expect the forthcoming XXL and XCS surveys to provide tighter constraints on the evolution of scaling laws thanks to the inclusion of cluster redshifts in both analyses.

	\subsection{CR--HR outliers}
Fig~\ref{fig_obs_compar} shows the good agreement between our best-fit model and the CR--HR distribution of clusters in the sample. However, we note the presence of sources outside the bulk of the diagram.
The top-left part\footnote{X--CLASS tags: 35, 86, 915, 997, 1032, 1655, 1741, 1886, 1947.} ($CR<0.06$, $HR>1.4$) contains regular, weak clusters for which measurement errors are large, particularly on HR (Fig.~\ref{fig_obs_crhr}). Such hardness ratios could be explained by, e.g. a high metallicity of the intra-cluster medium favoring emission from bright spectral lines.
The bottom-right region\footnote{X--CLASS tags: 102, 238, 541, 1020, 1218, 1480, 1906, 1937, 2046, 2048, 2162, 2321.} ($CR>0.1$, $HR<0.7$) contains 9 clusters whose morphology appeared strongly peaked at the centre, hence indicating a significant AGN or cool-core contribution; the net effect is to increase our integrated CR and modify the HR, with respect to normal clusters. One outlier (xclass~1937) is a compact group (HCG~057, \citealt{Hickson:1989p6385}) presenting a complex X-ray emission.
A complete model would enclose cluster spectral peculiarities in the calculation of the CR-HR diagram; however, we consider here that their number is sufficiently low to neglect their impact onto our results.

	\subsection{Predictions for eRosita}
In paper~I, we presented Fisher forecasts for an XMM  100\,sq.~deg. cluster survey at a 10\,ks depth, providing a sample of 570 clusters. The CR--HR method was predicted to yield the following accuracy: $\sigma(\Om)=0.09$, $\sigma(\sigma_8)=0.14$, $\sigma(\zevolmt)= 0.6$, $\sigma(\zevollt)=2.3$ and $\sigma(\xc)=0.04$. 
The present study has been conducted for similar conditions (comparable area, a somewhat more stringent selection function and partially deeper exposures).
The good agreement between the predicted uncertainties and the output of the MCMC runs shows that both analyses are consistent with each other. We note however that there is not a one-to-one correspondence between them as the Fisher analysis involved more free parameters (though constrained with stringent priors) and there are 347 clusters in the present study, which are on average better measured than assumed in paper~I.
The difference in the total number of clusters is explained by: i) the more stringent selection function, ii)~the actual 90~sq.deg. coverage of the X--CLASS survey, iii)~inhomogeneities in the survey depth in terms of pointing background and hydrogen column density and iv)~differences in the fiducial parameters of the Fisher matrix and those derived in the present work (e.g.~$\xc=0.1$ in paper~I and $0.2$ in this paper).

The overall good agreement between predicted and measured uncertainties allows us to propose  general predictions for the Rosita all-sky survey \citep{Predehl:2010p5342} to be obtained by the CR-HR method, following the formalism developed in paper~I. We assume a total area of 20,000~sq.deg. (extragalactic survey) and a custom selection function being a scaled-up version of the C1$^{+}$ selection (Fig.~\ref{fig_simulations}). Within our fiducial model, the survey is expected to yield $2.5$ clusters per sq.deg., hence a total of 50,000 detected clusters. In contrast to paper~I, we do not assume priors on local scaling laws. We allow their normalization, slope and intrinsic scatter to vary, and let them evolve with redshift as $(1+z)^{\gamma}$. We assume a prior Fisher matrix on $\Om$, $\sigma_8$, $\Ob$, $n_s$ and $h$ as will be available from {\it Planck} and calculated identically as in \citet{Pierre:2011p5484} (based on the {\it Planck} mission definition -- {\it Bluebook}\footnote{http://www.rssd.esa.int/index.php?project=Planck}).

\begin{figure*}
	\begin{tabular}{cc}
		\includegraphics[width=84mm]{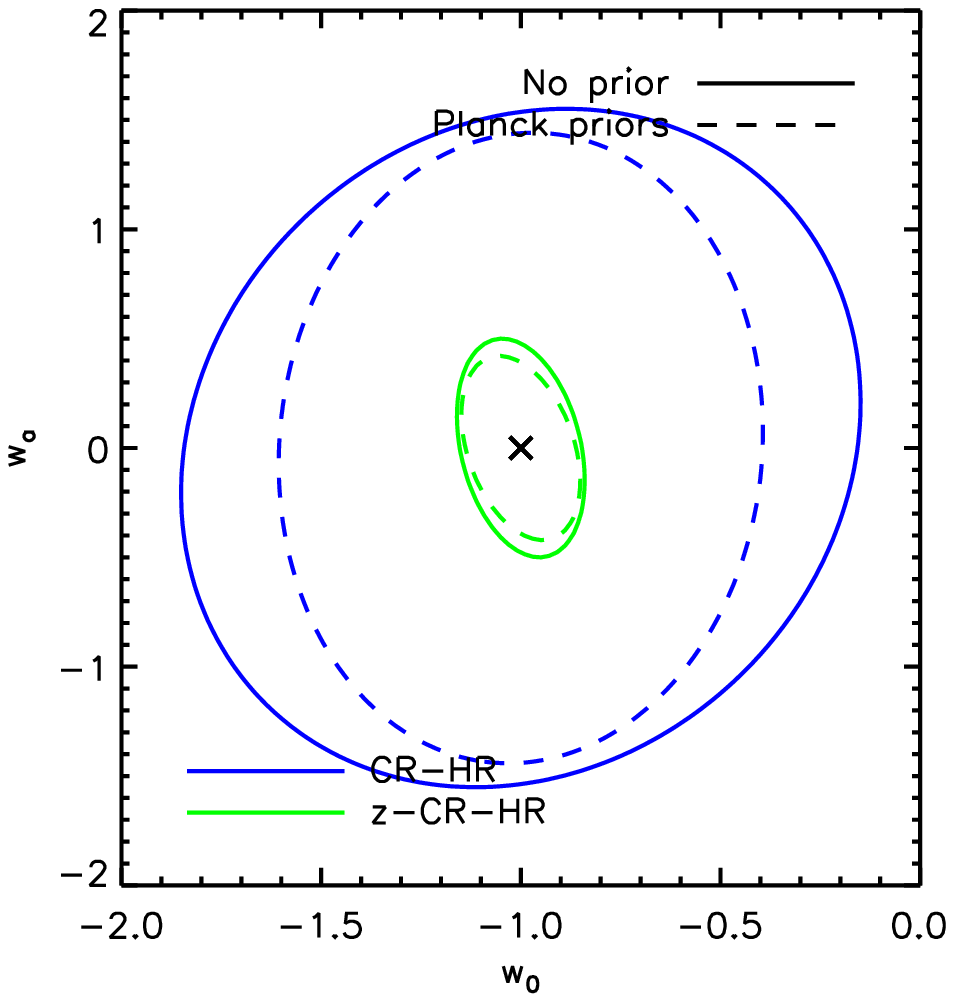} &
		\includegraphics[width=84mm]{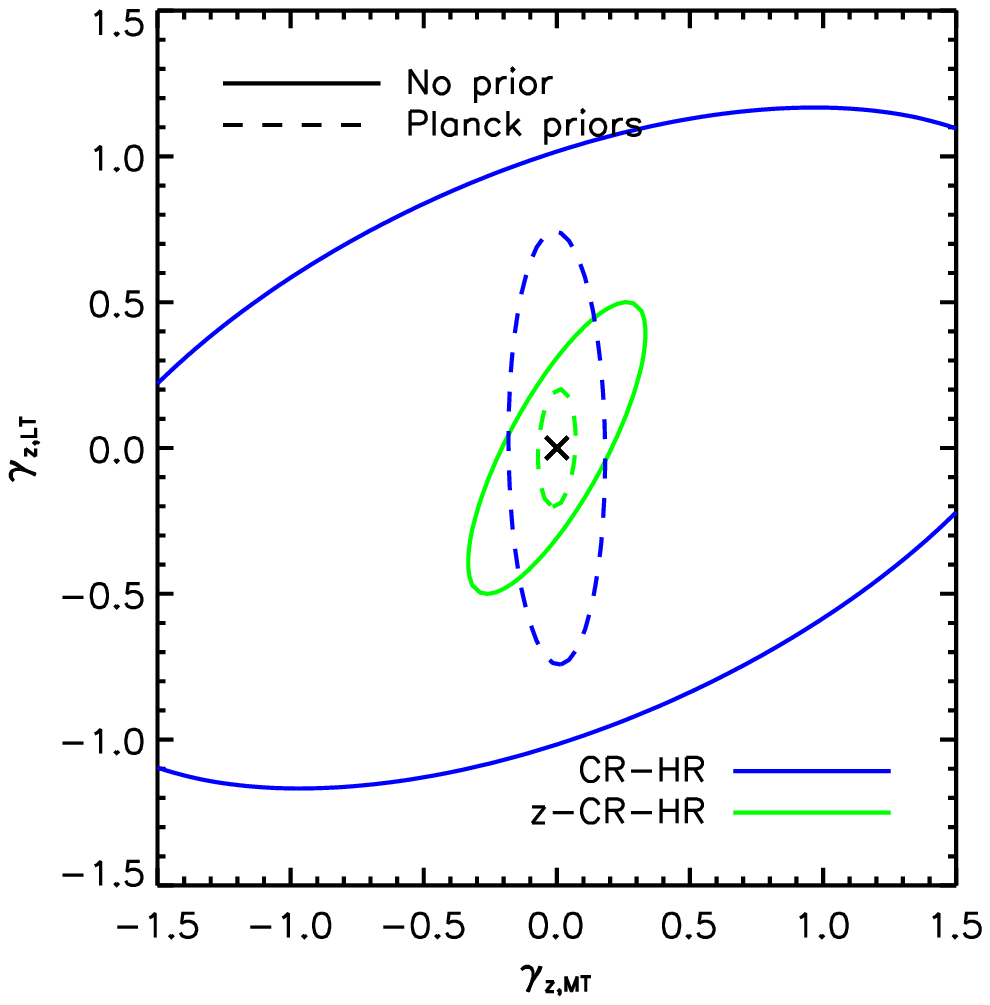} \\
	\end{tabular}		
	 \caption{{\em Left:} constraints on the dark energy parameters $w_0$ and $w_a$ for the eRosita 20000\,sq.deg survey as predicted by the CR-HR method (blue), possibly supplemented by photometric redshifts for all clusters (z-CR-HR, green).  No assumption has been made on scaling laws nor on their evolution and Planck priors were applied to $\Om$, $\sigma_8$, $\Ob$, $n_s$ and $h$. The {\em right} panel shows the predicted constraints for cluster evolution}
 \label{fig_erosita} 
\end{figure*}

We estimate measurement errors by assuming a mean exposure time of $2.5$\,ks and an effective area equal to that of XMM MOS+PN \citep{Predehl:2010p5342}, thus applying a factor 2 to the uncertainties quoted in paper~I for a 10\,ks XMM survey.
We considered two extreme situations: either no redshifts are available and we apply a simple CR--HR analysis, or all clusters do benefit from photometric redshifts and we use the more fruitful $z$--CR--HR analysis with bins of $\Delta z=0.03$. 
Corresponding results are shown on Fig.~\ref{fig_erosita} and quoted in Table~\ref{table_erosita}, for the dark energy parameters $(w_0,w_a)$ and for the parameters  governing the scaling-law evolution ($\zevolmt$, $\zevollt$).
We find that even without redshifts, the CR--HR method yields good constraints on the evolution of scaling laws, provided that the cosmological model is known at the accuracy expected from the Planck mission. The constraints on $w_0$ and $w_a$ are less informative, but could be enhanced by a joint study of the angular correlation function of the detected clusters. In particular, we notice that adding Planck priors to the analysis has a mild impact on the dark energy constraints, possibly because of the degeneracies within the scaling-relation parameters and the relatively high uncertainties on the count-rate measurements. 
Adding redshift information considerably improves the expected constraints on both sets of parameters. 
\begin{table}
	\centering
			\caption{\label{table_erosita} Expected marginalized constraints on dark energy parameters and parameters describing the  scaling-law evolution for the 20,000\,sq.deg. eRosita survey. In each case, the scaling relations are let free in the analysis (slope, normalization and scatter). Planck priors are applied to the five parameters $\Om$, $\sigma_8$, $\Ob$, $n_s$ and $h$.}
			
		\begin{tabular}{@{}lcccc@{}}
\hline
			&	\multicolumn{2}{c}{CR-HR}	&	\multicolumn{2}{c}{$z$-CR-HR}	\\
			&		No prior	&	Planck priors	&	No prior	& Planck priors	\\
\hline
$w_0$		&		0.6			&	0.4				&	0.1			&	0.1			\\
$w_a$		&		1.0			&	0.9				&	0.3			&	0.3			\\
$\zevolmt$	&		1.3			&	0.1				&	0.2			&	0.05		\\
$\zevollt$	&		0.8			&	0.5				&	0.3			&	0.1			\\
\hline
		\end{tabular}
\end{table}

%
%
\section{Conclusions and perspectives}
\label{conclusion}
We have presented the cosmological study of a sample of 347 clusters detected  in the full {\it XMM-Newton} archive  using X-ray criteria only (ancillary optical images were used to discard `extended' sources irrelevant for our analysis such as nearby galaxies, saturated point-sources etc...). The study relies on the sole   instrumental count-rates measured for each cluster in three X-ray bands.
The selection function of our sample has been thoroughly defined by means of extensive image simulations and we proposed a method to account for the presence of pointed clusters in the analysis.
We have then modeled the sample CR--HR diagram (whose properties are discussed in paper~I) by self-consistently including a $\Lambda$CDM cosmological model, X-ray scaling laws, selection effects and measurement errors. This allowed us to fit   $\Om$ and $\sigma_8$ along with the parametrized evolution of scaling laws plus a parameter $\xc$ characterizing the X-ray gas extent in clusters. We summarize below our main conclusions:
\begin{itemize}

\item When setting the cosmological parameters to their WMAP-5 values, we observe a preference for the `Non Cool Core' normalization and slope of the L--T relation of \citet{Pratt:2009p322}, if we assume an intrinsic scatter of $\scattlt \sim 0.7$. Our data then indicates a positive evolution of the M--T relation and a negative evolution of the L--T relation with respect to the self-similar expectation.
\item Fitting $\Om$ and $\sigma_8$ together with the evolution of scaling laws, we still find a preference for the `NCC' scaling law and find $\Om=0.24^{+0.04}_{-0.09}$ and $\sigma_8 = 0.88^{+0.10}_{-0.13}$, in agreement with the most recent cosmological studies. Again, the T--M and L--T scaling relations are found to evolve negatively with respect to the self-similar expectation.
\item Assuming the  `ALL' scaling relation, our data indicates $\Om \sim 0.15$ and $\sigma_8 \sim 0.96$ with a milder evolution of scaling laws.
\item The ad hoc parameter $\xc=r_c/\rccc$  giving the X-ray extent as a function of cluster mass is found to be well constrained within our framework, with a best value of 0.24, compatible with individual cluster studies.
\item The interpretation and use of the currently available local cluster scaling relations has proven one of the main hurdle of our study. Given that these relations do not agree with each other, it is probable that they have not been entirely corrected from the selection biases affecting the sample from which they are derived.
\item The scatter in the scaling relations plays an important role in the cosmological analysis and is probably degenerate with the slope and normalisation of the relations. It is likely that the scaling relations will  be reliably determined only  with very large cluster samples along with the simultaneous fit of cosmology and selection effects. We stress that the CR--HR method   is able to achieve this in a self-consistent manner, by-passing the tedious step that consists in determining individual cluster masses.  
\item As a logical follow-up of the present study, we propose predictions for the eRosita All-Sky survey. Assuming the Planck priors and letting all cluster scaling-law parameters free, we show that the z-CR-HR method will allow a determination of the equation of state of the dark energy at the level of  stage IV of the DETF \citep{Albrecht:2006p2502}.  In addition, the cluster scaling-law evolution will be well determined.
\item The X-CLASS serendipitous cluster catalogue extracted from the XMM archival data is available online at:  
http://xmm-lss.in2p3.fr:8080/l4sdb/
 \end{itemize}


\section*{Acknowledgements}
The authors thank the anonymous referee whose useful comments and suggestions improved the quality of this paper.
We acknowledge useful discussions with Gabriel Pratt, Kathy Romer, Martin Sahl\'en and Patrick Valageas.
The results presented here are based on observations obtained with XMM-Newton, an ESA science mission with instruments and contributions directly funded by ESA Member States and NASA.
The UK Schmidt Telescope was operated by the Royal Observatory Edinburgh, with funding from the UK Science and Engineering Research Council (later the UK Particle Physics and Astronomy Research Council), until 1988 June, and thereafter by the Anglo-Australian Observatory. The blue plates of the southern Sky Atlas and its Equatorial Extension (together known as the SERC-J), the near-IR plates (SERC-I), as well as the Equatorial Red (ER), and the Second Epoch [red] Survey (SES) were all taken with the UK Schmidt telescope at the AAO. 
This research has made use of the NASA/IPAC Extragalactic Database (NED) which is operated by the Jet Propulsion Laboratory, California Institute of Technology, under contract with the National Aeronautics and Space Administration.
The present study has been supported by a grant from the Centre National d'Etudes Spatiales.  TS acknowledges support from the ESA PRODEX Programme ``XMM-LSS", from the Belgian Federal Science Policy Office and from the Communaut\'e fran\c caise de Belgique - Actions de recherche concert\'ees. 
FP acknowledges support from Grant No. 50 OR 1003 of the
Deutsches Zemtrum f\"ur Luft- und Raumfahrt (DLR) and from the
Transregio Programme TR33 of the Deutsche
Forschungsgemeinschaft (DfG).


\appendix


\section[]{X-CLASS catalogue}
\label{catalogue_details}

		\subsection{Database}
The X-CLASS catalogue is accessible through a dedicated database at  http://xmm-lss.in2p3.fr:8080/l4sdb/

It contains 845 C1 clusters retained after data screening, in particular the 347 clusters used in the cosmological analysis. The public part of the database contains 422 clusters selected identically as the cosmological sample (C1$^{+}$, with high- and low-cuts in [0.5-2]\,keV count-rate and hardness-ratio, see Sect.~\ref{simulations}) but extended up to 13 arcmin off-axis distance (instead of 10 arcmin).
Redshifts were obtained from the NED and from the recent publication by \citet{Mehrtens:2011p4564} for the XCS survey.

\begin{figure}
	\includegraphics[width=84mm]{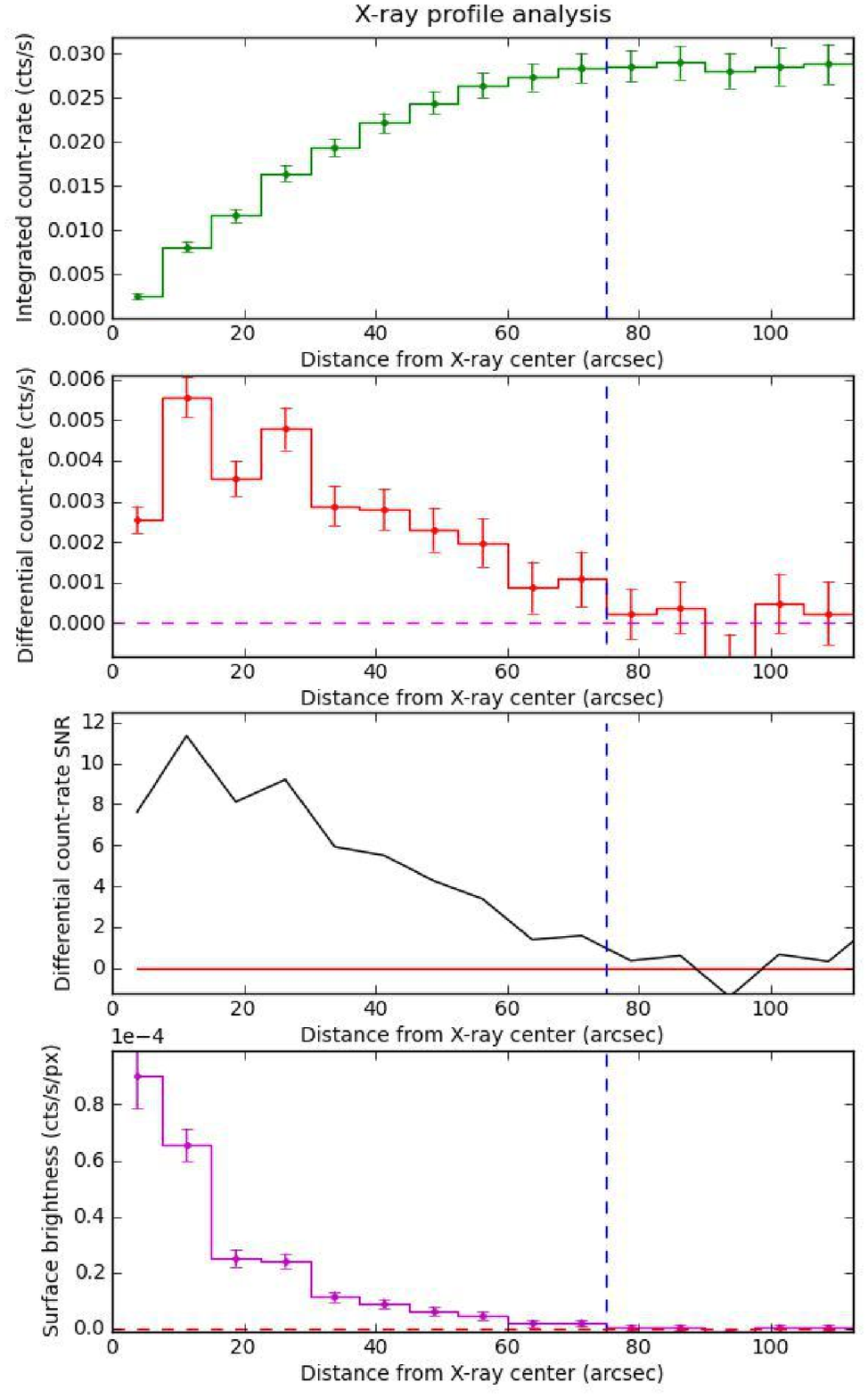}
 \caption{Screenshot of a webpage from the X-CLASS database (http://xmm-lss.in2p3.fr:8080/l4sdb/), for a particular cluster (tag 2094). Displayed are the cumulative, background-subtracted, count-rate profile ({\it top}) the differential count-rate profile ({\it 2nd panel}), the corresponding signal-to-noise curve ({\it 3rd panel}) and the surface brightness profile ({\it bottom}). The blue vertical lines shows the radius in which the measurement is performed.}
 \label{fig_sample_profil} 
\end{figure}

\begin{figure*}
	\includegraphics[width=\linewidth]{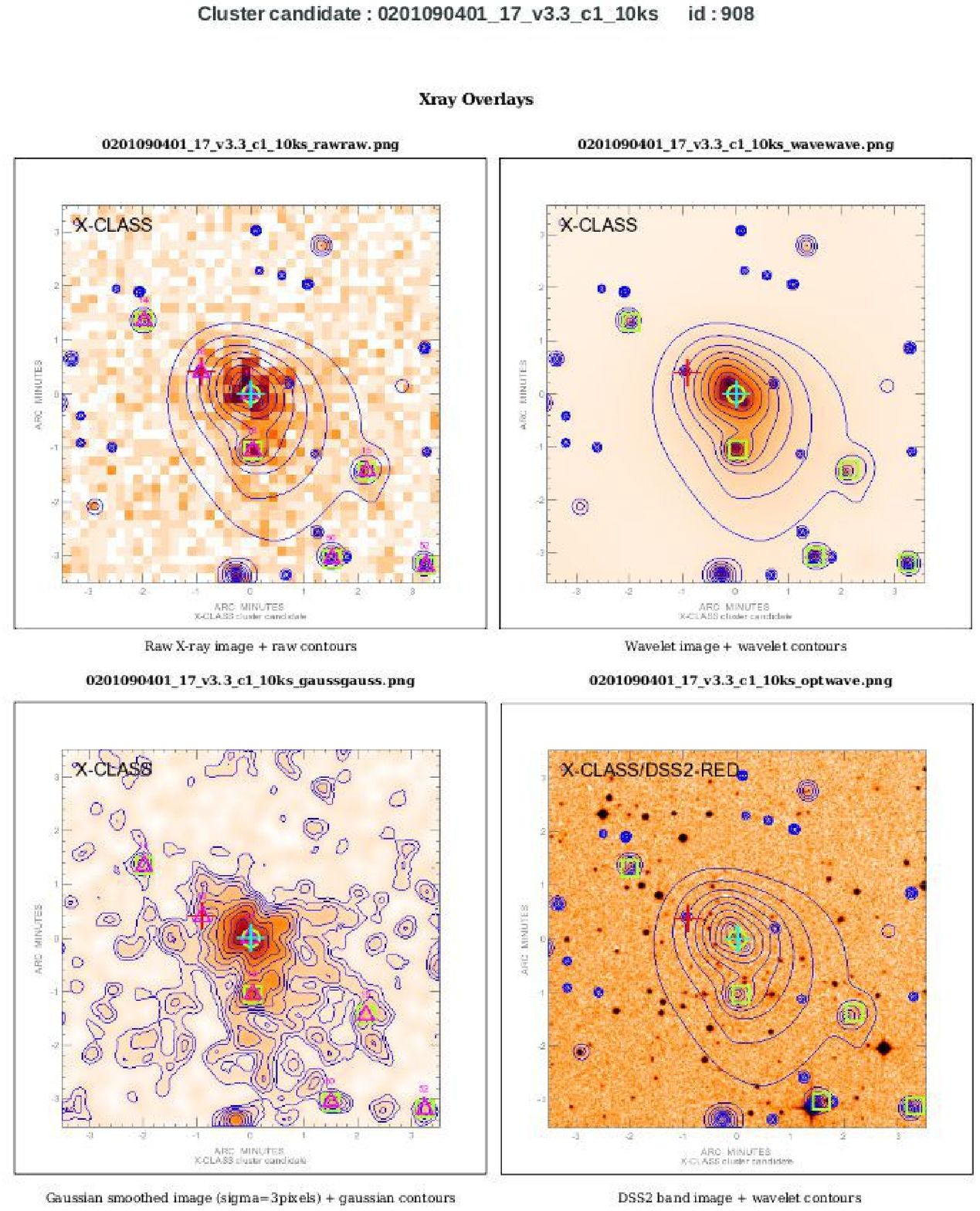}
 \caption{Screenshot of a webpage from the X-CLASS database (http://xmm-lss.in2p3.fr:8080/l4sdb/), for a particular cluster (tag 908). Displayed are the X-ray photon image ({\it top-left}) with associated contours, the wavelet-filtered ({\it top-right}) and gaussian-filtered ({\it bottom-left}) images and the optical (POSS-II) overlay ({\it bottom-right}).}
 \label{fig_sample_webpage} 
\end{figure*}

We describe below the informations contained in the database:
\begin{itemize}
\item {\bf Object name: }each cluster is referenced with a unique identifier (``tag''), a full name in the XMM-Newton format (XMMUJ) and a name output of the X-ray pipeline. The cluster name encloses the corresponding XMM ObsId (ex: 0502430101) and truncated exposure time (10ks or 20ks) at which it has been detected.
\item {\bf Object position: }the position of the cluster (right ascension and declination) as provided by the pipeline is given in addition to the position measured by hand in the course of the cluster count-rate measurement (Sect.~\ref{fluxmes}).
\item {\bf NED identifications: }column `NED' lists all sources (galaxies, galaxy clusters, groups, QSO, etc.) within 3 arcmin of the cluster centre having a redshift indication from the NED (photometric or spectroscopic). An illustrative example is given in Table~\ref{table_webNED}.
\item {\bf Redshift: }when a redshift indication is available, we provide a flag describing the current status of the redshift determination: `confirmed' for a cluster definitely confirmed, `tentative' if less than 3 concordant redshifts within 3 arcmin are available and `photometric'. Redshifts are quoted from NED first, then from the XCS-DR1 taking into account the provided flags.
\item {\bf X-ray properties: }basic X-ray properties output of the XAmin pipeline (Sect.~\ref{data_processing}) are given for each cluster, in the [0.5-2]\,keV detection band: number of counts, total count-rate, apparent Extent and Extent Likelihood as well as the distance to the centre of the pointing it belongs to (off-axis).
\item {\bf Count-rates: }count-rates measured manually in several energy bands are also available, in particular for the [0.5-2]\,keV band. In any case, count-rates are specified `on-axis', i.e. corrected from the local exposure map, and do not include the filter and aperture corrections as discussed in Sect.~\ref{fluxmes}. A webpage shows for each cluster the profiles generated for the count-rate measurement (see Fig.~\ref{fig_sample_profil} for a particular example).
\item {\bf Cluster images: }X-ray photon images, filtered images and optical cut-outs from the Palomar Observatory Sky Survey (POSS-II, \citealt{Reid:1991p6434}) have been produced and linked to each catalogue entry. A screenshot image is shown on Fig.~\ref{fig_sample_webpage}.
\item {\bf Database interface: }the electronic catalogue can be sorted according to any of entry of the database and can be downloaded as a machine-readable file.
\end{itemize}

\begin{table*}
			\caption{\label{table_webNED} Sample table attached to the `NED' column in the database, for a particular cluster (tag 1758). This table lists all objects from the Nasa Extragalactic Database (April 2011) within 3 arcmin of the cluster center, with an associated redshift. Values in the table are directly copied from the NED and originate from various surveys/follow-ups. In particular, the redshift accuracy is highly inhomogeneous and in some cases a flag indicates its reliability. The velocity of each object is expressed in km/s. The distance is quoted relative to the cluster centre (units arcmin).}
		\begin{tabular}{@{}llccrrcccc@{}}
\hline
N 	& 	Name 	& 	R.A. 	& 	Dec 	& 	Type 	& 	Velocity 	& 	Redshift 	& 	Z flag 	& 	Magnitude 	& 	Distance	\\
\hline
1 	& 	2MASX J10531862+5720438 	& 	163.32800 	& 	57.34570 	& 	G 		& 	101956 	& 	0.34009 	&		 	& 20.3g		& 	0.0	\\
2 	& 	SL J1053.4+5720 			& 	163.32700 	& 	57.34640 	& 	GClstr 	& 	101929 	& 	0.34 		& 			&			&	0.1	\\
3 	& 	SHADES J105319+572110 		& 	163.33000 	& 	57.35280 	& 	G 		& 	779460 	& 	2.6 		&	PHOT 	& 			&	0.4	\\
4 	& 	1EX 179 					& 	163.32899 	& 	57.36120 	& 	G 		& 	39872 	& 	0.133 		&		 	& 	17.1R 	&	1.0	\\
5 	& 	SDSS J105318.96+572140.5 	& 	163.32899 	& 	57.36130 	& 	G 		& 	39911 	& 	0.133 		&			& 	17.9g 	& 	1.0	\\
6 	& 	[MBC2005] 0086 				& 	163.28999 	& 	57.34990 	& 	G 		& 	1025290 & 	3.42 		&			& 			&	1.3	\\
7 	& 	SDSS J105329.42+572104.2 	& 	163.37300 	& 	57.35120 	& 	QSO 	& 	343262 	& 	1.145 		& 			& 	21.3	& 	1.5	\\
8 	& 	SDSS J105319.03+571851.8 	& 	163.32899 	& 	57.31440 	& 	G 		& 	213152 	& 	0.711 		&			& 	21.3g 	& 	1.9	\\
9 	& 	SDSS J105328.81+572205.1 	& 	163.37000 	& 	57.36810 	& 	G 		& 	232639 	& 	0.776 		&			& 	23.5g 	& 	1.9	\\
10 	& 	SDSS J105319.99+572251.1 	& 	163.33299 	& 	57.38090 	& 	G 		& 	158290 	& 	0.528 		& 			& 	22.6g	&	2.1	\\
11 	& 	[ZMF2005] 007 				& 	163.37601 	& 	57.37820 	& 	G 		& 	258721 	& 	0.863 		&			& 			&	2.5	\\
12 	& 	SDSS J105311.65+572305.6 	& 	163.29900 	& 	57.38490 	& 	G 		& 	462880 	& 	1.544 		& 			& 	23.2 	&	2.6	\\
13 	& 	SDSS J105330.86+572247.6 	& 	163.37900 	& 	57.37990 	& 	G 		& 	190068 	& 	0.634 		& 			& 	23.3g	&	2.6	\\
14 	& 	[ZMF2005] 006			 	& 	163.38800 	& 	57.37640 	& 	G 		& 	220347 	& 	0.735 		&			& 			&	2.7	\\
15 	& 	[ZMF2005] 005 				& 	163.40199 	& 	57.37150 	& 	G 		& 	144500 	& 	0.482 		&			& 			&	2.8	\\
16 	& 	Bolocam LE 1100.01 			& 	163.23801 	& 	57.35080 	& 	G 		& 	929357 	& 	3.1 		& 	PHOT 	& 	26.14 	& 	3.0	\\
\hline
\end{tabular}
\end{table*}

	\subsection{Comparison to the XCS survey}
\begin{figure*}
	\includegraphics[width=\linewidth]{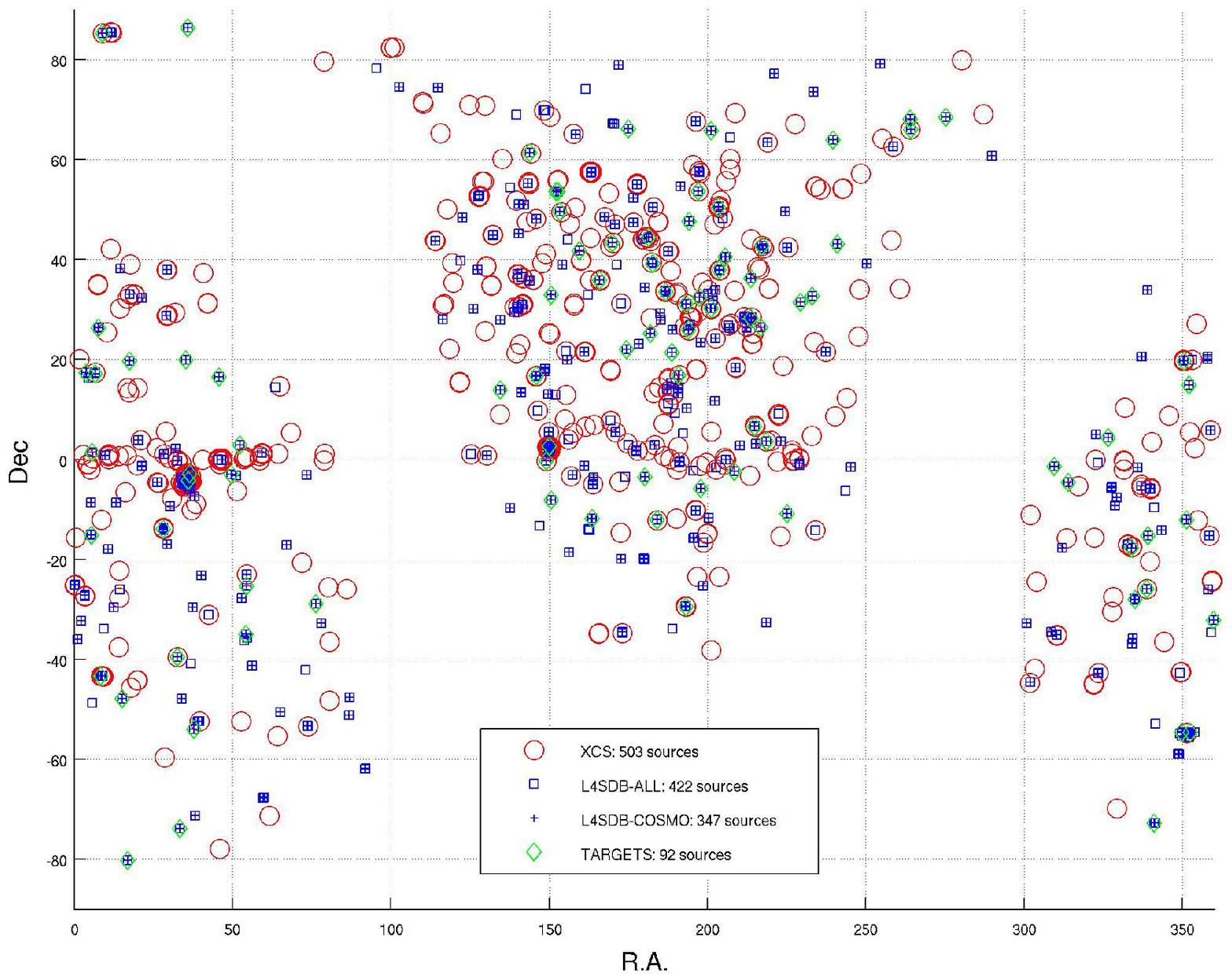}
 \caption{Comparison between the X--CLASS and the XCS-DR1 \citep{Mehrtens:2011p4564} catalogues. Blue crosses stand for the cluster sample used in the cosmological analysis, blue squares for the ``extended" cosmological sample (i.e. up to 13 arcmin off-axis distance on the XMM detectors). Green diamonds show the 92 sources of our sample located at the centre of the XMM field of view (i.e. less than 3 arcmin off-axis distance).}
 \label{fig_cluster_xcs} 
\end{figure*}

We compared our catalogue to the first release of the XCS survey XCS-DR1 \citep{Mehrtens:2011p4564} based on the analysis of all publicly available data in the XMM archive. For this comparison, we included our 347 clusters selected for the cosmological analysis (with off-axis distance below 10 arcmin) plus 75 clusters selected upon identical criteria, but with off-axis distances between 10 and 13 arcmin (thus 422 clusters in total).
The XCS-DR1 sample is composed of 503 clusters which are optically confirmed and detected with more than 300 counts in the [0.5-2]\,keV band. Clusters identified as targets of a particular XMM observation are not included in the XCS-DR1.
Fig.~\ref{fig_cluster_xcs} illustrates the comparison between the two samples. Out of our 422 clusters, 159 are new discoveries (i.e. not in the XCS-DR1 and without information from NED).
The differences between the two catalogues can be attributed to the different pointing selection  and to differences in the X--ray detection algorithms. A further comparison between the two analyses will provide useful insights into the different selection effects and possible systematics contained in both samples.

\section[]{Bias correction for pointed observations}
	\label{bias_details}

We detail in this Appendix our method for correcting from the presence of pointed clusters in the XMM archive.
As shown on Figures~\ref{fig_surface_density} and~\ref{fig_obs_bias}, more clusters are detected in the inner 5 arcmin   than expected from solely considering the sensitivity gradient on the detectors.
This sensitivity difference is   partly due to the EPIC vignetting function (loss of $\sim$\,60\% in effective area at 10\,arcmin off-axis ("XMM-Newton Users Handbook", Issue 2.9, 2011 (ESA: XMM-Newton SOC)) compared to the centre). The degradation of the telescope PSF at larger off-axis distances is the second most important cause of sensitivity variation as it dilutes the signal of faint sources and distorts their shape.

	\subsection{Bias model}
We want to correct the CR--HR distribution from the bias due to pointed clusters, directly on the predicted CR--HR distribution. We assume that this correction does not depend on HR and divide the count-rate distribution in several bins indexed by $j$. We call $n_j$ the underlying cluster surface density, i.e.~the value obtained after integrating the true cluster log($N$)-log(CR) in bin $j$. We introduce the survey selection function through the factor $\epsilon_{j}$ (comprised between 0 and 1) such that the net number of clusters detected in bin $j$ for an arbitrary region covering $A= \Omega\,f_{sky}$~sr. on sky is:
\begin{equation}
\label{equ_ntot_nobias}
N^{\rm tot}_j = A \, \epsilon_j \, n_j
\end{equation}
Here the sky area is $\Omega = 4\pi (1-\cos 70^{\circ})$ as we exclude the galactic plane ($\pm 20^{\circ}$) from the analysis.

We now divide the survey in two sub-surveys labelled 'in' and 'out'. The first one gathers all regions belonging to the inner [0-5]\,arcmin  and the latter corresponds to the [5-10]\,arcmin regions of the same pointings. The total area covered by the survey reads $A=A^{\rm in}+A^{\rm out}$.
Introducing $\mu^{\rm in}_{j}$ ($\geq 1$) the sensitivity of the inner sub-survey relative to the total survey we write (still for an unbiased sample):
\[
N^{\rm in}_j = A^{\rm in} \, \epsilon_j \, \mu_j^{\rm in} \, n_j
\]

We now consider that an unknown fraction $f_j$ of all existing clusters on sky has been observed and pointed in the central region of the detectors, thus augmenting $N^{\rm in}_j$ by a quantity:
\[
N^{\rm in, pointed}_j = \Omega \, n_j \, f_j \, \epsilon_j \, \mu_j^{\rm in}
\]

Because those clusters have been removed from the sky population of clusters, the remaining density on sky is $n_j^{\prime} = (1-f_j)\,n_j$ and the survey provides a total number of clusters:

\begin{eqnarray}\nonumber
\label{equ_ntot_bias}
N^{\rm tot, biased}_j &=& N^{\rm in, pointed}_j + A\,\epsilon_j\,n_j^{\prime} \\
\nonumber
	&=& \Omega \, n_j \, f_j \, \epsilon_j \, \mu_j^{\rm in} + A\,\epsilon_j\,(1-f_j) \, n_j \\
	&=& \left(\frac{1}{f_{sky}}\,f_j\,\mu_j^{\rm in} + (1-f_j) \right) N_{j}^{\rm tot}
\end{eqnarray}

Comparing equations~\ref{equ_ntot_nobias} and~\ref{equ_ntot_bias} we obtain the bias factor:
\begin{equation}
\label{equ_bias_factor}
F_j = N_{j}^{\rm tot, biased} / N_{j}^{\rm tot}= \frac{1}{f_{sky}}\,f_j\,\mu_j^{\rm in} + (1-f_j)
\end{equation}

	\subsection{Bias estimation}
Following previous equations, the expected number of clusters in the 'in' survey writes:
\begin{eqnarray}\nonumber
\label{equ_nin_bias}
N_{j}^{\rm in, biased} &=& \Omega \, n_j \, f_j \, \epsilon_j \, \mu_j^{\rm in} + A^{\rm in} \, \epsilon_j \, \mu_j^{\rm in} \, n_j^{\prime} \\
	&=& \left[ \frac{1}{f_{sky}} \, f_j \, + (1-f_j) \frac{A^{\rm in}}{A} \right] \, \mu_j^{\rm in} \, N_{j}^{\rm tot}
\end{eqnarray}
while the expected number of clusters in the 'out' survey is:
\begin{equation}
\label{equ_nout_bias}
N_{j}^{\rm out,biased} = (1-f_j) \left[1 - \mu_j^{\rm in}\,\frac{A^{\rm in}}{A} \right] N_{j}^{\rm tot}
\end{equation}

At this point, only two quantities are unknown, $f_j$ and $N_{j}^{\rm tot}$.
Parameters $A$ and $A^{\rm in}$ directly come from the survey geometric design.
The factor  $\mu_j^{\rm in}$ is obtained by comparing the results of the simulations (Sect.~\ref{simulations}) for clusters in the full [0-10]\,arcmin off-axis area and for clusters in the central [0-5]\,arcmin region.

We compute independently in each bin the joint likelihood for the observed (biased) quantities $\widehat{N_{j}}^{\rm out}$ and $\widehat{N_{j}}^{\rm in}$ on a fine bidimensional grid sampling values for $f_j$ and $N_{j}^{\rm tot}$.
We then compute the marginalized probability distribution $P(f_j)$ assuming flat priors for $N_{j}^{\rm tot}$ by numerical integration of the sampled likelihood.

The expectation value and variance for $F_j$ are finally derived by integrating Eq.~\ref{equ_bias_factor} against $P(f_j)$ and are displayed on Figure~\ref{fig_obs_bias_result} for the present sample. The predicted, unbiased CR--HR distribution is multiplied by the expectation value of $F_j$ linearly interpolated at each CR value.

This model relies on the hypothesis that all pointed clusters are centered onto the detectors. It thus neglects spatial correlation effects which may artificially boost the number of clusters in the surroundings of pointed clusters, in particular in the outer parts of the detectors.

\bsp

\label{lastpage}

\end{document}